%% file: main.tex
\documentclass[%
 reprint,
 amsmath,amssymb,
 aps,
floatfix,
]{revtex4-1}
\usepackage{float}

\usepackage{graphicx}
\usepackage{dcolumn}
\usepackage{bm}
\usepackage{color}
\usepackage{ulem}
\usepackage{mdframed}
\usepackage{framed,color}
\usepackage{threeparttable}
\usepackage{longtable}
\usepackage{afterpage}
\usepackage{multirow}
\usepackage{siunitx}
\usepackage{comment}
\usepackage{amsmath}
\usepackage[T1]{fontenc}
\AtBeginDocument{\usepackage{booktabs}}
\usepackage[colorlinks = true,
linkcolor = blue,
urlcolor  = blue,
citecolor = blue,
anchorcolor = blue]{hyperref}
\definecolor{shadecolor}{rgb}{1,0.8,0.3}
\usepackage{hyperref}

\DeclareUnicodeCharacter{2212}{-}
\DeclareUnicodeCharacter{0088}{-}
\DeclareUnicodeCharacter{0092}{-}
\begin{document}


\title{Computing and Memory Technologies based on Magnetic Skyrmions} 
\author{Hamed Vakili}
 \affiliation{Department of Physics, University of Virginia, Charlottesville VA 22904 USA}
 \author{Md Golam Morshed}
 \affiliation{Department of Electrical and Computer Engineering, University of Virginia, Charlottesville VA 22904 USA}
 \author{Timothy Q. Hartnett}
 \affiliation{Department of Materials Science and Engineering, University of Virginia, Charlottesville VA 22904 USA}
 \author{Wei Zhou}
 \affiliation{Department of Physics, University of Virginia, Charlottesville VA 22904 USA}
 \author{Mohammad Nazmus Sakib}
 \affiliation{Department of Electrical and Computer Engineering, University of Virginia, Charlottesville VA 22904 USA}
  \author{Samiran Ganguly}
 \affiliation{Department of Electrical and Computer Engineering, University of Virginia, Charlottesville VA 22904 USA}
 \author{Chung T. Ma}
 \affiliation{Department of Physics, University of Virginia, Charlottesville VA 22904 USA}
  \author{Jun-Wen Xu}
 \affiliation{Center for Quantum Phenomena, Department of Physics, New York University, New York, NY 10003 USA}
 \author{Yassine Quessab}
 \affiliation{Center for Quantum Phenomena, Department of Physics, New York University, New York, NY 10003 USA}
 \author{Kai Litzius}
 \affiliation{Center for Materials Science and Engineering, Massachusetts Institute of Technology, Cambridge MA 02139 USA}
 \author{S. J. Poon}
 \affiliation{Department of Physics, University of Virginia, Charlottesville VA 22904 USA}
 \author{Mircea Stan}
 \affiliation{Department of Electrical and Computer Engineering, University of Virginia, Charlottesville VA 22904 USA}
 \author{Prasanna Balachandran}
 \affiliation{Department of Materials Science and Engineering, University of Virginia, Charlottesville VA 22904 USA}
 \author{Andrew D. Kent}
 \affiliation{Center for Quantum Phenomena, Department of Physics, New York University, New York, NY 10003 USA}
 \author{Geoffrey Beach}
 \affiliation{Center for Materials Science and Engineering, Massachusetts Institute of Technology, Cambridge MA 02139 USA}
 \author{Avik W. Ghosh}
 \affiliation{Department of Electrical and Computer Engineering, University of Virginia, Charlottesville VA 22904 USA}
 \affiliation{Department of Physics, University of Virginia, Charlottesville VA 22904 USA}
\begin{abstract}
\textcolor{black}{Solitonic magnetic excitations such as domain walls and,  specifically, skyrmionics enable the possibility of compact, high density, ultrafast, all-electronic, low-energy devices, which is the basis for the emerging area of skyrmionics. The topological winding of skyrmion spins affects their overall lifetime, energetics, and dynamical behavior. In this review, we discuss skyrmionics in the context of the present-day solid-state memory landscape and show how their size, stability, and mobility can be controlled by material engineering, as well as how they can be nucleated and detected. Ferrimagnets near their compensation points are promising candidates for this application, leading to a detailed exploration of amorphous CoGd as well as the study of emergent materials such as Mn$_4$N and Inverse Heusler alloys.
Along with material properties, geometrical parameters such as film thickness, defect density, and notches can be used to tune skyrmion properties, such as their size and stability. Topology, however, can be a double-edged sword, especially for isolated metastable skyrmions, as it brings stability at the cost of additional damping and deflective Magnus forces compared to domain walls. Skyrmion deformation in response to forces also makes them intrinsically slower than domain walls. We explore potential analog applications of skyrmions, including temporal memory at low density - one skyrmion per racetrack - that capitalizes on their near ballistic current-velocity relation to map temporal data to spatial data and decorrelators for stochastic computing at a higher density that capitalizes on their interactions. We summarize the main challenges of achieving a  skyrmionics technology, including maintaining positional stability with very high accuracy and electrical readout, especially for small ferrimagnetic skyrmions, deterministic nucleation, and annihilation and overall integration with digital circuits with the associated circuit overhead.
}
\end{abstract}
{\maketitle}
\tableofcontents
\section{The Memory Landscape}
\subsection{Introduction: The Role of Nonvolatile Magnetic Memory }
Digital electronics has been driven for several decades by sustained hardware scaling and Moore's law. With the recent slowdown in advances in Complementary Metal Oxide Semiconductor (CMOS) hardware and rapid growth of software, \textcolor{black}{along with the migration from cloud computing towards edge devices}, there is a strong impetus to re-examine the limits of computing. In conventional Von Neumann computer architecture, the processor needs to access data stored in separate memory banks in order to perform logic operations. Unfortunately, the speed of memory scaling (1.1$\times$ in every two years) has not kept up with the drastically increased speed of processor scaling (2$\times$  in every two years). This increasing gap between memory and processor performance, the so-called {\it{memory wall problem}}, is considered one of the main bottlenecks to increasing computer performance.

To mitigate the memory wall problem, a multi-level memory hierarchy is used, with the most frequently invoked instructions and data sets stored in a high-speed on-chip cache so that they can be accessed and executed efficiently by the processor. On-chip cache memory is typically built out of static random access memory (SRAM). However, SRAM is expensive as it requires at least six transistors {per bit} (two sets of CMOS inverter pairs to form a memory latch plus two access transistors) in order to store binary data with acceptable reliability. The need for so many transistors per cell also means that the capacity of the on-chip cache cannot be too large, requiring a large degree of off-chip access to dynamic random access memory (DRAM). Each DRAM  cell consists of single capacitor-transistor pairs and is significantly smaller than an SRAM cell, but it is slow and also needs a regular refresh to replenish data that leaks from the capacitor. Consequently, the number of clock cycles to access data from the off-chip DRAM also increases, resulting in overall energy inefficiency. A lot of active research is thus focused on developing a memory technology that can match the speed performance of SRAM and the high density of DRAM. Magnet-based nonvolatile memory is an emerging candidate addressing that technology bottleneck. 

Magnet-based non-volatile memory has a rich history, with field-switched magnetic RAMs (MRAMs) and spin-transfer torque based RAMs (STT-RAMs) now commercialized~\cite{Kent2015,Thomas2017,Slaughter2017}. In magnetic memory, the different directions of magnetization can be used to store data as digital bits (e.g., up equals one, down is zero). \textcolor{black}{Magnetic states do not leak away like charges (i.e., magnetic states can be non-volatile).} 
The energy required to switch a magnet is fairly low, mainly because internal exchange forces make the magnet act as a single giant spin. However, energy is dissipated in the overhead/control circuitry due to the need for currents generated to create the switching fields. Field switched MRAMs are hard to scale because the scaling of the drive circuit requires increasing current densities with increasing energy dissipation. STT-RAMs have better-scaling properties, as the switching is due to spin-polarized currents directly injected into a magnet across the oxide of a magnetic tunnel junction (MTJ). 

STT-RAM has garnered a lot of attention both from academia and industry due to its compatibility with CMOS processes and voltages, zero standby leakage (non-volatility), scalability, high endurance and retention time, and overall reliability. In fact, STT-RAM has evolved from the discovery of giant magnetoresistance in the late 1980s~\cite{fert1988,Grunberg1988} and theoretical ideas soon after~\cite{Slonczewski1989,slonczewski_current-driven_1996,Berger1996}, to a dark horse in the 2000s~\cite{kiselev_microwave_2003,gilbert} to becoming now a commercially viable candidate for on-chip cache memory~\cite{45nmstt}, with both the integration density of DRAM and comparable performance to SRAM\cite{Shao2021}. In particular, its density advantage compared to volatile transistor-based SRAM can improve the on-chip cache capacity significantly. Even if just the last level cache capacity in the multi-level memory hierarchy increases, this reduces the number of off-chip DRAM accesses. In brief, there is significant motivation to develop a high-capacity non-volatile on-chip last-level cache.  

One of the significant challenges with the widespread adoption of STT-RAMs as a universal memory technology is the high write energy and low read speed of STT-RAM cache memory relative to CMOS on-chip cache. It is worth mentioning that the write latency and energy of STT-RAM can be decreased by trading against the retention time, the energy barrier between binary states. Longer access times, along with the high error rates, pose other challenges in using STT-RAM for on-chip cache (in short, there is an overall energy-delay-error trade-off) \cite{xie_materials_2017}. The error rate of STT-RAM can be mitigated by adopting intelligent array-level design techniques; however, they do not maximize STT-RAM benefits. The error rate problem gets exacerbated, and the robustness of STT-RAM suffers while it is scaled down, requiring stronger error correction schemes and a compromised energy efficiency benefit. Requirements like bidirectional write currents, asymmetry in the critical write currents, single-ended sensing operation, and shared current paths for read and write impose further challenges in establishing STT-RAMs as a replacement of last level cache.  

Part of the energy hungriness of STT-RAMs arises from the writing of information, which requires passing current through a high resistance tunnel barrier. The same current path is used to read, but much lower currents are used.
Structures with orthogonal read-write paths provide a possible way out of this impasse---a metallic path being used to write information separate from the read path (which can again be via a magnetic tunnel junction). \textcolor{black}{More details on the target specifications of SOT and STT RAMs is discussed in \cite{Shao2021}}

This brings us to the evaluation of bits encoded by topological (solitonic) excitations in thin magnetic films, such as line-like domain walls and circular skyrmions. Such excitations can encode information in ultrasmall volumes below the thermal superparamagnetic limit that constrains regular magnets. The solitons can be driven at high speeds along with magnetic racetracks by modest currents and energy dissipation in heavy-metal underlayers, generating entirely unique device applications. \textcolor{black}{The high density of ultrasmall skyrmions stabilized by their topological barriers, as well as their quasi-ballistic, tunable, and linear dynamics, are particular attributes that make skyrmionic devices potentially useful in a variety of applications.} 

The purpose of this review is to go over the physics of isolated skyrmions---the factors controlling their size, dynamics, lifetime, and switching, the material classes that support them, and how they may potentially be utilized for conventional or unconventional computing applications. 

The structure of this paper is as follows: In Section I, we go over the memory landscape; in Sections II and III, we describe the fundamental physics \textcolor{black}{based and material based limitations and parameter dependences of skyrmions and domain walls, along with experimental characterization}. In Section IV, we present skyrmion device applications. Lastly, in Section V, we discuss the challenges and opportunities with select skyrmion devices.

\begin{table*}[htp]
\centering
\caption{Comparison of proposed domain wall-based racetrack memory (RTM) \cite{Parkin190} with other memory technologies, including spin-transfer torque-based random access memory (STT-RAM), resistive random access memory (RRAM), phase-change memory (PCM), and magnetic random access memory (MRAM). {DW-RTM has lower predicted energy consumption, leakage power, and size, but DWs are typically limited by pinning at the racetrack edges, where skyrmions are purported to have a distinct advantage~\cite{parkin_review}. R. Bläsing et al., Proceedings of the IEEE, vol. 108, no. 8, pp. 1303-1321, Aug. 2020; licensed under a Creative Commons Attribution (CC BY) license.
}} 
\label{bias}
\begin{center}       
\begin{tabular}{p{1.7cm}|p{2cm} p{1.5cm} p{1.8cm} p{1.5cm} p{1.5cm} p{1.5cm} p{1.5cm} p{1.5cm}}
\toprule
 Technology & Write/Erase Times (ns)& Write Energy & Read\newline Times (ns) & Read Energy & Leakage Power & Endurance & Non-volatility & Cell\newline Size ($F^2$)\\
\hline \hline
 RTM & 3-250$^*$& Low & 3-250$^*$ & Low & Low & $\geq$ 10$^{16}$  & Yes & $\leq$ 2 \\
\hline
 RRAM & 20 & High & 10-20 & Low & Low & 10$^{11}$  & Yes & 4-10  \\
\hline
 PCM & >30 & High & 5-20 & Medium & Low & 10$^{9}$ & Yes & 4-12  \\
\hline 
 MRAM & 10-20 & High & 3-20 & Low & Low & 10$^{12}$ & Yes & 10-60  \\
\hline 
 STT-RAM & 3-15 & High & 3-15 & Low & Low & 4$\times10^{12}$ & Yes & 6-50 \\
\bottomrule 
\end{tabular}
\end{center}
\end{table*}

\subsection{Energy-Delay-Error Trade-off: Skyrmions as small stable mobile magnets} One of the performance metrics of any binary switch is its energy-delay product, arguing for schemes that lower both their power dissipation and overall latency. The energy-delay product can be written as $I^2R_\mathrm{on}t^2 = Q^2R_\mathrm{on}$ with $t=Q/I$, where $Q$ is the amount of charge needed to switch from off to on and $R_\mathrm{on}$ is the on-state resistance. The charge $Q$ needed to switch a small magnet is set by (a) the minimum critical charge $Q_c$ needed to switch magnetization, mandated by angular momentum conservation, and (b) an added overdrive $Q/Q_c \gg 1$ required to reduce the dynamic write error rate (WER) or switching time to application-specific targets. This charge gets smaller for small volume magnetic excitations, \textcolor{black}{but so does the energy barrier that sets the thermal stability. Technologies such as Heat Assisted Magnetic Recording (HAMR) and Bit Patterned Recording attempt to address this specific challenge. HAMR reduces the anisotropy barrier quickly through spot heating over a localized size and then allowing it to restore through fast cooling. Bit Patterned Recording breaks the magnet into tinier lithographically patterned ordered grains that can individually switch quickly but gain volume stability through strong inter-grain exchange interactions. A third way to get small stable magnetic excitations is by nucleating metastable} 
skyrmions, \textcolor{black}{essentially tiny mobile magnets,} held together against thermal fluctuations by their topological properties.

Let us start with conventional spin-transfer torque (STT), where a current injected across the oxide of a magnetic tunnel junction (MTJ) gets polarized by a fixed magnet, and then the spins enter a free magnet and start precessing incoherently around its magnetization until their transverse component gets fully absorbed. The transferred angular momentum generates an anti-damping torque that rotates the free magnetization by acting directly against damping.
We can work out the dynamics of these spins using the Landau-Lifschitz-Gilbert-Slonczewski (LLGS) \cite{slonczewski_current-driven_1996,gilbert} equation with a stochastic thermal torque, or equivalently a Fokker-Planck equation for the probability distribution function of the magnetization \cite{fpe,xie_materials_2017}. For a small magnet with perpendicular magnetic anisotropy in an MTJ, we can get the time-dependent WER from
an analytical solution to the 1-D Fokker Planck equation. For large overdrive current $i = I/I_c \gg 1$,  this works out to be \cite{munira_quasi-analytical_2012}
\begin{equation}
\mathrm{WER}_\mathrm{STT}(t) 
\approx  \frac{\pi^2\Delta_\mathrm{th}}{4}e^{\displaystyle -2Q/Q_c}, ~~~Q = It,
\end{equation}
where the critical current $I_c$ is set in turn by a critical charge that satisfies angular momentum conservation
between electron and flipped magnetization
\begin{eqnarray}
\underbrace{\Biggl(\displaystyle\frac{Q_c}{q}\Biggr)}_{\displaystyle N = n\Omega}\frac{\hbar}{2}P \gamma = 
M_s\Omega(1+\alpha^2)  \nonumber
\end{eqnarray}
where $P$ is the spin polarization of the current, $n$ is the electron density, $q$ is the charge of an electron, $\gamma = g\mu_B/\hbar$ is the electron gyromagnetic ratio, $\hbar$ is the reduced Planck's constant, $g$ is the electron dimensionless magnetic moment {or g-factor}, $\mu_B$ is the Bohr magneton, $\mu_0$ is the permeability of free space, $M_s$ is the saturation magnetization density, $\alpha$ is Gilbert damping, $\Omega$ is the volume of the flipped magnetic layer and $\Delta_\mathrm{th}$ is the thermal stability factor, $\Delta_\mathrm{th}=E_B/(k_BT)$, the energy barrier $E_B$ for magnetization reversal divided by the thermal energy, Boltzmann's constant $k_B$ times the temperature $T$. The critical current density is given by \cite{SUN1999157,Jsun}:
\begin{equation}
j^{STT}_c = \displaystyle\frac{2q}{\hbar}\frac{\alpha}{P}  \mu_0 M_s H_kt_\mathrm{F},
\end{equation} 
where $t_\mathrm{F}$ is the magnetic layer thickness, and $H_k$ is the anisotropy field. The interesting observation is that the {critical switching time}, set by the ferromagnetic resonance frequency and the damping $\tau_D = (1+\alpha^2)/(\alpha\gamma \mu_0 H_k)$, increases as we reduce the damping. However, the critical current $I_c= Q_c/\tau_D = 2q\alpha E_B/(\hbar P)$ decreases, such that $Q_c$ is independent of anisotropy and weakly dependent on damping, and is {\it{set by angular momentum conservation alone}}. Note also that the critical spin current (angular momentum per second), can be written simply as $I_s = (I_c/q) \times P\hbar/2 = \alpha E_B$.

The equations above set the charge $Q$ required for writing information at a given speed and error rate. The demand for a low error rate, $\sim 10^{-9}$ for memory and $\sim 10^{-15}$ for logic, will require a significant overdrive $Q \gg Q_c$. 
In a $100\times 20\times 1$ nm$^3$
Fe magnet of saturation magnetization $\mu_0 M_s \approx 1.9$ T  with about $10^4$ spins, assuming a Gilbert damping of $\alpha = 0.01$
and polarization $P = 0.7$, we get $Q_c/q$ about $5 \times 10^5$ electrons for destabilization. 
Assuming an anisotropy energy $\sim 55$ kJ/m$^3$, 
we get an energy barrier $E_b \approx 30 k_BT$, which implies an overdrive with $Q/q \approx 6.5 \times 10^6$ electrons at an error rate of $10^{-9}$. At $1$ ns switching time, the current density $\approx 50$~MA/cm$^2$. 

While the critical current is set by the magnet's energy barrier $E_B$ and the charge by angular momentum conservation, the dissipated energy $E_\mathrm{diss} = I^2R_\mathrm{on}t$ is several orders of magnitude higher. This is because the voltage needed to produce the necessary current is set by the resistance of the oxide tunnel barrier $R_\mathrm{on}$, whose energy barrier is much larger than that of the switching magnet.  

The energy dissipated can be reduced in all metallic devices such as proposed race track memories (Table I), where a spin current is injected into a metallic magnet from a heavy metal underlayer and the magnetization state read with a vertical current (e.g., a current passing through a tunnel barrier between the magnet and a fixed magnet with much higher anisotropy barrier), thus separating the read and write paths. Spin-orbit interactions in the heavy metal inject a spin perpendicular to the metal-magnet interface. 
The corresponding spin-orbit torque (SOT) can be used to flip a magnet. 
 The current density can once again be obtained from energy considerations, while the write error rate obtained using a 2-D Fokker Planck equation with in and out-of-plane fields. Assuming small in-plane field $H_x$ compared to the anisotropy field $H_k$, i.e., $h_x = H_x/H_k \ll 1$ and a magnetization perpendicular to the heavy metal-magnet interface, the equations once again simplify \cite{lee_threshold_2013,lee_thermally_2014,xie_materials_2017} 
\begin{eqnarray}
\mathrm{WER}_\mathrm{SOT}(t) &\approx& e^{\displaystyle -f_0t\exp{(-E_B/k_BT)}}, \nonumber\\
E
_B &=& U_0\left[(1-h_x)^2-2h_s(\pi/2-h_x-h_s)\right],\nonumber\\ 
h_s &=& c_J/H_K, ~~~ c_J = (\hbar/2q)(J\theta_{SH}/M_s t_F),\nonumber\\
j^{SOT}_c &\approx&  \displaystyle\frac{2q}{\hbar}\frac{\mu_0 M_st_F}{\theta_{SH}}\Biggl(\frac{H_k}{2}-\frac{H_x}{\sqrt{2}}\Biggr),
\end{eqnarray}
where $c_J$ is the coefficient of the spin-orbit torque term. 

{Once again, the current burden is higher for smaller error rates that need higher anisotropy barriers and for larger flipped magnetization volumes set by the film thickness}. In fact, $j_c^{SOT}$
has a similar structure as $j^{STT}_c$ except we replace polarization $P$ with spin Hall angle $\theta_{SH}$, and remove the damping term $\alpha$, as the torque involves precessional rather than anti-damping switching (for perpendicularly magnetized films). $f_0$ is the attempt frequency relating to the overall energy landscape through an entropy term. Plugging typical numbers ($P\approx0.7,\theta_{SH}\approx0.2,\alpha\approx0.3$, same film thickness), we find $j_C^{SOT}/j_C^{STT} \approx P/\theta_{SH}\alpha \approx 10$, larger primarily because the STT anti-damping current is proportional to Gilbert damping $\alpha$ ($\alpha \ll 1)$, while the SOT in this example acts perpendicular to the damping torque, like a field-induced torque. However, the switching time is correspondingly reduced, and once again, the energy-delay product depends primarily on the switching charge $Q$, set in essence by angular momentum conservation and acceptable error rates. A further energy reduction happens as current flows through the lower resistance $R_\mathrm{on}$ of the metal underlayer, which can be more than an order of magnitude lower than that of the MTJ's tunnel barrier.    
Note that the critical charge $Q_c$ depends on the volume of the switching free layer magnet $\Omega$. It is, therefore, useful to scale the magnet down to reduce the energy-delay product. However, the energy barrier $E_B$ and the thermal stability factor $\Delta_\mathrm{th}$ also decrease, as a result, making the magnet thermally unstable (this is called the {\it{superparamagnetic limit}}), \textcolor{black}{a challenge which the aforementioned HAMR and bit patterned recordings approaches address head-on}.
Topological excitations such as skyrmions have added protection coming from their spin texture, which adds an extra barrier to their erasure, in other words, to continuously deform them to alternate textures with different topological numbers---for instance---uniformly oriented spins in their ferromagnetic background.
    
It is, therefore, intriguing to consider information encoded in these topological excitations, how we can move them around with an in-plane current, and read information encoded in them with a vertical read operation. In this review, we will discuss the physics of isolated skyrmions, for instance, how to shrink them to a high density while maintaining adequate stability, and the tradeoffs involved in driving them deterministically and ballistically. We will argue, for example, that ferrimagnets can be ideal for generating small N\'eel skyrmions.
We will also show that adjusted for inertial effects when a current is switched off, the effective speed of skyrmions is less than that of domain walls for the same set of magnetic parameters. This is because skyrmions tend to distort during flow, making them ultimately move more like domain walls. However, domain walls have more opportunity for pinning---in particular at edges, for moderate speeds, while at higher speeds, the angular momentum tends to transfer to azimuthal precession of its individual spins. \textcolor{black}{We discuss potential uses for small, fast skyrmions - isolated skyrmions as temporal memory bits for unary wavefront-based computing, high density cascaded skyrmions for reconfigurable logic, skyrmion gases as decorrelators in stochastic computing, as well as neural and reservoir computing.}

\subsection{Road to a Skyrmionic Device}
\textcolor{black}{For a skyrmionic device to be competitive with the existing and emerging technologies, the size of the skyrmion needs to be small. The small size of the skyrmion will enable dense memory and logic circuit design. Furthermore, the device should have properties like low energy cost, high frequency, and overall material stability with temperature and external magnetic fields. To elaborate on the various needs:
\\
\textbf{Size:} In a magnetic material, the competition between the different energy terms (Eq. \ref{eq:continuous_Hamil}) can stabilize an isolated skyrmion. The phase space of the material parameters can be calculated numerically and analytically (Section \ref{phasespace})). Depending on the specific memory technology, sub-50 nm skyrmions need to be stable enough over seconds (e.g., cache memory) to years (e.g., hard drive) at room temperature (Section~\ref{phasespace}~E). Larger skyrmions tend to diffuse more aggressively, while smaller skyrmions tend to last shorter, pin easier, and are harder to read electrically. We will show that the saturation magnetization $M_s$ can set the skyrmion size, while the film thickness can be used to control their lifetime. 
}
\begin{figure}[ht!]
    \includegraphics[width=\columnwidth]{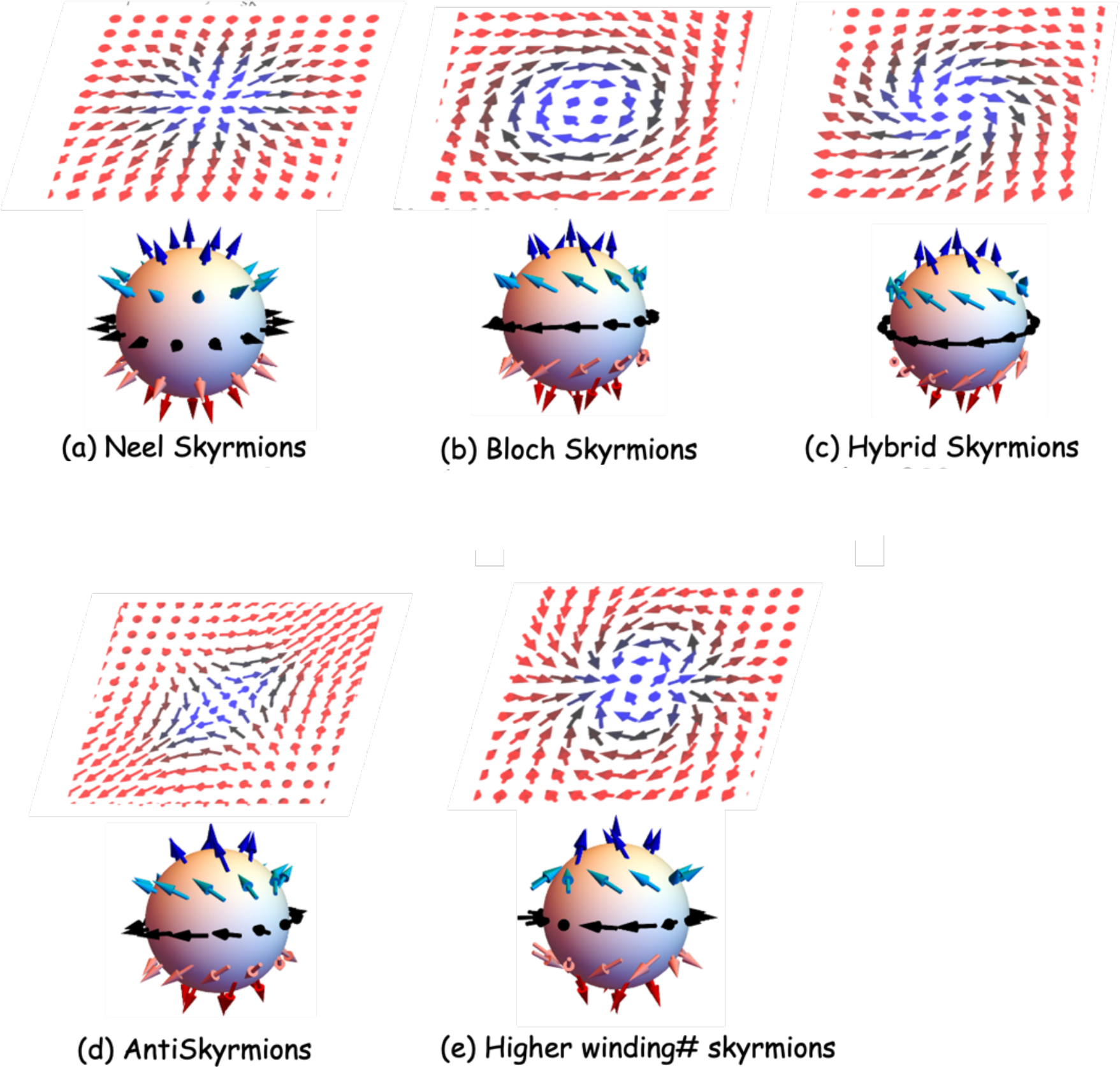}
    \caption{3-D spherical arrangement of spins that generate 2-D skyrmions through stereographic projection. The north pole becomes the skyrmion core while the south pole maps to the background spin texture. The figure shows skyrmions of various winding numbers $N_{sk}$ \textcolor{black}{with same polarization $p=1$} (Eq.~\ref{eqnsk}) and domain angles/helicities $\psi$, along with typical materials supporting them. (a) Neel skyrmions (e.g. CoGd on Pt), $\psi = 0,N_{sk} = 1$ (b) Bloch skyrmions (B20 solids, e.g. FeGe), $\psi = \pi/2,N_{sk} = 1$ (c) Hybrid skyrmions (e.g. B20 on Pt), $\psi = \pi/4, N_{sk} = 1$ (d) AntiSkyrmions (D2d tetragonal inverse Heuslers e.g. MnPtSn), $\psi = \pi/2,N_{sk} = -1$ (e) Higher winding number skyrmions (Frustrated FM, merging of two skyrmions with opposite vorticities), $\psi = 0, N_{sk} = 2$. These excitations map onto the usual classification of 2-D linear excitations-- stars for N\'eel skyrmions, cycles for Bloch skyrmions, spirals for hybrids and saddle points for antiskyrmions. In this picture, skyrmion spins wind anticlockwise as we move anticlockwise around a circle, while antiskyrmion spins wind in the opposite way. For higher $m$ values skyrmions create cycloids along the circle while antiskyrmions trace out astroids. In general, a skyrmion of winding number $N_{sk}$ has $N_{sk}-1$ kinks (lines of 180$^0$ spin reversal) while an antiskyrmion of winding number $-N_{sk}$ has $|N_{sk}|+1$ lines of singularity. }
    \label{fig:top}
\end{figure}

\textcolor{black}{\textbf{Speed and operation time:} For a skyrmionic racetrack device, the rate of data processing is set by the speed at which skyrmions can travel along the racetrack. The high-frequency operation of a skyrmionic device enables low energy logic and memory operations. Thus, the speed of the skyrmion is a key factor for skyrmionic devices to become a competitive technology. In order to achieve a high speed for skyrmions, small $M_s$,  low Gilbert damping and topological damping are required (Section \ref{sectionSkyrmionDynamics}, Eq.~\ref{eqvsk}). Small topological damping and size make nearly compensated ferrimagnets promising candidates for skyrmions. When a very large current is applied to skyrmions, it has been reported \cite{litzius2bePublished} that skyrmions can be distorted or even get annihilated, which puts an upper limit on skyrmion operating current. A speed between 100-700 m/s is probably a suitable target for the nearly linear, quasi-ballistic operation of skyrmion devices.}

\textcolor{black}{\textbf{Material Stability:}
Due to the thermal dissipation from the applied electric current when a hypothetical skyrmionic device is operating, the temperature of the magnetic layer can increase well above the room temperature. In order to have a reliable operation, the magnetic layer needs to have a sufficiently high Curie temperature so that the changes in temperature do not affect skyrmions too much. Additionally, if the device is expected to work under real-life conditions, it has to be resistant with respect to external magnetic fields. In the case of ferri- or antiferromagnets, the contribution of external magnetic fields is negligible.} 
All of these factors will make nearly compensated ferrimagnets and antiferromagnets particularly important to skyrmionics.

\subsection{Skyrmions and Topological Stability}  \label{subsectionSKstability}
\begin{figure}[t]
    \includegraphics[width=\columnwidth]{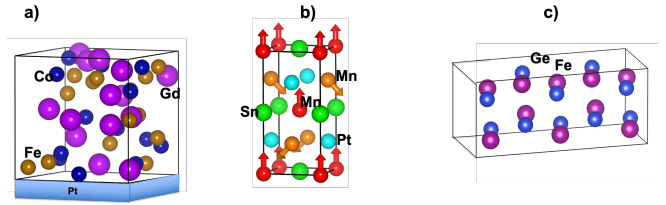}
    \caption{Structures with broken spatial inversion symmetry: (a) An amorphous ferrimagnets (CoFeGd) on top of a heavy metal (Pt) with interfacial DMI supporting N\'{e}el skyrmions; (b) $D_{2d}$ tetragonal unit cells like Mn$_2$PtSn supporting antiskyrmions; (c) B20 FeGe supporting Bloch skyrmions.}
    \label{fig:xals}
\end{figure}

Skyrmions are topological excitations in thin magnetic films and can be viewed as  circular domains with an inverted core relative to the surroundings (Fig.~\ref{fig:top})\cite{Skyrmion_Electronics, Finocchio_2016, everschor,kang_unconv}. Unlike small magnets that tend to get their spins randomized by thermal fluctuations, a skyrmion is stabilized by 
spatial inversion symmetry breaking, generating the Dzyaloshinskii-Moriya interaction (DMI) \textcolor{black}{of the form $E_{DMI} = \sum_{ij}D_{ij}\cdot(\vec{S}_i\times\vec{S}_j)$} between lattice sites labeled by $(i,j)$. 
The symmetry breaking (Fig.~\ref{fig:xals}) can come from bulk inversion asymmetry, such as in B20 solids like MnSe and FeGe, giving rise to Bloch skyrmions whose spins progressively flip spatially from up outside to down inside the skyrmion moving in a plane perpendicular to the radial axis. An alternate way to break the inversion asymmetry is with a heavy metal underlayer, where the interfacial spin-orbit coupling creates the DMI resulting in N\'eel skyrmions, whose spins flip in the plane containing the radial axis. We can also get antiskyrmions in solids with $D_{2d}$ symmetry, such as tetragonal Heuslers. 

The topological index for a skyrmion is given by its winding number 
set by the orientation of the magnetization unit vector $\mathbf{m}$ in the 2-D $(x,y)$ plane
\begin{equation}
    N_{sk} = \frac{1}{4\pi}\int dxdy~\mathbf{m}\cdot\Bigl(\partial \mathbf{m}/\partial x \times \partial\mathbf{m}/\partial y\Bigr)
    \label{eqnsk}
\end{equation}
A detailed discussion on the physical origin can be found in the Appendix. The magnetization vector $\hat{m}$ is characterized by 3-D direction cosines $(\theta,\Psi)$ while their spatial locations are described in 2-D polar coordinates $(r, \phi)$. If we assume there is a vortex-like motion, then the azimuthal angle $\theta$ must not vary around the circumference for a skyrmion and depend on radial coordinate alone, $\theta = \theta(r)$, while the in-plane tilt angle $\Psi$ should increase linearly around the circle until it covers an integer multiple of $2\pi$, so that $\Psi = N_{sk}\phi + \psi$, with vorticity  $N_{sk}$ an integer and $\psi$ the domain angle {or helicity}. The integral $N_{sk}$ (Eq.~\ref{eqnsk}) then simplifies to $[m_z(r=\infty)-m_z(r=0)]\times [\Psi(\phi=2\pi)-\Psi(\phi=0)] = pN_{sk}$, {where $m_z(r) = \cos{\theta(r)}$ and $[m_z(r=\infty)-m_z(r=0)]$ is the polarity $p = \pm 1$}. {The $m_z$ difference relates to the radial evolution of magnetization tilt from core to surrounding (e.g. skyrmions vs merons), while the $\psi$ difference is the overall winding}. \textcolor{black}{For simplicity, in this paper we take the polarization $p = 1$. }\\
\indent It would be useful to see how $\psi$ and $N_{sk}$ can be determined from the energetics. The DMI energy density can be written in the continuous limit as:
\begin{eqnarray}
    \displaystyle\frac{\epsilon_\mathrm{DMI}}{2\pi t_F} = 
    \begin{cases}
      D\boldsymbol{m\cdot[(z\times\nabla)\times m]}, ~\text{ Interfacial DMI}\\
      D\boldsymbol{m\cdot(\nabla\times m)}, ~\text{B20 DMI}\\
     D\Biggl[\displaystyle\frac{\partial{m}_z}{\partial y}m_x+\frac{\partial{m}_z}{\partial x}m_y\Biggr.\\
     ~~~~\Biggl.-m_z\left(\displaystyle\frac{\partial{m}_y}{\partial x}+\frac{\partial{m}_x}{\partial y}\right)\Biggr], ~\text{$D_{2d}$ DMI}\\
    \end{cases}
\end{eqnarray}
Using $\mathbf{m} = \left(\sin\theta\cos\psi,\sin\theta\sin\psi, \cos\theta\right)$, the DMI energy terms can be simplified to:
\begin{eqnarray}
    \epsilon_\mathrm{DMI} = 
    \begin{cases}
     D_0\cos\Bigl[(N_{sk}-1)\phi+\psi\Bigr]\Bigl(\partial_r \theta  +\displaystyle\frac{N_{sk}}{r}\sin{\theta}\cos{\theta}\Bigr)\\
     ~~~~~~\text{(Interfacial~DMI,~N\'{e}el~skyrmion)}\\
     D_0\sin\Bigl[(N_{sk}-1)\phi+\psi\Bigr]\Bigl(\partial_r \theta  +\displaystyle\frac{N_{sk}}{r}\sin{\theta}\cos{\theta}\Bigr) \\
     ~~~~~~{\text{(B20~bulk~DMI,~Bloch~skyrmion)}}\\
     D_0\sin\Bigl[(N_{sk}+1)\phi+\psi\Bigr]\Bigl(-\partial_r \theta \cos{2\theta}\\
     ~~~~~~~~~~~~~~~~~~~~~~~~~~~~~~+\displaystyle\frac{N_{sk}}{r}\sin{\theta}\cos{\theta}\Bigr)\\
     ~~~~~~{\text{($D_{2d}$~DMI,~antiskyrmion)}}
    \end{cases}
\end{eqnarray}
with $D_0 = 2\pi t_F D$ and $t_F$ the film thickness. The ${N_{sk}}\sin{\theta}\cos{\theta}/r$ term, as we will see in the next section, has a much smaller contribution to the energy than the other terms. This would make $N_{sk}>1$ unfavorable, as the term in exchange energy related to the winding number changes as $N_{sk}^2$. We can then see from the integrals \textcolor{black}{in the Appendix} that for interfacial and B20 DMI $N_{sk}=1$ gives the minimum energy for $\psi = 0(\pi),\pi/2(3\pi/2)$ respectively (depending on DMI sign). For the antiskyrmion, $N_{sk}=-1$ and $\psi = \pi/2(3\pi/2)$.

\begin{table*}[t]
    \centering
    \vspace{-1.5cm}
    \renewcommand{\arraystretch}{1.3}
    \caption{Selection of skyrmionic materials in bulk, thin films, and multilayers. Taken with permission from Phd thesis of K. Litzius, Johannes Gutenberg-Universität Mainz (2018) \cite{litziusPhD}\\}
    \makebox[\columnwidth]{
    \begin{tabular}{ l|*{7}{c}|l}
        \toprule
        Material & Sample & Conduction & T$_\mathrm{sky}$/K & $\lambda\mathrm{H}$/nm & $d\mathrm{SiSky}$/nm & $|D|$ /$\frac{\mathrm{mJ}}{\mathrm{m}^2}$ & Type & Refs. \\ \hline \hline
        MnSi                     & bulk               & metal       & 28 - 29.5          & 18       & ---    & N.A.   & Bloch  & \cite{artcl:bulk_skyrmion0,neubauer2009topological,artcl:Adams2011,artcl:Kindervater2016} \\ \hline
        MnSi (press.)            & bulk               & metal       & 5 - 29             & 18       & ---    & N.A.   & Bloch  &
        \cite{artcl:Lee2009,artcl:Ritz2013,artcl:Chacon2015} \\ \hline
        MnSi                     & film ($\sim$ 50 nm)  & metal       & <5 - 23            & 18       & ---    & N.A.   & Bloch  &
        \cite{artcl:Tonomura2012,artcl:Yu2015} \\ \hline
        Fe$_{1-x}$Co$_{x}$Si     & bulk               & semi-metal  & 25 - 30            & 37       & ---    & N.A.   & Bloch  &
        \cite{artcl:Adams2010,artcl:Munzer2010} \\ \hline
        Fe$_\mathrm{0\text{.}5}$Co$_\mathrm{0\text{.}5}$Si   & film ($\sim$ 20 nm)  & semi-metal  & 5 - 40             & 90       & $\sim$ 75    & N.A.   & Bloch  &
        \cite{yu2010real} \\ \hline
        FeGe                     & bulk               & metal       & 273 - 278          & 70       & N.A.    & N.A.   & Bloch  &
        \cite{huang2012extended}\\ \hline
        FeGe                     & film ($\sim$ 75 nm)  & metal       & 250 - 270          & 70       & N.A.    & N.A.   & Bloch  &
        \cite{artcl:Yu2011,huang2012extended,artcl:McGrouther2016}\\ \hline
        FeGe                     & film ($\sim$ 15 nm)  & metal       & 60 - 280           & 70       & ---    & N.A.   & Bloch  &
        \cite{artcl:Yu2011}\\ \hline
        Cu$_2$OSeO$_3$           & bulk               & insulator   & 56 - 58            & 60       & ---    & N.A.   & Bloch  &
        \cite{artcl:Adams2012,seki2012observation,artcl:Okamura2016,artcl:Zhang_resonant,artcl:Zhang2016_imaging,artcl:Zhang2016_multidomain} \\ \hline
        Cu$_2$OSeO$_3$           & film ($\sim$ 100 nm) & insulator   & <5 - 57            & 50       & ---    & N.A.   & Bloch  &
        \cite{seki2012observation,artcl:Rajeswari2015} \\ \hline
        Co$_8$Zn$_8$Mn$_4$       & bulk               & metal       & 284 - 300          & 125      & ---    & N.A.   & Bloch  &
        \cite{artcl:Karube2016} \\ \hline
        Co$_8$Zn$_9$Mn$_3$       & bulk               & metal       & 311 - 320          & >125     & ---    & N.A.   & Bloch  &
        \cite{artcl:Tokunaga2015} \\ \hline
        Co$_8$Zn$_9$Mn$_3$       & film ($\sim$ 150 nm) & metal       & 300 - 320          & >125     & ---    & N.A.   & Bloch  &
        \cite{artcl:Tokunaga2015} \\ \hline
        GaV$_4$S$_8$             & bulk               & semi-metal  & 9 - 13             & 17.7 & ---    & N.A.   & N\'eel &
        \cite{gav4s8} \\ \hline
        Fe/Ir(111)               & monolayer          & N.A.         & $\sim$ 11          & $\sim$ 1 - 20  & <5    & N.A.   & N\'eel &
        \cite{heinze2011spontaneous,artcl:Bergmann2014,artcl:Wiesendanger2016} \\ \hline
        PdFe/Ir(111)             & bilayer            & N.A.         & $\sim$ 4.2   & $\sim$ 1 - 20  & <5    & N.A.   & N\'eel &
        \cite{artcl:Bergmann2014,artcl:Wiesendanger2016,romming2013writing,artcl:romming2015,artcl:Schmidt2016,artcl:Hanneken2016} \\ \hline
        (Ir/Co/Pt)$_{10}$        & multilayer         & metal       & $\lessgtr$300      & 30 - 90& 100    & 2.0   & N\'eel &
        \cite{MoreauLuchaire2016} \\ \hline
        (Pt/Co/MgO) & single layer  & metal       & $\lessgtr$300      & $\sim$ 500 & 70 - 130 & 2.0   & N\'eel &
        \cite{Boulle2016} \\ \hline
        Pt/Co/Ta                                      & multilayer         & metal    & $\lessgtr$300& 480   & 75 - 200    & 1.3   & N\'eel &
        \cite{woo2016observation} \\ \hline
        Pt/Co$_\mathrm{60}$Fe$_\mathrm{20}$B$_\mathrm{20}$/MgO       & multilayer         & metal    & $\lessgtr$300& 344  & <250    & 1.35   & N\'eel & \cite{woo2016observation,litzius2017skyrmion,Lemesh2018} \\ \hline
        Ir/Co/Pt                                      & multilayer         & metal    & $\lessgtr$300& ---   & 25 - 100    & N.A.   & N\'eel & \cite{litziusPhD}
        \\ \hline
        W/Co$_\mathrm{20}$Fe$_\mathrm{60}$B$_\mathrm{20}$/MgO        & multilayer         & metal    & $\lessgtr$300& 460  & $\sim$ 250    & 0.3 - 0.7   & N\'eel &
        \cite{artcl:samridh_WCoFeB} \\ \hline
        Pd/Co$_\mathrm{60}$Fe$_\mathrm{20}$B$_\mathrm{20}$/MgO       & multilayer         & metal    & $\lessgtr$300& 300   & $\lesssim$ 200    & 0.78   & N\'eel & \cite{litziusPhD}
        \\ \hline
        Ta/Co$_\mathrm{20}$Fe$_\mathrm{60}$B$_\mathrm{20}$/(Ta)/MgO       & single layer  & metal    & >300    & 4200   & 1000 - 2000    & 0.33   & N\'eel &
        \cite{artcl:jakub_diffusion} \\ \hline
        Ta/Co$_\mathrm{20}$Fe$_\mathrm{60}$B$_\mathrm{20}$/MgO       & multilayer    & metal    & $\lessgtr$300    & <900   & $\sim$ 300    & 0.33   & N\'eel & \cite{litziusPhD}
         \\ \bottomrule
    \end{tabular}
    }
    
\end{table*}
The topological property of a skyrmion is characterized by its winding number $N_{sk}$ (Eq.~\ref{eqnsk}). The 2-D chiral skyrmions can be generated by stereographic projection of continuous arrows spanning every point of a sphere onto the 2-D plane {(Fig.~\ref{fig:top})}, except a singularity at one of the poles, which is the so-called Bloch point.
As long as the continuous approximation to the magnetic texture holds, the skyrmions are topologically protected because the winding number acts as an invariant.
Continuous deformation of a skyrmion preserves its winding number and creates a barrier to phases with alternate winding numbers (e.g., a homogeneous ferromagnetic background with winding number zero), as such a deformation amounts simply to rotations on the surface of the projecting sphere and preserves the winding number. This would allow the skyrmions to deform or shrink to unpin from the defects at small currents without undergoing annihilation. \textcolor{black}{Recently, the topological stability of skyrmion has been directly demonstrated by Je \textit{et al.} \cite{je_direct_2020}, with longer lifetimes observed for magnetic skyrmions than trivial bubble structures.}

Topological protection has two limitations---one mechanism (intrinsic annihilation) is due to thermal effects, which can randomly flip spins, shrinking the skyrmion until it reaches a Bloch point, at which point the atomic structure of magnets starts to matter, and the skyrmion winding number can change to zero. The other limit (edge annihilation) is from finite-size effects when the skyrmion gets too close to the edges of the magnets or to non-magnetic defects, whereupon the texture of the confined/defect-adjacent skyrmion no longer maps onto the whole sphere, leading to a similar annihilation. The difference between edge and intrinsic annihilation is that for edge annihilation, the texture of the magnet does not pass through the Bloch point and the change in winding number from 1 to 0 is gradual compared to intrinsic annihilation. This suggests that as a skyrmion gets larger, the barrier to intrinsic annihilation becomes larger but not that to edge annihilation.

Historically, skyrmions were experimentally observed in the form of a periodic lattice in B20 solids like MnSi. These skyrmion lattices are ground state configurations that have a strong degree of topological protection characterized by a fixed winding number in k-space. Due to their topological nature, skyrmions can shrink to line singularities but are typically unable to unwind into their background continuum. As we will now see, skyrmions exist as a result of a competition between exchange anisotropy, demagnetization, and chiral DMI. Only the DMI part of the energy, specifically its sign, is responsible for the chirality of skyrmions. At low temperature in an infinite continuum solid, this imposes a very high barrier that keeps this skyrmion lattice topologically stable, even while its critical current density ${j_c}$ for unpinning and initiating motion is quite small, orders of magnitude less than a domain wall that tends to snag onto edges.  Unfortunately, it is not obvious how to encode information into a periodic lattice. The alternate is to work with individual skyrmions nucleated around defects. However, the thermal stability barrier for these metastable non-periodic skyrmions is much smaller, allowing them to naturally `melt' into the surrounding magnetic background unless they are properly stabilized. 

Table II shows a list of materials where skyrmions have been seen so far. Let us now discuss the phase space attributes in these materials that are needed to stabilize skyrmions.

\section{Phase-Space for static skyrmions}\label{phasespace}
\begin{figure}
    \includegraphics[width=\columnwidth]{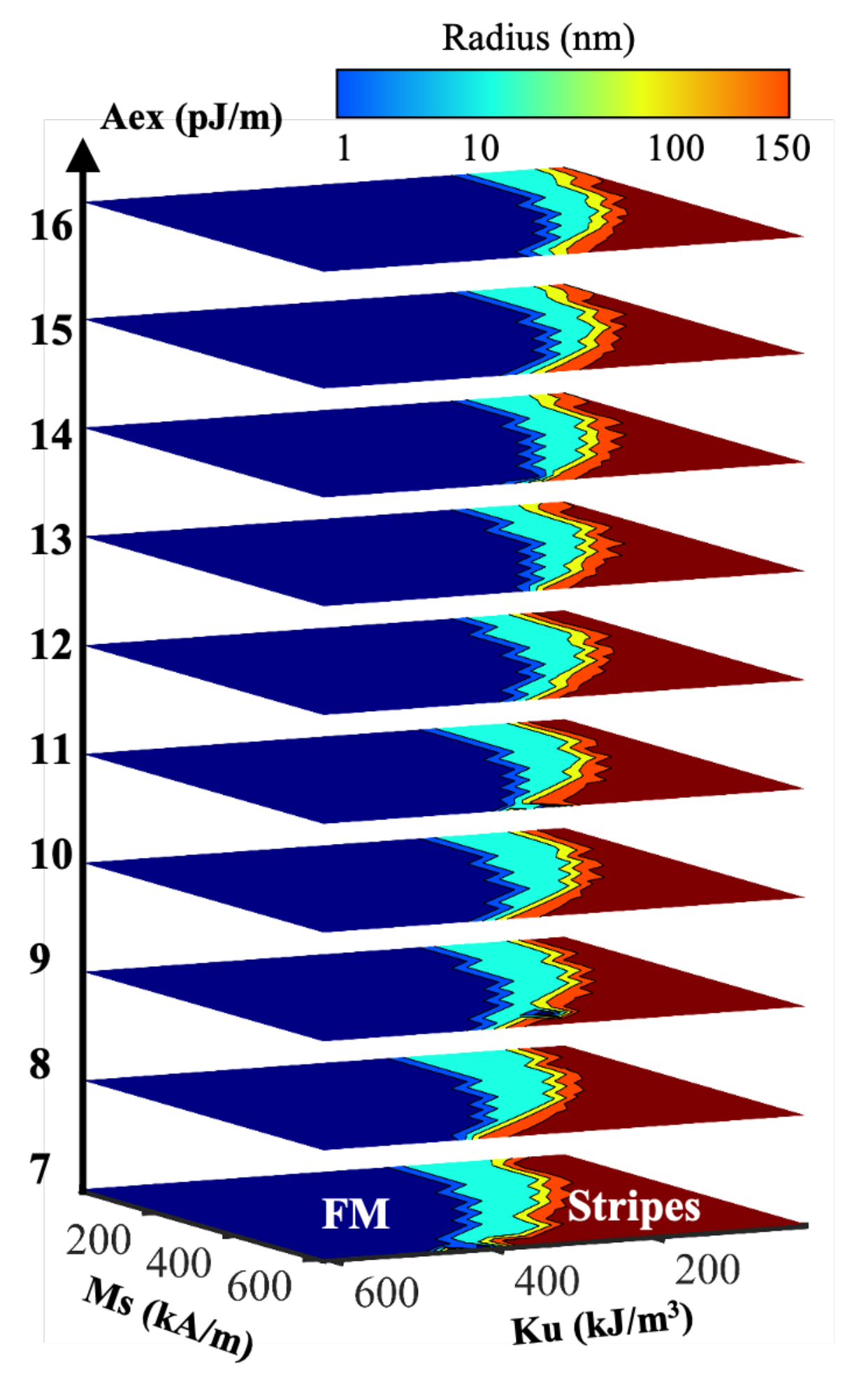}
    \includegraphics[width=\columnwidth]{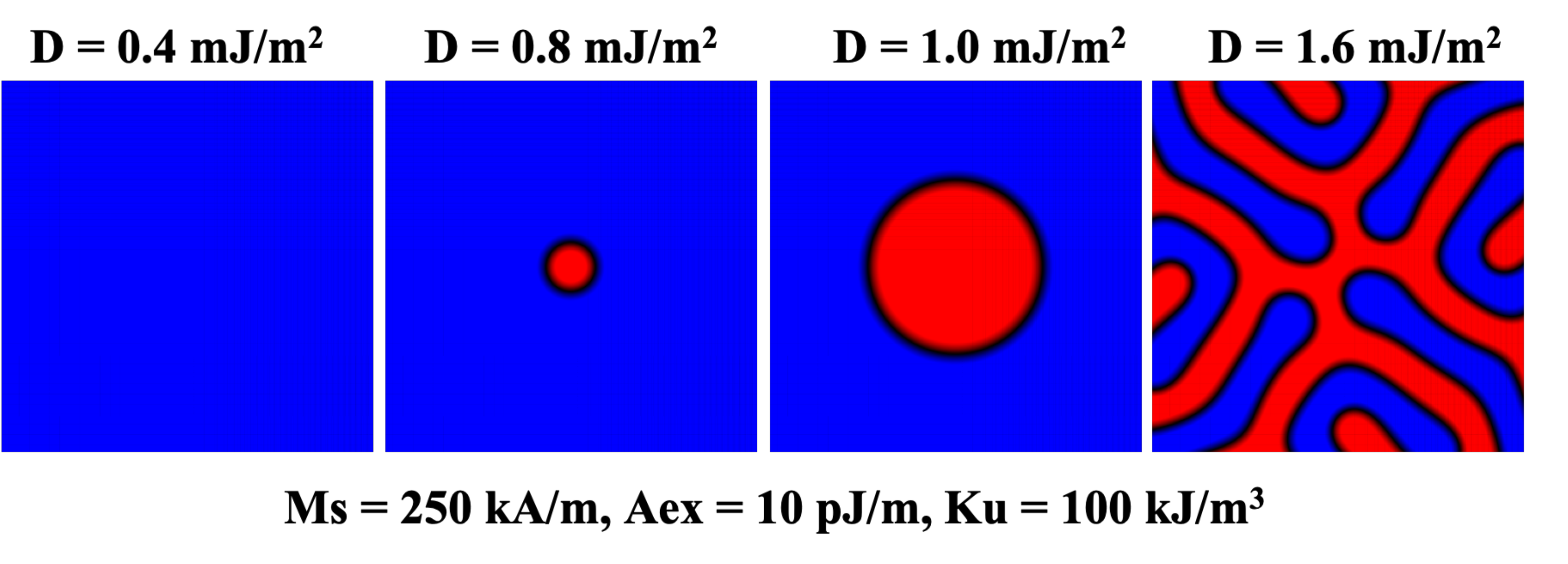}
   \caption{A sequence of {zero-temperature} phase space diagrams for various exchange stiffnesses with interfacial DMI, $D = 1$ mJ/m$^2$ and thickness $t_{FM}=3~nm$. As stiffness increases, the favorable region that can host isolated metastable $\sim 10$ nm skyrmions shrinks, and the borders separating them from competing phases (ferro/ferrimagnetic background, vs. stripe phase) move towards lower anisotropy. This is expected, as a smaller anisotropy compensates for larger exchange. The pictures below the phase diagram show the evolution of a skyrmion stabilized at zero temperature with parameters listed above, with subsequently increasing the DMI. Going from the ferromagnetic phase (left) to the stripe phase (right).} 
   \label{fig:barrier}
\end{figure}

\textcolor{black}{Several parameters set the skyrmion static and dynamic properties - specifically saturation magnetization $M_s$, anisotropy field $H_k$, damping $\alpha$, external magnetic field $H_\mathrm{ext}$, DMI $D$ and exchange stiffness $A_\mathrm{ex}$. The phase space can guide us in choosing materials that can host small skyrmions with long enough lifetimes. Additionally, a high Curie temperature, set primarily by exchange, is required to have the required material stability. Figure~\ref{fig:barrier} shows the skyrmion phase space sandwiched between the background ferro/ferrimagnetic ground state and the stripe phase.}

\subsection{What determines Skyrmion Size?}\label{size}
\begin{figure}[htp]
    \includegraphics[width=\columnwidth]{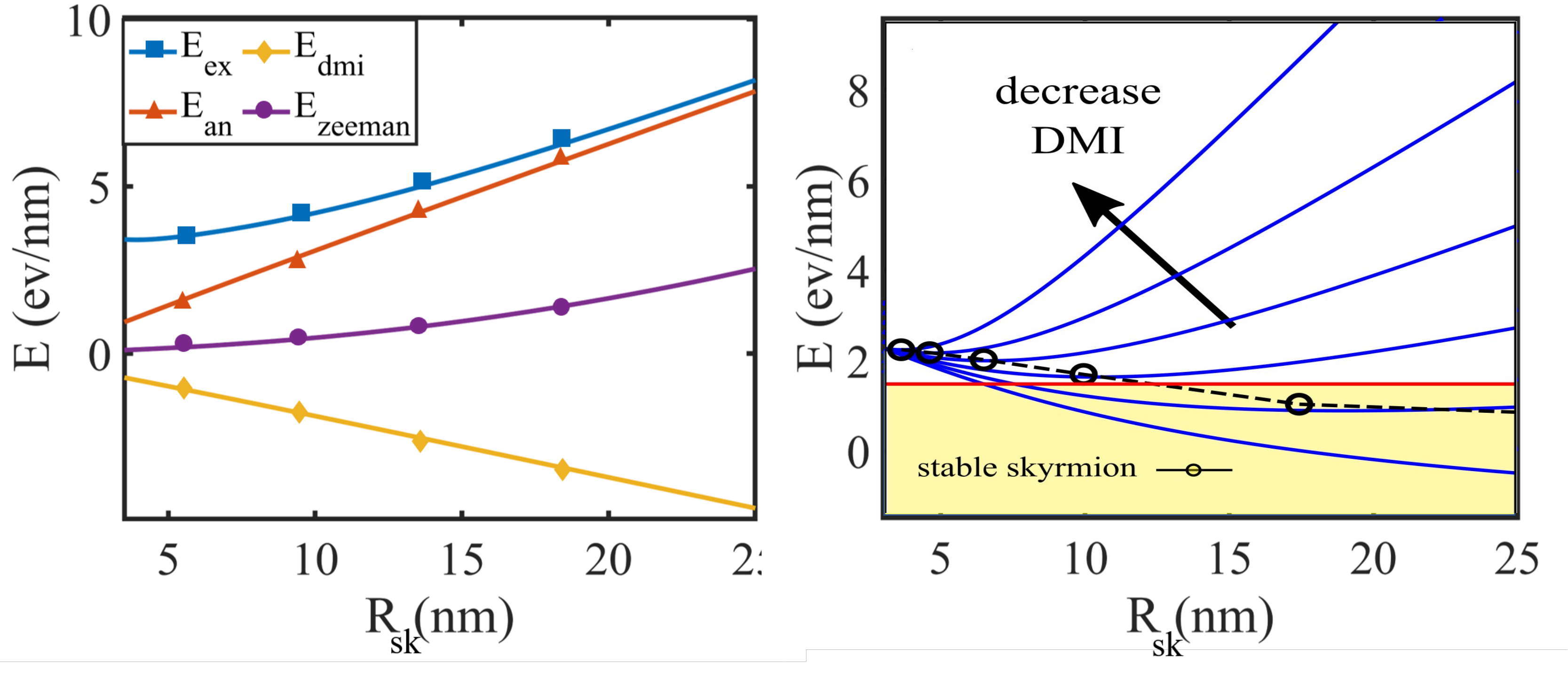}
    \caption{{(Left) Schematic of energy terms for a skyrmion. The 2-D topological contribution to exchange $\propto N_{sk}^2\Delta/R$ and the negative DMI energy $\propto -N_{sk}\pi D$ together produce a local energy minimum for an isolated metastable skyrmion. (Right) Reducing DMI shifts the minimum to a lower radius, creating smaller isolated skyrmions while at the same reducing the energy well and thus their lifetime to dissolve into the background continuum. Increasing DMI increases skyrmion size, until at $D = D_c$ the skyrmion destabilizes into a stripe phase, indicated in yellow.}}
    \label{fig:en}
\end{figure}

A skyrmion has a circular core and a transition region (domain wall) to the background spin texture on the outside. By minimizing energy terms with respect to skyrmion radius and domain wall transition width, we get an equation for skyrmion size. The skyrmion energy barrier then can be calculated from the skyrmion radius and width, {as shown in Fig.~\ref{fig:en}}. For a constant interfacial DMI, there is an optimized thickness that gives the maximum energy barrier. The points to note are that the energy barrier denoting skyrmion stability/lifetime is maximized for specific skyrmion radius and film thickness and that this barrier maximizing (and local energy minimizing) radius increases with DMI until a destabilization point where there is a phase transition to a skyrmion lattice. Lowering DMI for an isolated skyrmion lowers its size but also reduces thermal stability (Fig.~\ref{fig:en}) by making the metastable wells shallower and allowing the skyrmion to melt into the homogeneous magnetic background. 

The energy landscape of a ferromagnet is described by the Heisenberg energy, written in the continuous form as:
\begin{eqnarray}
    E&=t_{F}\int \big[A_\mathrm{ex}\left(\boldsymbol{\nabla}\mathbf{m}\right)^2-K_u m_z^2-\mu_0 M_s\mathbf{m}\cdot\mathbf{H}_\mathrm{ext}\nonumber\\ 
    &- \frac{1}{2}\mu_0 M_s\mathbf{m}\cdot \mathbf{H}_\mathrm{d} + \epsilon_\mathrm{DMI}\big]d^2\mathbf{r},
\label{eq:continuous_Hamil}
\end{eqnarray}
where $A_\mathrm{ex}$ is the exchange stiffness related to the exchange energy by a length unit (e.g. for a simple cubic unit cell of length $a$, $A_\mathrm{ex} = J/a$), $K_u$ is the uniaxial anisotropy, $\mathbf{H}_\mathrm{ext}$ is the external magnetic field, $\mathbf{H}_d$ is the demagnetization field, $t_F$ is the thickness and $\epsilon_\mathrm{DMI}$ is the DMI energy. 
Minimizing the energy for an isolated skyrmion, we get the skyrmion size.  The result for large skyrmions at zero magnetic field is \cite{Thiaville,bogdanov2001chiral,Buttner2018,wang_theory_2018}
\begin{equation}
    R_{sk} =  
    \displaystyle \Delta\sqrt{\frac{A_\mathrm{ex}}{A_\mathrm{ex}-\Delta^2K}} = \frac{\Delta}{\sqrt{1-(D/D_c)^2}}, ~~~~\Delta = \frac{\pi D}{4K},
    \label{radius}
\end{equation}
where $D$ is the DMI. The critical DMI is $D_c = 4\sqrt{A_\mathrm{ex}K}/\pi$. Here the effective anisotropy includes contributions from the demagnetization field $K \approx K_u - \mu_0M_s^2/2$. Equation\ref{radius} shows that a perpendicular magnetic anisotropy is required to stabalize skyrmion in zero magnetic field.

Note that these equations change a bit with size.
For ultra small skyrmions ($R_{sk} \lessapprox$ 10 ~nm, $\rho = R_{sk}/\Delta ~\lessapprox$ 2), additional form factors $f(\rho)$ (Eq.~\ref{eneqnsfit}) must be taken into consideration, at the cost of more numerical complexity. In general, the radius takes the form
\begin{equation}
    R_{sk} =  
    \displaystyle \Biggl(\frac{D}{D_c}\Biggr)^a\frac{C_1\Delta}{\sqrt{1-C_2(D/D_c)^b}}
    \label{radius2}
\end{equation}

It is worth emphasizing that the two terms essential for stabilizing the skyrmion, both involving $N_{sk}$, are the $ \sim -t_F DN_{sk}$ Dzyaloshinskii-Moriya term that gives a negative slope for the energy vs. radius curve, and the $\sim N_{sk}^2\Delta/R_{sk}$ term in the exchange energy that provides a curvature to its energy curve. Together these two terms allow the creation of a metastable minimum (Fig.~\ref{fig:en}). The last term appears solely due to the circular nature of the skyrmion, i.e., added terms in the cylindrical gradient arising from variations in cylindrical radial and angular unit vectors from point to point along the azimuthal direction. 
\subsection{Ferrimagnets for Small, Fast Skyrmions} \label{subsectionFIM} 
A ferrimagnet is a magnetic material with two sublattices of oppositely oriented but unequal magnetic moments. Calculations have shown that several ferrimagnets are promising materials for skyrmion device applications, \textcolor{black}{primarily because we can reduce their saturation magnetization $M_S$ near compensation, which is predicted to reduce the stray field and thus the skyrmion size while pushing up their speed.} These ferrimagnets include amorphous rare-earth-transition-metal (RE-TM), Mn-based inverse Heuslers (IHs), and Mn$_4$N. Their small $M_{s}$ and low $K_{u}$, in the order of $10^4$ J/m$^3$, fall within the parameter space of hosting ultrasmall ($\sim$10 nm) and fast skyrmions ($>$100 m/s) at room temperature. Among the parameters determining skyrmion radius, the exchange constant is usually constrained by the need for an adequate ordering temperature, an ordered state well above room temperature.

\begin{figure}
\centering
    \includegraphics[width = \columnwidth]{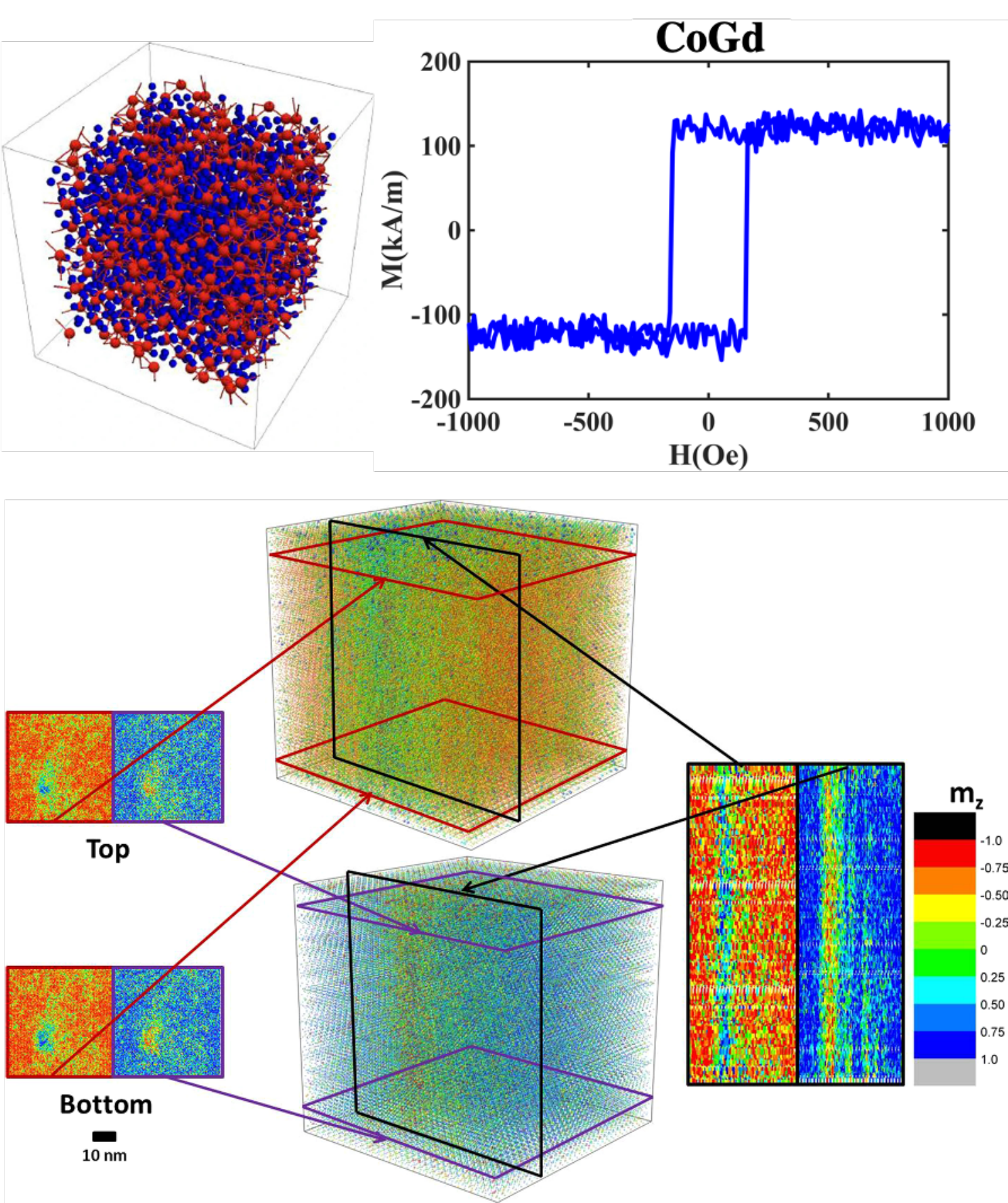}
    \caption{(Top) Amorphous RE-TM structure from a molecular dynamics simulation. The calculated RE-TM amorphous structure is a 1.6 nm cube. Red spheres represent RE atoms, and blue spheres represent TM atoms. The composition of this structure is RE$_{25}$TM$_{75}$  ~\cite{Ma2019}. Out-of-plane M-H loop of 10 nm CoGd at 300 K. The out-of-loop has a squareness of 1, demonstrating robust PMA. $M_s$ is $\sim$100 kA/m. (Bottom) A simulated cross-sectional tomograph of an ultrasmall room temperature skyrmion shows a columnar distribution throughout a 10 nm CoGd sample. Co-sublattice spins are in the top box, Gd-sublattice in the bottom box, in-plane cross-sections near the top and bottom interface (left), and out-of-plane cross-sections (right). C.T. MA \textit{et al.} Sci Rep 9, 9964 (2019); licensed under a Creative Commons Attribution (CC BY) license. \cite{Ma2019}.}
    \label{fig:a-RE-TM}
    \end{figure}

One of the well-studied classes of ferrimagnets hosting skyrmions is amorphous rare earth-transition metal (RE-TM) ferrimagnets like CoGd. The structure of an amorphous RE-TM alloy is shown in Fig. \ref{fig:a-RE-TM}. We would normally expect an amorphous alloy to be isotropic in the out-of-plane and in-plane directions, resulting in zero perpendicular magnetic anisotropy (PMA). However, amorphous RE-TM thin films are found to have structural anisotropy, leading to PMA \cite{Harris1992}. In RE-TM, magnetic moments in the RE and TM species arrange to form two antiferromagnetically-coupled sublattices. The magnetization of RE-TM ferrimagnets varies with composition and temperature. At the compensation temperature, moments in the RE and TM sublattices cancel each other, leading to a vanishing magnetization.  Simulations \cite{Ma2019} and experiments \cite{Caretta2017, Quessab2020} have reported skyrmions as small as 10 nm in RE-TM with heavy metal interfaces at room temperature. In atomistic simulations with the Landau-Lifshitz-Gilbert (LLG) equation, ultrasmall skyrmions are found to be stable at room temperature in 5 to 15 nm thick CoGd with a compensation temperature near 250 K. For 10 nm thick CoGd, a sub 10 nm skyrmion is predicted for an interfacial DMI of 0.8 to 1.0 mJ/m$^2$, which is within the range of measured DMIs in Pt/CoGd/Pt\textsubscript{1-x}W \cite{Quessab2020}. Furthermore, a columnar growth of skyrmions across the film thickness is seen to arise (Fig.~\ref{fig:a-RE-TM}) through computational tomography \cite{Ma2019}.  Such robustness of structure is critical to the stability of skyrmionic information-carrying bits. \par
    
\begin{figure}
\centering
    \includegraphics[width=\columnwidth]{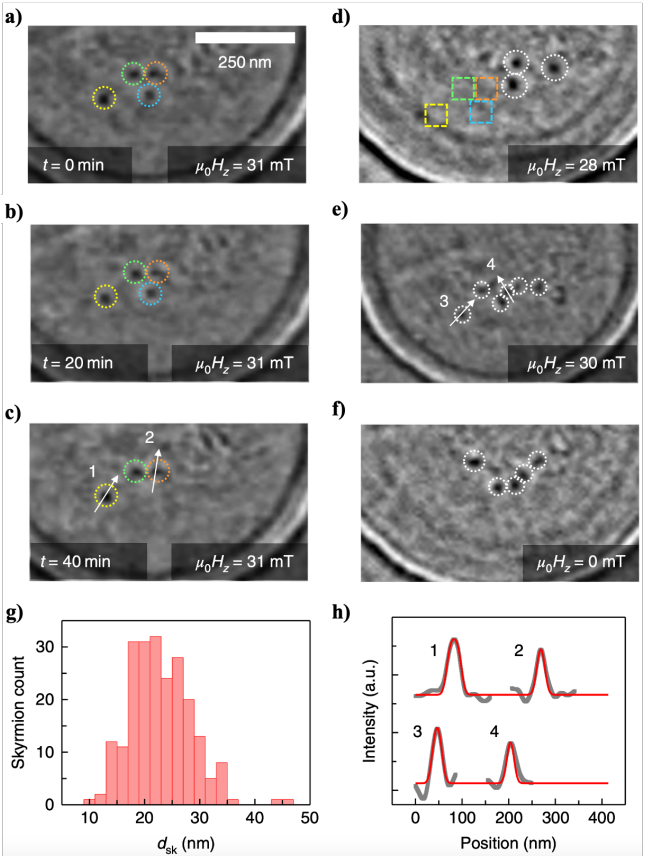}
    \caption{Imaging of ultra-small skyrmions in ferrimagnets. (a–c) X-ray holography images of Pt/Co$_{44}$Gd$_{56}$/TaO$_x$ immediately after injecting a nucleation current pulse (a), after 20 min (b), and after 40 min (c). (d–e) Image of skyrmions at different positions obtained after saturating the film and injecting a nucleation current pulse. The colored squares indicate the absence of skyrmions in locations present in a and b, respectively. (f) Skyrmions under zero out-of-plane field $\mu_0H_z$ (nucleated in a bias field which was subsequently reduced to zero). In all the images, the light (dark) contrast indicates magnetization out of (into) the plane. The ring-like features in some of the images are not a magnetic feature but simply an artifact from the circular field of view. (g) The histogram shows the distribution of skyrmion diameters. (h) Magnetic contrast line scans of the numbered skyrmions in (c) and (e) (grey-filled circles) and fits (red lines). Reproduced with permission from L. Caretta \textit{et al.}, Nature Nanotech 13, 1154–1160 (2018); Copyright 2018, Nature Publishing Group. \cite{Caretta2017}.}
    \label{fig:ultra_small_gdco}
\end{figure}

It should be noted that in experimental studies, the ultra-small skyrmions are much more difficult to image than larger, bubble-like skyrmions. Only a few techniques are capable of imaging in the nanometer regime and simultaneously also recording time-resolved image sequences and applying stabilizing fields. The most prominent techniques currently are X-ray holography, scanning transmission X-ray microscopy (STXM), full-field transmission soft X-ray microscopy and X-ray magnetic circular dichroism (XMCD-PEEM) \cite{Schutz2011, wang_thermal_2020}. All these methods utilize the X-ray magnetic circular dichroism (XMCD) effect that leads to a magnetization-dependent absorption of photons in the material. An additional advantage of this technique is the element-specific resolution of recorded data, i.e., the absorption of only one atomic species can be selected to image sublattices and otherwise compensated materials such as ferrimagnets. While X-ray imaging can also investigate antiferromagnets directly, this feature also allows us to obtain data on synthetic antiferromagnets. A holography image showing our reported 10 nm skyrmions in ferrimagnetic CoGd is presented in Fig. \ref{fig:ultra_small_gdco} \cite{litzius2bePublished}.

In experiments, an M-H loop of 10 nm CoGd in Fig. \ref{fig:a-RE-TM} shows a robust PMA, with an $M_s$ close to 100 kA/m. $K_u$ is measured to be about 30 kJ/m$^3$. As discussed earlier, these parameters are very promising for ultrasmall skyrmions at room temperature. Indeed, $\sim$10 to $\sim$ 100 nm skyrmions have been reported in RE-TM. $\sim$150 nm skyrmions are reported in Pt/GdCoFe/MgO through scanning transmission X-ray microscopy (STXM) \cite{Woo2018}. { Magnetic force microscopy (MFM) finds sub $\sim$100 nm skyrmion in Pt/CoGd/W \cite{Quessab2020}.} The smallest skyrmions in RE-TM are imaged by X-ray holography, see Fig. \ref{fig:ultra_small_gdco}. Ultrasmall skyrmions of close to 10 nm are observed in Pt/CoGd/TaO\textsubscript{x} at room temperature \cite{Caretta2017}.
Through current injection,  a distribution of skyrmion sizes, from 10 to 50 nm, can be nucleated in CoGd, as shown in Fig. \ref{fig:ultra_small_gdco}. The distribution in sizes arises from local variations in magnetic properties. Despite these variations, the skyrmions are seen to be stable in zero applied field---a critical attribute for commercial applications. High-speed skyrmions of several hundred m/s are also reported in RE-TM, which will be discussed in detail in Section \ref{sectionSkyrmionDynamics}. The observed small and fast skyrmions make CoGd one of the most promising materials in realizing skyrmion-based devices. \par
\begin{figure}
\centering
    \includegraphics[width=\columnwidth]{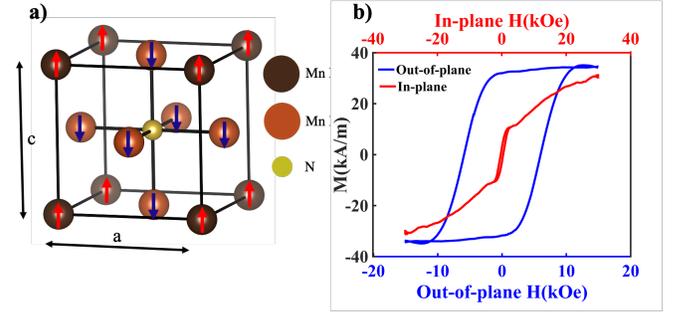}
    \caption{(a) Anti-perovskite Mn$_4$N. Atoms at the corner are Mn-I, and atoms at the face center are Mn-II. Mn-I and Mn-II atoms have nonequivalent magnetic moments and form the two ferrimagnetic sublattices in Mn$_4$N. (b) M-H loop of 15 nm Mn$_4$N at 300 K. The loops show robust PMA in Mn$_4$N. $M_s$ is  $\sim$ 40 kA/m and in-plane anisotropy field is larger than 1.5 T. Reproduced from W. Zhou, \textit{et al.}, AIP Advances 11, 015334 (2021), with the permission of AIP Publishing. \cite{zhou2020}}
    \label{fig:Mn4N}
\end{figure}

\begin{figure}
\centering
    \includegraphics[width=\columnwidth]{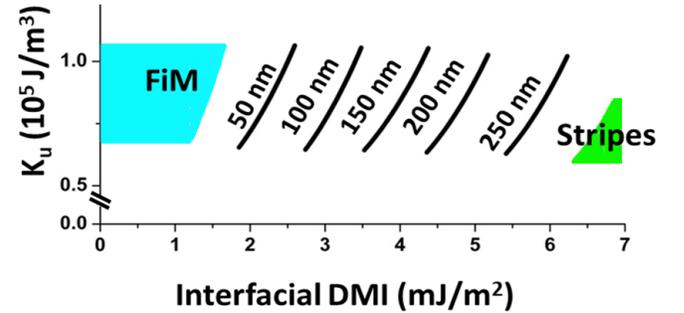}
     \caption{Simulated skyrmion interfacial DMI-$K_u$ phase diagram of Mn$_4$N at 300  K. With $K_u$ from 0.75 to 1 kJ/m$^3$, and interfacial DMI from 0 to 7 mJ/m$^2$, various sizes of skyrmion are stabilized as indicated by the isolines for skyrmion radius in the phase diagram.}
    \label{fig:Mn4N_phase}
\end{figure}

\begin{figure}
\centering
    \includegraphics[width =0.4\columnwidth]{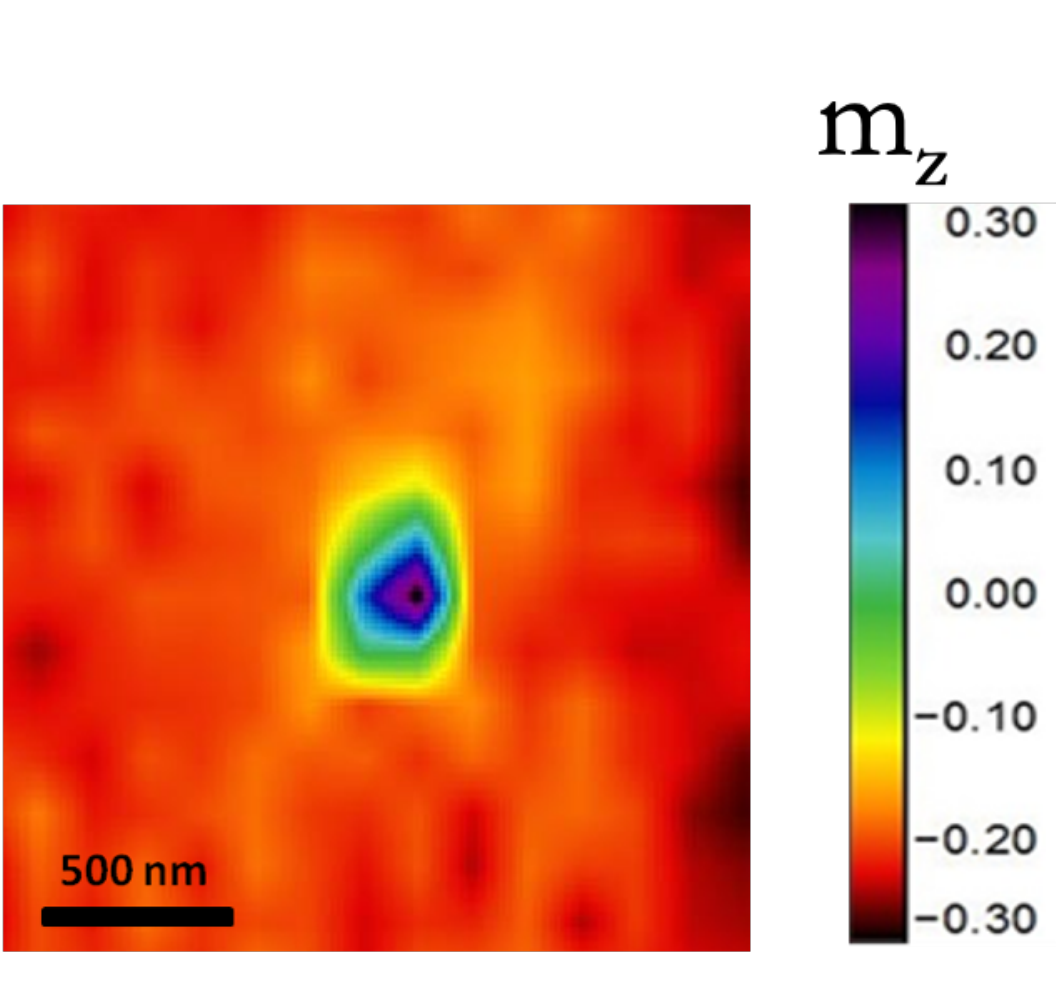}
    \caption{Magnetic force microscopy (MFM) image reveals $\sim$ 300 nm skyrmion in Mn$_4$N (15 nm)/Pt on MgO substrate. The red region corresponds to down magnetic moments; the dark blue region corresponds to up magnetic moments, which make up the core of the skyrmion.}
    \label{fig:Mn4N_MFM}
\end{figure}

Alternatively, epitaxial Mn$_4$N is a RE-free ferrimagnet that seems to be a good candidate material for skyrmion-based devices. Unlike amorphous RE-TM, Mn$_4$N is an anti-perovskite with intrinsic perpendicular magnetic anisotropy. With a deposition temperature greater than 400$^{\circ}$C, Mn$_4$N is thermally stable and will remain robust during device processing. In comparison, the thermal stability of amorphous RE-TM remains a question. It has been shown that in GdCo, PMA is lost abruptly after annealing above 300$^{\circ}$C temperature\cite{gdcoannealing}. As shown in Fig. \ref{fig:Mn4N}(a), the corner Mn atoms (Mn I) and face-centered Mn atoms (Mn II) couple antiferromagnetically to form the two magnetic sublattices. Since the two sublattices have a similar but unequal moment, Mn$_4$N  has a small $M_s$ of less than 100 kA/m. The intrinsic perpendicular magnetic anisotropy of Mn$_4$N originates from unequal in-plane (a) and out-of-plane (c) lattice constants. The experimental $c/a$ ratios of Mn$_4$N thin films are close to 0.99, and $K_u$ less than 100 kJ/m$^3$ \cite{Isogami2020Mn4N}. The {measured} M-H of 15 nm film in Fig. \ref{fig:Mn4N}(b) shows robust PMA, $M_s$ of 40 kA/m, and an estimated $K_u$ of 80 kJ/m$^3$. While $K_u$ of Mn$_4$N is about three times larger than $K_u$ of CoGd, it still falls within the parameter space for room temperature skyrmions. Although no experiments so far, to our knowledge, have reported skyrmions in Mn$_4$N, simulation of Mn$_4$N in Fig.~\ref{fig:Mn4N_phase} {(atomistic for $< 50$ nm skyrmions, micromagnetic for $> 50$ nm)}, based on these experimental values of $M_s$ and $K_u$, find stable room temperature skyrmions. With the experimental value of anisotropy and DMI ranges from 0 to 7 mJ/m$^2$, the predicted skyrmion size varies from 10 to 500 nm. \textcolor{black}{Our MFM measurements Fig.~\ref{fig:Mn4N_MFM} show $\sim$ 300 nm  skyrmions in 15 nm Mn$_4$N with Pt cap layer.} DMI engineering, discussed in the next section (Sec.~\ref{subsectionParameters}), will further allow the tuning of skyrmion sizes in Mn$_4$N.

\begin{figure}
\centering
    \includegraphics[width=\columnwidth]{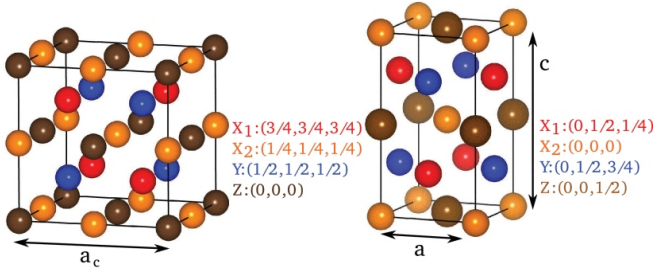}
    \caption{Schematic representation of (a) cubic inverse-Heusler $\mathrm{XA}$ structure and (b) tetragonal inverse-Heusler structure. In the inverse-Heusler structure, $\mathrm{X_{1}}$ and $\mathrm{X_{2}}$ are the same transition metal element but they have different environments and magnetic moments.}
    \label{fig:ih_structure}
\end{figure}

Besides Mn$_4$N, Mn-based \textcolor{black}{Inverse Heuslers} (IH) are also RE-free ferrimagnets, and are predicted to host ultrasmall skyrmions at room temperature \cite{xie2020computational}. While no experiments have investigated skyrmions in IH to our knowledge, recent computational works have shown IH to be candidate materials for skyrmion applications. {In addition, the Heusler family of solids are good candidates for Slater-Pauling half-metals, which suppress interband spin-flip scattering and support ultra-low Gilbert damping} \cite{galanakis_electronic_2006}. A schematic diagram of cubic and tetragonal IH is shown in Fig.~\ref{fig:ih_structure}. It is worth noting that two X atoms are sitting at non-equivalent sites in IH. In Mn-based IH, Mn is element X. From calculations, several Mn-based IHs are predicted to host ultrasmall skyrmions at room temperature. 

\begin{figure*}[t]
    \centering
    \includegraphics[width=\textwidth]{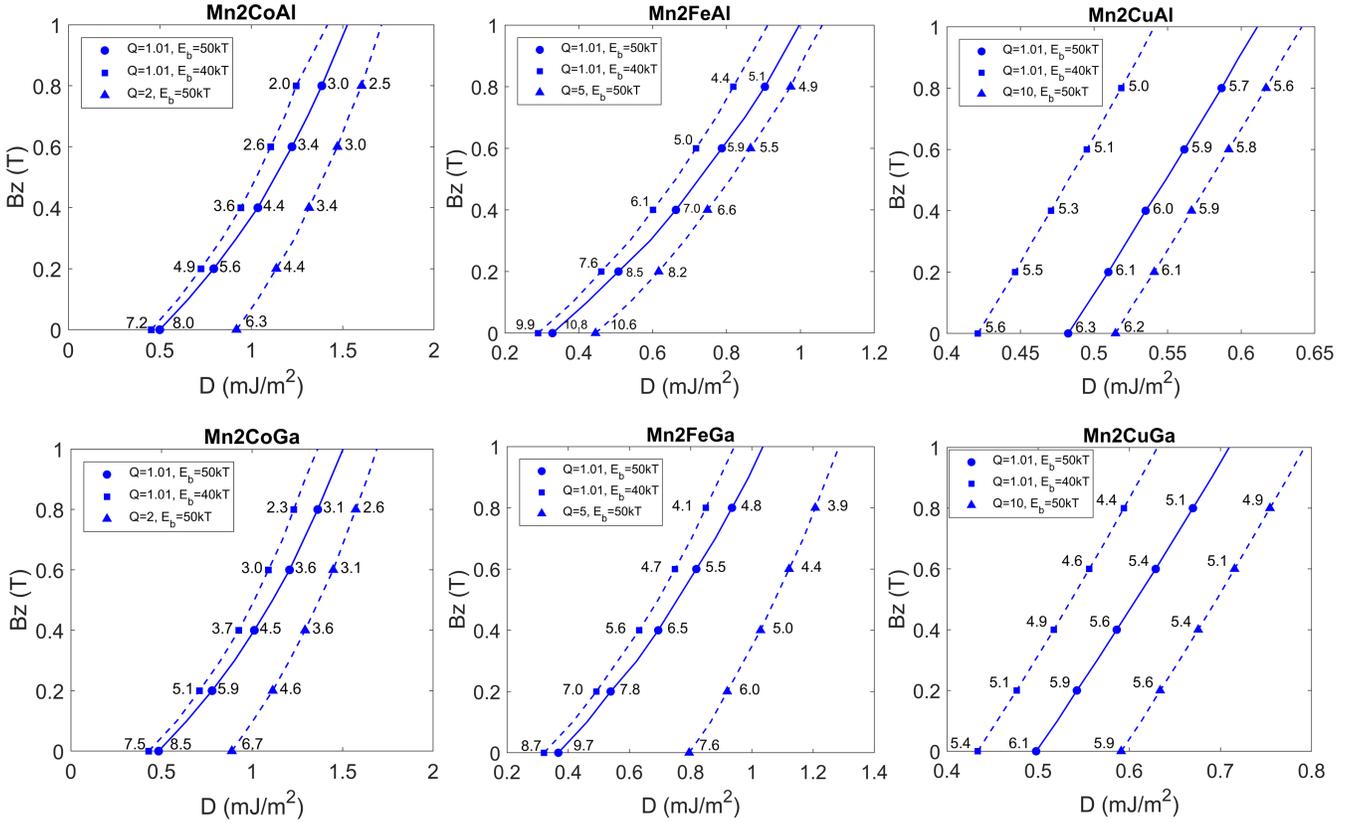}
    \caption{Smallest stable skyrmion boundary for inverse Heuslers \textcolor{black}{(zero-temperature LLG simulations of skyrmion sizes in a confined $100\times 100$ nm magnetic layer shown here alongside dots on the lines)}. The lines indicate the smallest skyrmions with an energy barrier $E_b=50 k_BT$ or $E_b=40 k_BT$ between the ferromagnetic state and skyrmion state. The scatter plot samples the skyrmion radius {in nm} along the boundary. The film thickness is assumed $5\,\mathrm{nm}$ in all calculations. $Q=2K/\mu_0M_s^2$ defines the effective anisotropy compared to the demagnetization, and $Q>1$ is needed for a perpendicular system. To show the effect of anisotropy on skyrmion size and stability, different $Q$ values are chosen for different materials with different saturation magnetization. Reproduced with permission from IEEE Transactions on Magnetics, vol. 56, no. 7, pp. 1-8, 2020 \cite{xie2020computational}. Copyright 2018 IEEE. \textcolor{black}{The magnetic field $B_z$ is parallel to the background, so the skyrmion shrinks with increasing $B$ field.} }
    \label{fig:ih_phase_space}
\end{figure*}

Figure~\ref{fig:ih_phase_space} illustrates the smallest stable skyrmions predicted in 5 nm thick Mn-based IHs {using LLG, assuming varying uniaxial anisotropies (scaled by the corresponding saturation magnetization, $Q = 2K_u/\mu_0M_s^2$) for a given energy barrier $E_b$}. Out of these IHs, Mn$_2$CuAl, and Mn$_2$CuGa are predicted to have $M_s$ less than 100 kA/m {(Appendix A)}. In zero applied field, the smallest stable skyrmions with a 50 k$_B$T energy barrier are predicted to be 6.3 nm in Mn$_2$CuAl and 6.1 nm in Mn$_2$CuGa. In addition to small $M_s$, the Gilbert damping coefficient of Mn-based IH is predicted to be in the order of $10^{-4}$ to $10^{-3}$ \cite{half-heusler}. This is significant because in addition to small $M_s$, low damping $\alpha$ is also important in obtaining high skyrmion speed, as per Eq.~\ref{eqvsk}. The small $M_s$ and small damping together make Mn-based IHs promising candidates to achieve stable and fast skyrmions at room temperature. \textcolor{black}{Mn$_2$CuAl , Mn$_2$CuGa, Mn$_2$CoAl and Mn$_2$CoGa have very large Curie temperature (above 700 K) which would is critical for reliable room temperature skyrmionics.}

In this section, we have discussed several promising ferrimagnets to achieve small and stable skyrmions at room temperature. Small $M_s$, {$\lessapprox 100$ kA/m} and {small to moderate, }$K_u$ $\lessapprox 300$ kJ/m$^3$ in these ferrimagnets are keys to stabilizing ultrasmall and fast skyrmions at room temperature (Table III). Computational models predict ultrasmall skyrmions in RE-TM, Mn-based Inverse Heusler, and Mn$_4$N. Experiments have measured close to 10 nm skyrmions in CoGd. For N\'eel skyrmions, while intrinsic properties of the magnetic layer are important for skyrmion size and stability, the DMI from the adjacent metal layers is equally crucial. Tuning of DMI will, in fact, have a strong influence on the properties of skyrmions \cite{soumyanarayanan_tunable_2017}, and will be discussed in the next section.\\
\begin{table}[htp]
\caption{Materials for hosting skyrmions. \textcolor{black}{CoGd and Mn$_4$N are promising materials due to their low $M_s$. Mn$_4$N has a very large Curie temperature which would make it stable at room temperature. \\}}
\begin{tabular}{c|c c c c}
\toprule
    Material & $M_s$(kA/m) & $K_u$ (kJ/m$^3$) & Size (nm) & $T_C$ (K)\\
    \hline\hline
    FeGe & 300-330 & 14 & Lattice & 280 \cite{FeGe,artcl:Yu2011}\\
    \hline
    CoGd & 100 & 50 & 10-150 & 450 \cite{Caretta2017,Quessab2020,Woo2018}  \\
    \hline
    Mn$_{2.5}$Ga & 250 & 1239 & N.A. & 600 \cite{Mn2.5Ga} \\
    \hline
    MnGa & 247.3 & 1000 & N.A. & 600 \cite{MnGa}\\
    \hline
    Mn$_4$N & 87-100 & 88-100 & 100-300 & 710 \cite{JmaxNeel1}\\
\bottomrule
\end{tabular}

\label{tab:promise materials}
\end{table}

\subsection{Parameter Engineering of Skyrmion Size and Stability } \label{subsectionParameters}

To achieve ultra-high density data storage and memory devices, it is desirable to scale down the size of skyrmions. In addition, the stability and lifetime of the skyrmion are other important aspects to consider for technological applications (e.g., volatile or non-volatile memories).\\

The topological phase in which the formation of skyrmions is possible highly depends on the competition of the different energies in the magnetic system as discussed in Sec.~\ref{subsectionSKstability}. Similarly, the radius of the skyrmion, as defined in Eqs. \ref{radius} and \ref{radius2}, is determined by \textcolor{black}{the saturation magnetization}, the DMI, the \textcolor{black}{effective} anisotropy \textcolor{black}{including exchange and stray fields} and the exchange interaction. Therefore, the tunability of the skyrmion size and stability is set by these parameters. A straightforward approach to this problem, as discussed in Sec.~\ref{subsectionFIM}, is to choose materials with a low saturation magnetization (to minimize the stray fields) \textcolor{black}{plus} a sizable DMI, such as in ferrimagnetic alloy thin films or multilayers \cite{Woo2018,Caretta2017,Quessab2020} and synthetic antiferromagnets \cite{Dohi2019,Legrand2020}.\\

Alternatively, several methods have been developed to engineer the DMI. Before we jump into the various approaches to engineer DMI, it is worth mentioning the physics of DMI briefly. In systems where magnetic thin films are grown on heavy metal layers, the broken spatial inversion symmetry leads to the antisymmetric interaction DMI \textcolor{black}{as mentioned earlier in Section I.D}. It is well established that the strong spin-orbit coupling (SOC) of HM layers adjacent to the magnetic layers has a dominant effect on the DMI. For Pt/Co bilayers, first-principles calculations demonstrate that DMI is almost purely interfacial, and the source of the DMI is found to be the SOC energy in the adjacent HM layer~\cite{Yang}. While HM SOC is crucial, it is not the only ingredient that contributes to the DMI. In Au-based systems, for instance, SOC alone fails to explain the smaller magnitude of the DMI despite the large SOC of Au. The other important ingredient that determines the DMI is the presence of HM metal d states around the Fermi level, which in turn controls the relative overlap between the d states of HM and magnetic elements, in other words, the orbital hybridization. Theoretical study based on a series of 3d transition metals monolayer deposited on several 5d HM metal substrates shows that the 3d-5d orbital hybridization plays a significant role in both the sign and magnitude of the DMI~\cite{Belabbes2016}. In the case of Au, the 5d states are far below the 3d states leading to a smaller degree of orbital hybridization that produces a smaller DMI value. Moreover, a recent study demonstrates that in HM/FM bilayers, the DMI magnitude and sign depend both on the orbital hybridization and the transitions between HM orbitals close to the K points in the BZ, namely, $d_{yz} \rightarrow d_{xy}$ and $d_{xz} \rightarrow d_{x^2-y^2}$ ~\cite{Banerjee}. To summarize, at HM/FM interfaces, the DMI emerges from the complex interplay between (i) inversion symmetry breaking, (ii) SOC of the HM metal layers, (iii) orbital hybridization between d states of HM layers and magnetic layers, and (iv) transitions between the HM d orbitals.

In metallic magnetic systems, interfacial DMI is commonly used to create skyrmions \cite{Boulle2016,woo2016observation}, so that we can tune the DMI by controlling the interfaces. A wide range of heavy metals with large spin-orbit coupling in proximity with a magnetic layer can induce DMI, as reported in first-principles calculations \cite{Belabbes2016}. A common approach to tuning the DMI relies on the overall chirality. Indeed, the bottom and top interfaces of a magnet both contribute to the net DMI. To maximize the net DMI, a magnetic layer can be grown on a heavy metal with a strong DMI, e.g., Pt/Co, and capped by a layer that has a weak DMI, such as oxides (e.g., MgO \cite{woo2016observation, Boulle2016} or TaO$_x$ \cite{Caretta2017, Arora2020}), or a strong negative DMI, e.g., Co/Ir. Using this additive method, an interfacial DMI close to 2 mJ/m\textsuperscript{2} was reported in Ir/Co/Pt multilayers, allowing the nucleation of sub-100 nm skyrmion \cite{MoreauLuchaire2016, Legrand2017}. Similarly, one of the highest surface DMI constant of about 2 pJ/m was measured in an ultrathin Pt/Co(0.5 nm)/MgO stack, which was shown to host 130-nm skyrmions \cite{Boulle2016}.

\begin{figure*}[t]
    \centering
    \includegraphics[width=0.7\textwidth]{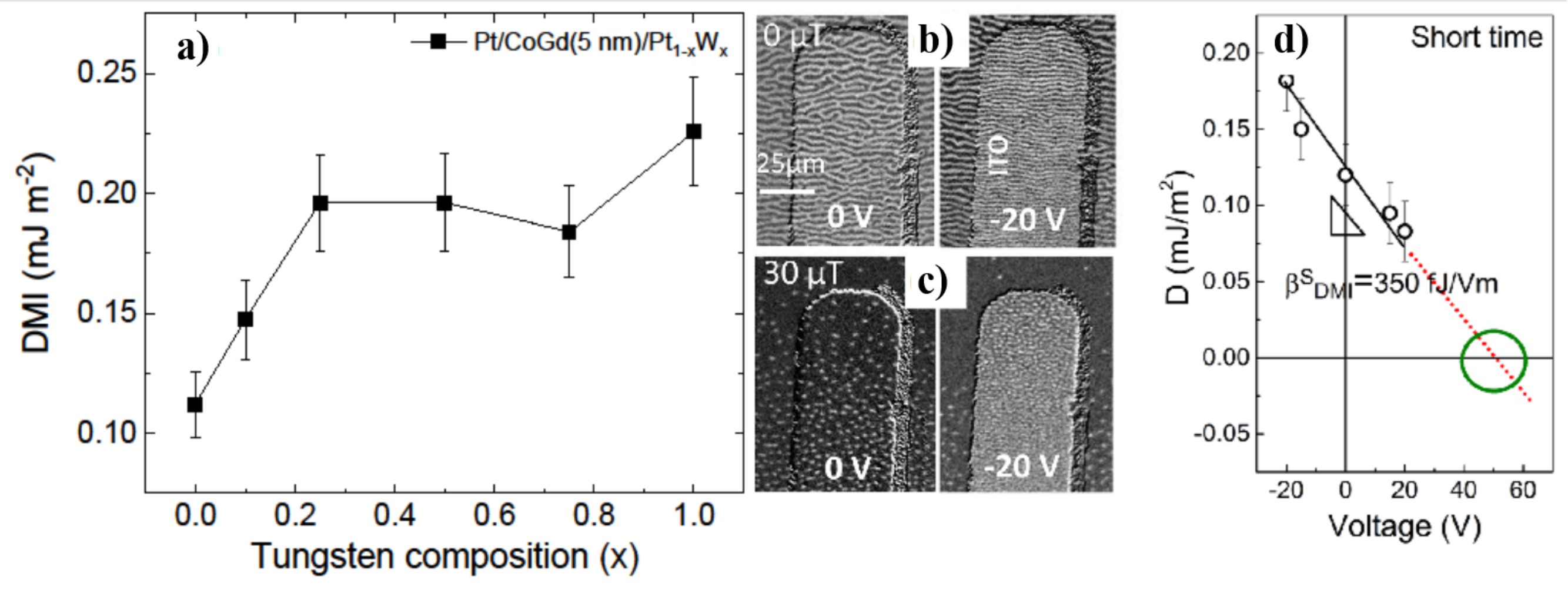}
    \caption{(a) Evolution of the DMI energy as a function of the W composition in Pt/CoGd(5 nm)/Pt\textsubscript{1-x}W\textsubscript{x} (Y. Quessab \textit{et al.} Sci Rep 10, 7447 (2020); licensed under a Creative Commons Attribution (CC BY) license.  \cite{Quessab2020}). (b-c) magneto-optical Kerr images of FeCoB/TaO$_x$ before and after applying a negative gate voltage at 0 $\mu$T and 30 $\mu$T, (d) variation of the DMI energy in FeCoB/TaO$_x$ as a function of the gate voltage (Reproduced with permission from Nano Letters. 2018, 18, 8. Copyright 2018 American Chemical Society 4871–4877\cite{Srivastava2018}).}
    \label{FigDMIControl}
\end{figure*}

Another method to control the DMI involves the use of heavy metal alloys, providing a finely tuned inversion symmetry breaking mechanism at the interface \cite{Quessab2020,LZhu2019}. In a recent study, Quessab \textit{et al.} demonstrated the ability to control the interfacial DMI in a thin ferrimagnetic alloy by varying the capping layer composition \cite{Quessab2020}. By gradually changing the W composition (x) in Pt/CoGd(5 nm)/Pt\textsubscript{1-x}W\textsubscript{x}, the stack symmetry was progressively broken to induce DMI, as seen in Fig. \ref{FigDMIControl}(a). The DMI energy was tuned over a large range, from a weak DMI to an interfacial DMI of 0.23 mJ/m, which allowed the formation of sub-100 nm skyrmions. Recently, Morshed \textit{et al.} investigated Pt/CoGd/$\text{Pt}_{1-x}W_{x}$ theoretically using first-principles calculations and found that the reduction of SOC at the metal layer near the capping layer interface and the simultaneous constancy at the bottom interface as a function of W composition are responsible for the saturating behavior of the DMI as shown in Fig.~\ref{FigDMIControl}(a) \cite{Morshed}. They also predicted tuning of the DMI by introducing a series of HM in the capping layer of Pt/CoGd/X where X=Ta, W, Ir, and found that W in the capping layer favors higher DMI while Ir produces lower DMI \cite{Morshed}. In another study, Zhu \textit{et al.} have also reported the tuning of the interfacial DMI  in a heavy metal/ferromagnet structure Pd\textsubscript{1-x}Pt\textsubscript{x}/FeCoB/MgO \cite{LZhu2019}. The DMI increased a factor of 5 as $x$ varied from 0 to 1.
Control of the spin-Hall efficiency was achieved as well, and a maximum spin-Hall angle of 0.60 was observed for Pd\textsubscript{0.25}Pt\textsubscript{75}, which could be of use for energy-efficient skyrmion dynamics induced spin-orbit torque.

Electric field control of the interfacial DMI has been recently demonstrated  \cite{Srivastava2018, HerreraDiez2019a}. Srivastava \textit{et al.} reported a 130\% variation of the DMI energy via voltage gating in Ta/FeCoB/TaO$_x$\cite{Srivastava2018} (see Fig. \ref{FigDMIControl}(d)). By applying a negative voltage (-20 V), the authors were also able to shrink the size of stripe domains and skyrmion bubbles, as depicted in Fig. \ref{FigDMIControl}(b) and \ref{FigDMIControl}(c), respectively. It was observed that ionic gating modifies not only the DMI but also the anisotropy and the saturation magnetization \cite{HerreraDiez2019a}. Interestingly, ion (He\textsuperscript{+}) irradiation was also shown to lead to an increase of the interfacial DMI and domain wall velocity in Ta/CoFeB/MgO due to disorder introduced at the interface~\cite{HerreraDiez2019b}.

 Computational methods, such as density functional theory (DFT) \cite{espresso:2009} and atomistic simulations \cite{Evans_2014}, can play an important role in parameter engineering and tuning. These methods can not only be used to rationalize an observation and uncover insights but also have predictive capabilities. We demonstrated the potential of DFT to calculate the saturation magnetization ($M_S$), magnetocrystalline anisotropy ($K_U$), and exchange coupling ($J$) for bulk ferrimagnetic Mn$_4$N.  The parameters calculated from DFT then feed the atomistic spin dynamics code, which in turn predicts the N\'{e}el temperature ($T_N$) for this system. \textcolor{black}{The saturation magnetization ($M_s$) is the total magnetic moment per unit volume, which can be calculated $M_s$ using the formula, $M_s = \mu_B/V$, where $\mu_B$ is the atomic magnetic moment (in Bohr magnetons) and $V$ is the unit cell volume (in cm$^3$). Both $\mu_B$ and $V$ can be obtained from spin-polarized collinear DFT calculations. We can also calculate $K_u$ using the formula, \begin{equation}\label{eq:ku} K_u = E^{[001]}-E^{[100]}\end{equation} where $E^{[001]}$ and $E^{[100]}$ are the total energies using non-collinear DFT calculations with the inclusion of spin-orbit coupling, where the spins are constrained in the [001] and [100] directions, respectively. The classical Heisenberg model suggested by K\"{u}bler {\it{et al.}} \cite{PhysRevB.28.1745} for the Heusler alloys was used to calculate $J$.}

\begin{table}
\footnotesize
\caption{The DFT calculated $M_s$, $K_u$ and $J$ parameters for the bulk ferrimagnetic Mn$_4$N using the experimental lattice constant value of 3.868~{\AA}. Our 0~K DFT calculations predict an in-plane [100]-direction as the easy-axis magnetocrystalline anisotropy.\\\\}
\vspace{-6mm}
\begin{center}
\begin{tabular}{c|c c c }
\toprule
Compound  & $M_s$(kA/m) & $K_u$(kJ/m$^3$) & $J$(meV)   \\
\hline
Bulk Mn$_4$N          & 121.94 & 1376.11  & 74.2    \\
\bottomrule
\end{tabular}
\end{center}
\label{tab:mn4n_dft}
\end{table}

\begin{figure}
\includegraphics[width=0.45\textwidth]{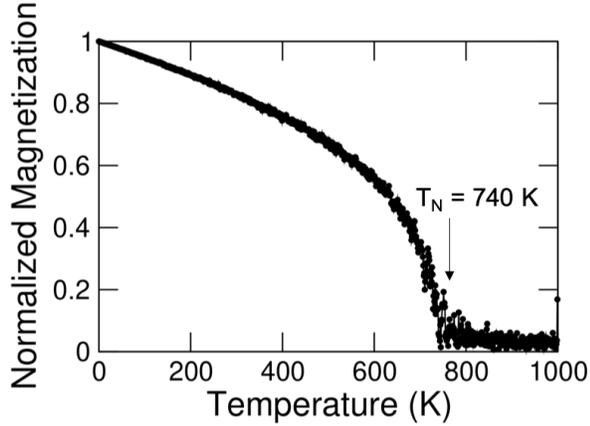}
\caption{\label{fig:mn4n_tn} Prediction of $T_N$ {for bulk Mn$_4$N} from atomistic simulation using the  DFT-calculated atomic magnetic moments, $K_U$, and $J$ as input (see Table \ref{tab:mn4n_dft}).}
\end{figure}

The calculated values for $M_s$, $K_u$ and $J$ using the experimental unit cell parameter value of 3.868~{\AA} is given in Tab. \ref{tab:mn4n_dft}. The predicted normalized magnetization {vs} Temperature curve is shown in Fig. \ref{fig:mn4n_tn}. The DFT parameters yielded a $T_N$ of 740~K from our atomistic simulation, which is in excellent agreement with the experimental $T_N$ of 745~K \cite{PhysRev.125.1893}.

\subsection{Nucleating isolated metastable skyrmions} 
The emergence of skyrmions in a material depends on the stability of the skyrmion phase. In some materials, the DMI stabilizes skyrmionic lattices that coincide with the ground state and are already present without nucleating them explicitly. Indeed, in epitaxial B20 FeGe films, it has been shown that skyrmions can appear at zero field, as evidenced by remanent topological Hall resistivity over a large range of temperatures \cite{Gallagher2017} or by electron holography in a FeGe nanodisk at 95 K \cite{Zheng2017}. Another example is in Fe/Ni/Cu/Ni multilayers, for which the interlayer exchange coupling allows the stabilization of zero-field and room-temperature skyrmions \cite{Chen2015}. However, for technological applications, it is usually desirable to create, {move} and annihilate skyrmions as a way of writing information. Therefore, materials where the skyrmion is a metastable state are more attractive, and in this case, several methods can be employed to nucleate multiple skyrmions, {with proposed device applications} \cite{Muller2017,finizio_deterministic_2019,je_targeted_2021}.

{\bf{Magnetic Fields:}} A common nucleation method relies on the use of a magnetic field that can be applied externally or locally (e.g., via an Oersted loop around a nanostructure \cite{woo2016observation}). In ferro- and ferri-magnetic films and multilayers with large demagnetization field, the ground state is formed by stripe domains that can be further shrunk into skyrmion bubbles \cite{woo2016observation, Lemesh2018,Woo2018, woo_deterministic_2018,juge_2019_current}. The application of an external magnetic field has also been shown to lead to the formation of antiferromagnetic skyrmions in synthetic antiferromagnets \cite{Legrand2020, Dohi2019}. Nevertheless, external magnetic fields are not compatible with the implementation of data storage or memory devices. 

{\bf{Spin Currents:}} On the other hand, several studies have reported the nucleation of metastable skyrmions by injecting an in-plane current in the presence of an out-of-plane magnetic field to provide stability \cite{Caretta2017, Lemesh2018, Song2020, Legrand2017}. Caretta \textit{et al.} demonstrated the nucleation of skyrmions in a nearly compensated ferrimagnetic CoGd alloy with a diameter as small as 10 nm at room temperature with a train of bipolar nanosecond pulses of a maximum amplitude of 1.7 $\times$10\textsuperscript{12} A m\textsuperscript{-2} \cite{Caretta2017}. Creation and annihilation of a single magnetic skyrmion was achieved in a ferrimagnetic Pt/GdFeCo/MgO multilayer using 10 ns current pulses with a current density of 2.5  $\times$10\textsuperscript{10} A m\textsuperscript{-2}
\cite{woo_deterministic_2018}. Furthermore, some studies have reported that current pulses can be used to nucleate skyrmions without applying an external magnetic field \cite{Lemesh2018, SWoo2017}. Lemesh \textit{et al.} found that in a Pt/CoFeB/MgO multilayer at 379~K, zero-field skyrmions could be nucleated upon injection of bipolar nanosecond current pulses \cite{Lemesh2018}. In a similar material, Woo \textit{et al.} reported the emergence of zero-field skyrmions at room temperature after sending 20~ns current pulses for 5~s with a repetition rate of 3.33 MHz and a current density of 1.6 $\times$10\textsuperscript{11} A m\textsuperscript{-2} \cite{SWoo2017}.

{\bf{Joule Heating:}} So far, the mechanism for the nucleation of metastable skyrmions has not yet been fully elucidated. Nevertheless, the use of current pulses, which are often applied over an extended period and with a large current density, suggests a nucleation mediated by thermal effects due to the Joule heating. This was suggested by Legrand \textit{et al.} \cite{Legrand2017} and later confirmed by the temperature-dependence study of current-induced skyrmion nucleation by Lemesh \textit{et al.} \cite{Lemesh2018}. Note that these studies report the nucleation of skyrmion arrays in the hottest area of the used wire devices and usually do not reliably nucleate single skyrmions. A connection to single skyrmion nucleation processes has been provided by B{\"u}ttner \textit{et al.} \cite{buttner_field-free_2017} that were able to produce similar results by application of millions of low current-density pulses that led to the formation of skyrmions on material inhomogeneities. \textcolor{black}{Wang \textit{et al.} have shown thermal generation and manipulations of skyrmions \cite{wang_thermal_2020}.}

\begin{figure*}[t]
    \includegraphics[width= \textwidth]{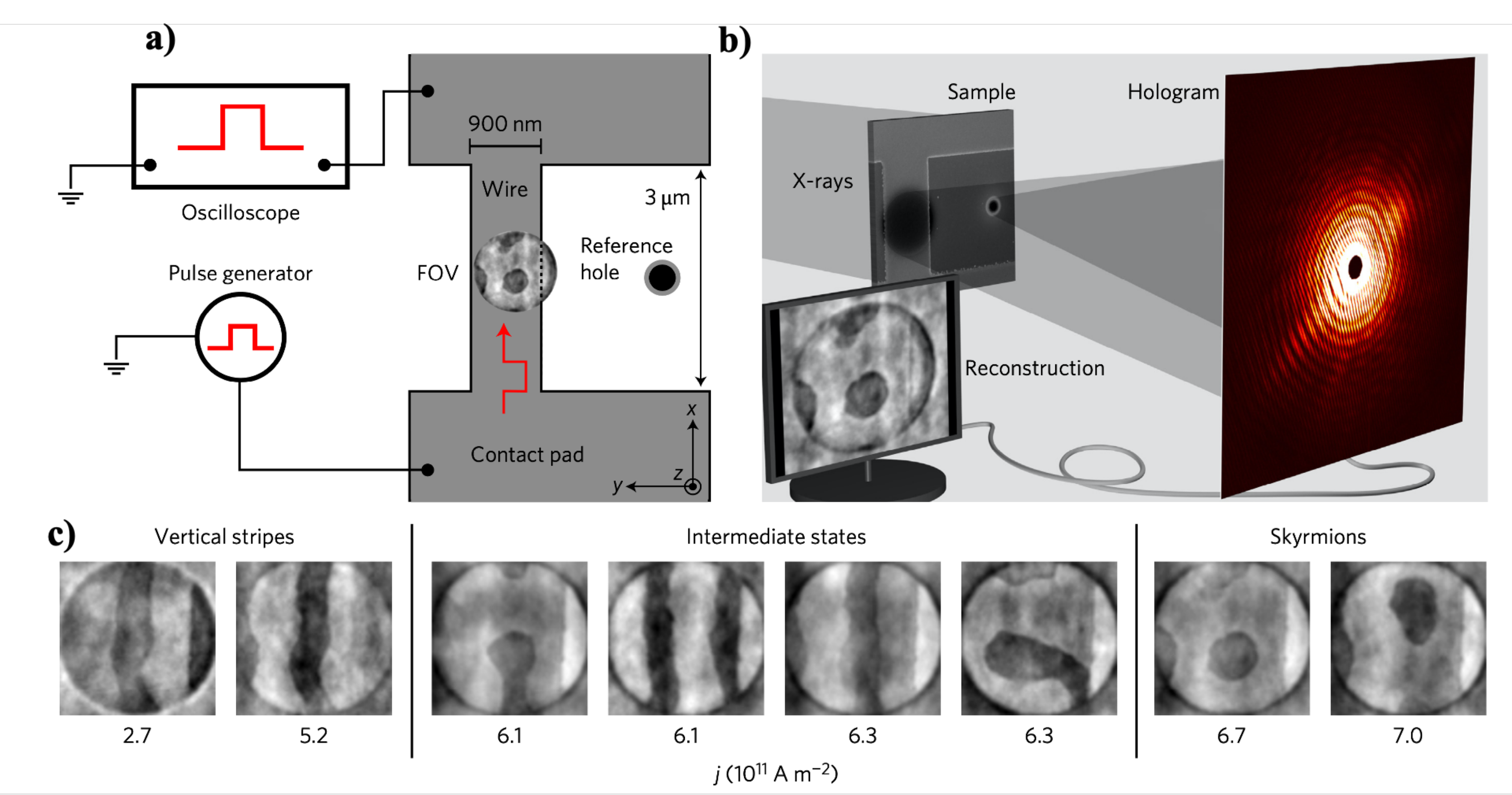}
    \caption{Observation of current-induced stripe and skyrmion nucleation in X-ray holography. (a) Schematic overhead view of the sample and connection to the pulse electronics. The sample consists of a magnetic wire that is directly connected to the pulse set-up. The center of the wire is imaged holographically (see (b)) with a circular FOV. (b) Imaging set-up. The magnetic wire was grown on a transparent Si$_3$N$_4$ membrane. The backside of the membrane is covered in gold, and two holes are cut by focused ion beam exposure to provide an interference pattern. The structure is exposed to circularly polarized X-rays. The interference of the transmitted beam of the two holes forms the hologram, from which an image of the out-of-plane magnetization in the wire can be reconstructed. (c) Nucleation of skyrmions and stripes with unipolar pulse trains. The injection of millions of pulses leads to the formation of stripe domains at low and skyrmions at high current densities. The black-and-white contrast denotes magnetization pointing in the +z and –z directions, respectively. Reproduced with permission from F. Büttner \textit{et al.} Nature Nanotech 12, 1040–1044 (2017). Copyright 2016, Nature Publishing Group  \cite{buttner_field-free_2017}.}
    \label{fig:nucleation}
\end{figure*}

\indent The general appearance of a current-induced skyrmion nucleation set-up is shown in Fig. \ref{fig:nucleation}. The current pulses first modulate the alignment and formation of stripe domains before breaking those into skyrmions at higher current densities. The involvement of thermal effects leads to a random contribution to the splitting of domains so that larger numbers of current pulses at lower current densities may suffice to nucleate a single skyrmion at a defect while at higher current densities require far fewer pulses to be applied to the sample, resulting in a lower nucleation latency.

{\bf Laser Pulses:} Alternatively, ultrafast nucleation of skyrmions using single ultrashort laser pulses was theoretically predicted \cite{Koshibae2014} and demonstrated in a CoFeB thin film \cite{Je2018} and a ferrimagnetic TbFeCo alloy \cite{Finazzi2013} as well as TmIG \cite{Buttner2020}. \textcolor{black}{Although the relatively thick films in the  TmIG study \cite{Buttner2020} show negligible Dzyaloshinskii-Moriya interactions, the results from \cite{prbtmig} suggest that in ultrathin rare earth iron garnet films, in which interfacial DMI has recently been found, fast thermal excitations might be used to controllably nucleate chiral magnetic skyrmions.} Interestingly, Je \textit{et al.} and B{\"u}ttner \textit{et al.} did not find any dependence on the light polarization; in fact, nucleation occurs even with linearly polarized light, which does not carry any angular momentum, thus demonstrating that heating could be a sufficient stimulus to create skyrmions \cite{Je2018}. Furthermore, it was demonstrated in a Pt/Co/Ir ultrathin multilayer that skyrmion bubbles in the micrometer scale could be formed using surface-acoustic waves (SAWs) in the presence of a small out-of-plane magnetic field \cite{Yokouchi2020}. It was found that the bubble size was independent of the {SAW} wavelength. The skyrmion density increased with the RF power and was maximum for SAWs with a wavelength of 16 $\mu$m, which is roughly twice the size of the observed bubbles. Remarkably, the authors argued that the SAWs did not generate a significant temperature increase that could explain the formation of the skyrmion bubbles \cite{Yokouchi2020}, which still keeps the mechanism for laser-induced skyrmion nucleation an open question.
\begin{figure}
    \centering
    \includegraphics[width=\columnwidth]{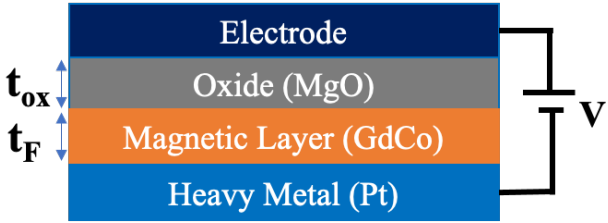}
    \caption{A schematic illustration of the VCMA gate.}
    \label{fig:VCMA}
\end{figure}

{\bf{Voltage Pulses:}}
\textcolor{black}{Another method to nucleate or annihilate skyrmions is by using the Voltage Controlled Anisotropy (VCMA) \cite{vcma_nucleation,schott_skyrmion_2017,bhattacharya_creation_2020}  (Fig.~\ref{fig:VCMA}). In this method a skyrmion is nucleated or annihilated by applying electric field pulses. The applied electric field can change the perpendicular magnetic anisotropy (PMA) according to the following equation}:
\begin{align}\label{vcma}
    K_u(V) = K_u (0) -{\xi V}/{t_{ox}t_{F}} 
\end{align}
\textcolor{black}{where $\xi$ is the VCMA coefficient, $V$ is the applied voltage, $t_{ox}$ is the oxide thickness and $t_F$ the film thickness. By lowering the anisotropy, the overall barrier for flipping of magnetic moments is also lowered, which would nucleate the skyrmion. Along a racetrack, the preferred locations of skyrmion nucleation would be at the defect sites, which already have lower anisotropy compared to their surrounding. Since the VCMA coefficient is not very large, the required voltage for skyrmion nucleation and annihilation is expected to be large ($\sim$7-20 V) \cite{schott_skyrmion_2017,bhattacharya_creation_2020} which would make its integration with an electric circuit challenging. This problem may be alleviated by building memory elements out of reshuttling skyrmions as in wavefront-based temporal memory, where occasional nucleation events can be used to replenish their finite lifetime at an energy cost that is readily amortized over multiple compute cycles. By using ion irradiation \cite{zhang_creation_2016} it is possible to engineer locations of the aforementioned defect sites in a controlled way, which, as we discuss in Section V, is critical for skyrmionic device reliability. Alternatively, voltage control for magnetic exchange bias has also been proposed for skyrmion nucleation \cite{li_electric-field_2017,yang_perspectives_2017}. The exchange bias (EB) is a property of a coupled antiferromagnetic (AFM)–ferro(ferri)magnetic (FM, FiM) system that occurs due to exchange coupling at the interface \cite{wu_reversible_2010,wu_full_2013}. Electric field control of EB has been used to achieve a deterministic and reversible switching of the magnetization in the FM (FiM) layer which can be used for the precise creation of single skyrmions \cite{guang_creating_2020}.}

\subsection{Skyrmion Stability and Lifetime}
As reported earlier, a number of different methods \cite{Caretta2017,Woo2018,Quessab2020} including STXM, MFM, and X-ray holography, have imaged 10 to 150 nm skyrmions in CoGd. The time needed for these measurements suggests lifetimes of at least several hours. Since decade-long lifetimes are needed for many applications, computational estimates are crucial to providing guidance on skyrmion lifetimes.

\begin{figure}
    \includegraphics[width=\columnwidth]{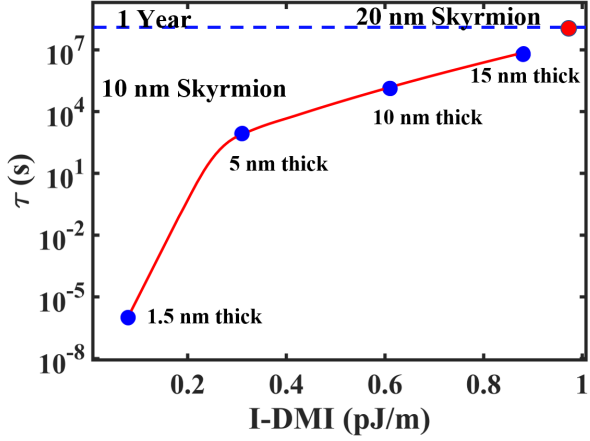}
    \caption{Simulated skyrmion lifetime by GNEB in CoGd films with $M_s$ of 50 kA/m. For 10 nm skyrmions, as indicated by the red curve, lifetimes range from microseconds in 1.5 nm thick CoGd film to near 10 days in 15 nm thick CoGd film. For larger skyrmions (20 nm), a year long  lifetime is found in 15 nm thick CoGd film (red circle).}
    \label{fig:skm-life}
\end{figure}

To calculate skyrmion lifetimes, the transition-state theory (TST) for spin transition \cite{NEB} has been employed. To use the TST, the minimum energy path needs to be calculated. Several studies have used the geodesic nudged elastic band (GNEB) method \cite{Bessarab2012,Bessarab2015} to compute the minimum energy path for skyrmion annihilation. From the MEP, the energy barrier and transition rate can be calculated. In the GNEB method, the two endpoints of the annihilation process along the energy landscape are set as the uniformly magnetized ferromagnetic and skyrmion states. These two states are then connected via a path which is divided into replicas (or `images') of the system.  Each image is connected to its neighboring images by a spring force. The total force from the springs as well as the force from the gradient of the energy landscape in the magnetization space is then calculated for each image. The GNEB approach matches are a more conventional nudged elastic band (NEB) approach, except in the former, the spring constants are also modified so that the highest energy image
climbs to the saddle point of the energy-magnetization landscape where the spring force disappears, and only the energy minimization force remains. The system is iterated based on the calculated dynamical equation until the path converges to the minimum energy path (MEP) with the desired accuracy. From the calculated MEP, the attempt frequency $f_0$ and the energy barrier $E_b$ are calculated and thence the lifetime is extracted, $\tau = f_0^{-1} e^{E_b/kT}$. \textcolor{black}{In the study by  Wild \textit{et al.} \cite{wild_entropy-limited_2017} a variation of more than 30 orders of magnitude in the attempt frequency was reported, which could suggest a dominant role for the entropy effects compared to the energy barrier.}
\textcolor{black}{Recently, spin-polarized scanning tunneling microscopy has been used to locally probe skyrmion annihilation by individual hot electrons \cite{muckel_experimental_2021}. Such measurements can be quite useful for a better understanding of skyrmion lifetime under practical conditions.}
Recently, room temperature lifetimes of up to 1 s were predicted for a 4 nm skyrmion in only a 7 layer ferromagnetic film \cite{Hoffmann2020}. This result is very promising for long-lifetime skyrmions at room temperature. According to the analytical model on skyrmions \cite{Buttner2018}, thicker films and larger skyrmions increase the energy barrier to annihilate a skyrmion, which leads to longer lifetimes. This is especially encouraging for ferrimagnets, such as CoGd and Mn$_4$N. As discussed earlier in Sec.~\ref{subsectionFIM}, intrinsic PMA allows the use of thicker films in these materials, so a lifetime of a decade may have already been achieved---and is certainly realistic. Our simulations with GNEB suggest that for 10 nm skyrmions, long lifetimes of up to 10 days are possible with CoGd thin films (Fig.~\ref{fig:skm-life}). For larger skyrmions $> 20$ nm with thick films ($> 15$ nm), years-long lifetimes seem feasible. Ultimately, there is a trade-off between film thickness and drivability because spin-orbit torque acts near the magnet-heavy metal interface. This brings us to a critical evaluation of skyrmion dynamics.

\section{Skyrmion Dynamics}
\label{sectionSkyrmionDynamics}
\begin{figure}
    \includegraphics[width= 8cm]{sot_skm.pdf}
    \includegraphics[width= \columnwidth]{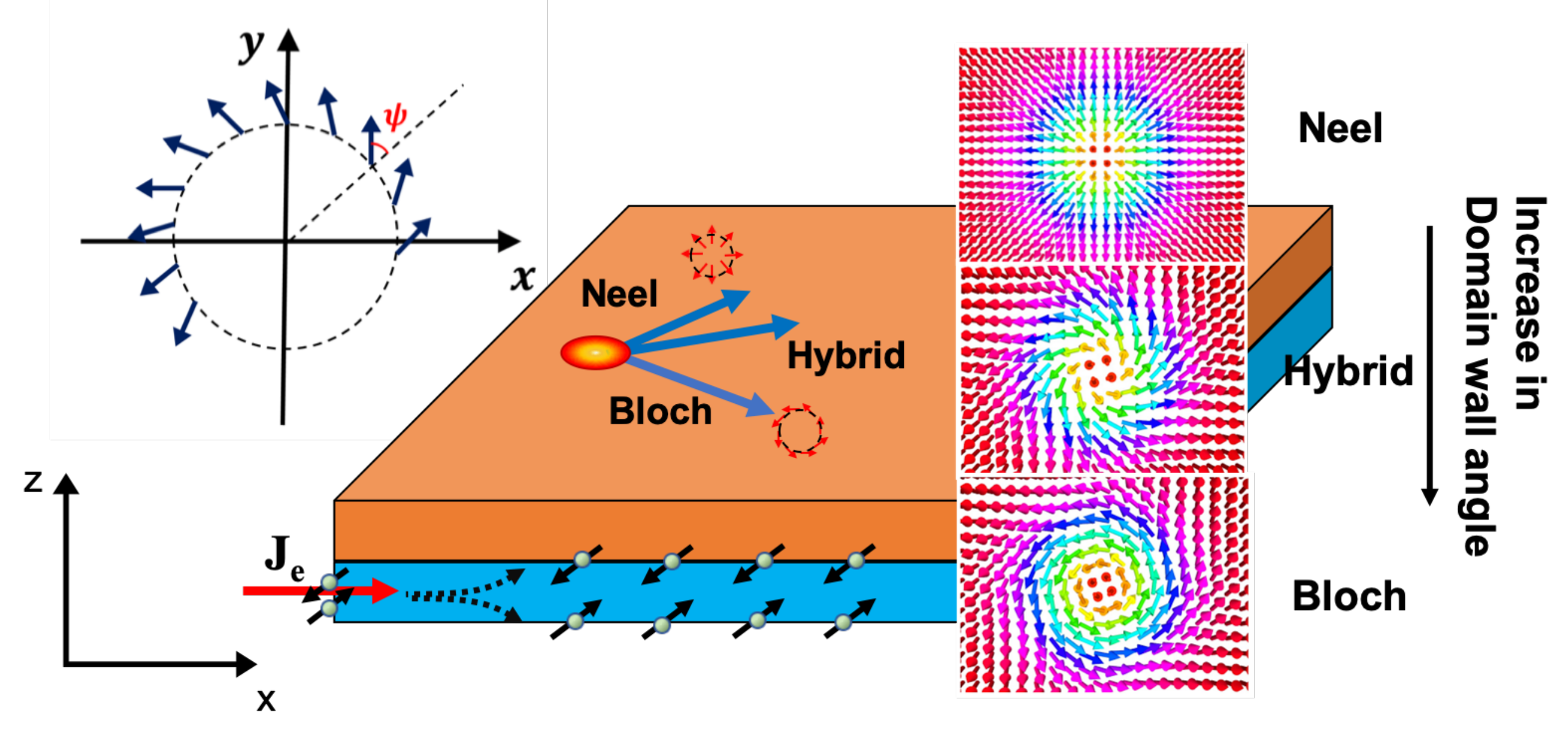}
    \caption{Current in a heavy metal underlayer separates spins through spin-orbit torque, and the injected spins lead to precessional torque on individual spins that drives a domain wall or a skyrmion forward. A finite tilt in domain angle $\psi$ causes the topological excitation to see a transverse Magnus force as well depending on whether it is N\'eel ($\psi = 0, \pi$), Bloch ($\psi = \pm \pi/2$) or hybrid.}
    \label{fig:sot_skm}
\end{figure}

\textcolor{black}{To have a low latency skyrmion-based device, fast skyrmions at small energy costs are required.}  
An efficient way to drive a skyrmion or a domain wall is using spin-orbit torque. As the picture shows (Fig. \ref{fig:sot_skm}), a flowing charge current in a heavy metal underlayer creates a vertical separation of spins along the axis perpendicular to the metal-magnet interface.
The injected spins create an effective magnetic field causing precession of the skyrmion spins around it, leading to a flow of the skyrmion whirring forward like a buzz saw. For a N\'eel skyrmion, there will also be a precessional phase lag in the transverse direction, leading to a net Magnus force that drives a N\'eel skyrmion at an angle with respect to the charge current. The actual expressions can be readily obtained from the Thiele approximation to the spin dynamics~\cite{thiele1973steady,hrabec_current-induced_2017, skmrace}.

\subsection{What determines skyrmion speed?}
We start with the LLG micromagnetic equation for the normalized magnetic moment, in presence of an applied SOT applied to different units $i$ (sublattices or layers)~\cite{slonczewski_current-driven_1996}
\begin{align}
    t_{F_i}\frac{M_{s_i}}{\gamma_i}\frac{d\boldsymbol{m}}{dt} &=
 - M_{s_i}\boldsymbol{m}\times \boldsymbol{H}_\mathrm{eff} +\alpha t_{F_i}\frac{M_{s_i}}{\gamma_i}\boldsymbol{m} \times \frac{d\boldsymbol{m}}{dt}\nonumber \\&  -a_j\: \boldsymbol{m}\times (\boldsymbol{m} \times \boldsymbol{P})-\eta \: a_j\: (\boldsymbol{m} \times \boldsymbol{P})
\end{align}
where $\boldsymbol{P}$ is the polarization of the injected spins from the heavy metal to the magnet, $a_j =  (\hbar/2q)\theta_\mathrm{SH}j_\mathrm{HM}$ is the adiabatic spin-orbit torque {coefficient} \textcolor{black}{that arises when conduction electrons follow the local texture of the spatially varying magnetization (i.e., the inertial term of a hydrodynamic equation)}, $\alpha$ is Gilbert damping, $\eta$ is the non-adiabaticity parameter, and $\theta_\mathrm{SH}$ is the spin Hall angle that relates transverse spin current density to longitudinal charge current density $j_\mathrm{HM}$, related through the strength of the spin-orbit coupling in the heavy metal.  

Using the Thiele approximation (see Appendix Subsection D) we get the following equation for i\textsuperscript{th} layer in a multi sublattice magnet~\cite{Buttner2018}
\begin{align}
    \boldsymbol{F}_{ex_i}+\Sigma_j F_{ij}+4\pi N_i\;t_{F_i}\;M_{s_{i}}(\hat{z}\times \boldsymbol{v}_{sk})/\gamma_i\nonumber\\ -t_{F_i}\;M_{s_{i}}\mathcal{D}.(\alpha\boldsymbol{v}_{Sk})/\gamma_i-\frac{4\hbar}{2q}\;I\; \theta_{{SH}} 
R(\psi)\boldsymbol{}{j}_\mathrm{HM}=0
\end{align}
We have taken $\boldsymbol{P}=\boldsymbol{j}_\mathrm{HM}\times\hat{z}$ to generalize to any $\boldsymbol{j}_{hm}$ direction. $\boldsymbol{F}_{ex_i} = -\mu_0  M_{s_i}\int \boldsymbol{H}_\mathrm{eff}.\nabla\boldsymbol{m}$ denotes forces from edges, defects, skyrmion-skyrmion repulsion and varying fields (for example anisotropy gradient), and is zero when $\boldsymbol{H}_\mathrm{eff}$ is uniform.
$F_{ij}$ on the other hand arises from interactions between neighboring layers {or sublattices}. The third term arises from the inertial term to the right of the LLG equation, with
$N_i = \pm 1$ the skyrmion winding number. The fourth term arises from the Gilbert damping term in the LLG equation, and involves $\mathcal{D}$ the dissipation tensor---assumed isotropic with diagonals ${\cal{D}}_{xx}$ set by the dissipation tensor, ${\cal{D}}_{ij} = \int dxdy(\partial_i\boldsymbol{m}\cdot\partial_j\boldsymbol{m}$). For a 2$\pi$ model this term matches the exchange integral to within a constant {(Eq.~\ref{eneqnsfit})}, ${\cal{D}}_{xy}={\cal{D}}_{yx}=0$, ${\cal{D}}_{xx} \approx R_{sk}/2\Delta \times f_{ex} = \rho f_{ex}(\rho)/2$. $\gamma_i$ is the gyromagnetic ratio,  $t_{F_i}$ is the thickness of sublattice i. $f_{ex}$ is the fitting term coming from the exchange integral. $M_{s_{i}}$ saturation magnetization of the ith film. The final term arises from the adiabatic spin orbit torque, where $I=\frac{1}{4} I_d (\rho,\Delta)$ is the integral of DMI energy Eq.~\ref{eneqns}, $I_d = \int_0^\infty dr[r\partial_r\theta + N_{sk}\sin{\theta}\cos{\theta}] = \pi R_{sk} f_\mathrm{DMI}$. This integral is, in effect, a shape factor of the skymion. $R(\psi)=\begin{pmatrix} 
\cos{\psi} & \sin{\psi} \\
-\sin{\psi} & \cos{\psi} 
\end{pmatrix}$ is the rotation matrix acting on the 2-D unit vectors and involves the domain angle $\psi$ ($\psi = 0, \pi$ for N\'eel skyrmions, $\pm \pi/2$ for Bloch, and in between for hybrid), the current density $\boldsymbol{j}_\mathrm{HM}$ in the HM layer assumed to be along the $x$ direction (polarization in $-\hat{y}$ direction). The non-adiabatic SOT term vanishes upon volume integration.

Solving for $\boldsymbol{v}_{sk}$, we get
\begin{equation}
    \boldsymbol{v}_{sk} = \displaystyle\frac{4\pi B\theta_{SH} R(\psi-\theta_0)}{\sqrt{\alpha^2S(T)^2{\cal{D}}_{xx}^2 + (4\pi S_N(T))^2}}\bold{j}_{hm},
    \label{eqvsk}
\end{equation}
where $B=\pi\hbar I/{2q}$, $S(T) = \Sigma_i{ M_{s_i}(T)t_{F_i}}/{\gamma_i}$ is the spin angular momentum summed over volume, $S_N(T) = \Sigma_i{N_{{sk}_i} M_{s_i}(T)t_{F_i}}/{\gamma_i}$ is the topological spin angular momentum and $\tan{\theta_0} = 4\pi\langle N_{sk}\rangle /\alpha{\cal{D}}_{xx}$.
\begin{figure*}[t]
    \centering
    \includegraphics[width=\textwidth]{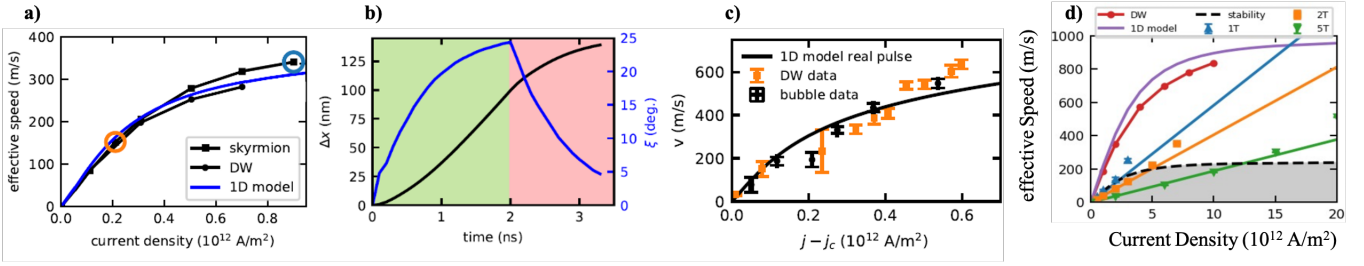}
    \caption{Non-linearity of drive and deformation as the origin of skyrmion speed reduction. (a) Simulated skyrmion and DW speeds as a function of applied current density. In contrast to Thiele model descriptions considering skyrmions as rigid particles (which expect skyrmion motion to follow a linear trend), the speeds of DWs and skyrmions are, in fact, practically identical and follow the 1-D DW model. (b) DW and skyrmion displacement (black line) increase even after the current is switched off due to inertial effects caused by the canting of spins in the wall (blue curve). This effect coupled with non-rectangular current pulses can {prolong} the displacement of skyrmions in an experiment. (c) Experimental observation of skyrmion and DW speeds and 1-D model taking into account the pulse shape. No significant difference is visible \cite{litzius2bePublished}.}
    \label{fig:nonlinear}
\end{figure*}

The velocity equation has a simple interpretation---we can write it as
\begin{eqnarray}
    qn_e{v}_{sk} = R(\psi-\theta_0){j}_{hm},
\end{eqnarray}
where the electron density $n_e$ is again set by angular momentum conservation {of the form}:
\begin{eqnarray}
    n_e \displaystyle\frac{\hbar}{2}  \theta_{SH} &=& \frac{M_s}{\gamma}{t_F}\sqrt{(4\pi N_{sk})^2 + \alpha^2{\cal{D}}_{xx}^2}, 
    \nonumber\\
    &=& \frac{M_st_F\alpha {\mathcal{D}}_{xx}}{\gamma}
    \sec{\theta_0}.
\end{eqnarray}
The new rotation matrix $R(\psi-\theta_0)$ tells us that there is a deviation $\Theta_{SkH}$ from the current path due to the Magnus force or the skyrmion Hall effect:
\begin{equation}
    \frac{v_{sk,y}}{v_{sk,x}} \equiv \tan{\Theta_{SkH}} = \Biggl(\frac{4\pi \langle N_{sk}\rangle \; \cos{\psi} - \alpha \: \mathcal{D}_{xx} \;\sin{\psi}}{4\pi \langle N_{sk}\rangle\; \sin{\psi}+\alpha  \:\mathcal{D}_{xx}\; \cos{\psi}}\Biggr),
    \label{eq:skh}
\end{equation}
{With the average topological index $\langle N_{sk}\rangle = S_N/S$. We see that even in the absence of dissipation, there is a topological damping term proportional to $N_{sk}$ that limits the skyrmion speed {compared to a domain wall}. which depends on the magnitude of the topological term $N_{sk}$ relative to the effective damping} $\alpha{\cal{D}}_{xx}$ (Gilbert damping times shape factor)---one term tries to move the skyrmion perpendicular to the current, while the other tries to bring it back to the current. The rotation matrix depends on the domain wall angle $\psi$, trying to move along the current for N\'eel skyrmions ($\psi = 0, \pi$) and perpendicular for Bloch skyrmions ($\psi = \pm \pi/2$).

Let us now compare these skyrmion velocities with the simpler 1-D DW equations within the Thiele approximation: 
\begin{equation}
    v_\mathrm{DW} = \frac{\pi}{2}\frac{D_{int}j_\mathrm{HM}}{\sqrt{(S_0(T)j_{hm})^2+(\alpha S(T)j_0)^2}}.
    \label{eqdw}
\end{equation}
With $S_0(T)= |S_1(T)-S_2(T)|$, $S= |S_1(T)+S_2(T)|$ and $j_0 = 2qt_FD_{int}/\hbar\theta_{SH}\Delta_0$. For speeds close to the magnon speeds, relativistic effects have to be taken into account. In Appendix. B, we explain the relativistic dynamics of the skyrmions and DWs and the relativistic corrections to Eqs.~\ref{eqvsk},\ref{eqdw} in detail.

{Note that Eq.~\ref{eqdw} suggests that \textcolor{black}{ferromagnetic} 
domain wall speeds saturate to a constant $\propto D_{int}/M_\mathrm{eff}$ for large current densities $j_\mathrm{HM}$ that sit in both the numerator and the denominator of the velocity term $\boldsymbol{v}_\mathrm{DW}$, leading to an overall damped motion ({Fig.~\ref{fig:nonlinear}}). This may imply that
skyrmions could quickly overcome the speeds of a corresponding straight, 1-dimensional domain wall in the same material}. However, it is important to realize that skyrmion speeds are furthermore dependent on internal spin dynamics, {which start to dominate at high currents and are not captured by the Thiele equations}. A simple example of this is the change in size and shape of a skyrmion, which occurs under the influence of a current \cite{litzius2017skyrmion, Litzius2020}. Especially at higher current drives, these effects can dominate the skyrmion dynamics, leading to an elongation of the skyrmion and the formation of a domain wall-like section within the otherwise round skyrmion. These, in turn, cause them to be driven by the same dynamics as domain walls. Together with other effects such as added topological damping, it becomes apparent that a skyrmion cannot move faster than a domain wall under identical conditions.

\indent{It is also important to point out that by applying a proper 1-D domain wall model that accounts for the drive’s pulse shape and inertial effects caused by domain wall spin canting (see Fig.~\ref{fig:nonlinear}), skyrmions near the relativistic speed limit (Appendix Subsection D) move with the same speeds as domain walls. This has been observed both in micromagnetic simulations as well as current-driven experiments on ferrimagnetic CoGd (Fig. \ref{fig:nonlinear}). The relativistic speed limit increases strongly for antiferromagnets {($M_\mathrm{eff} = 0$)} with zero topological damping, while the limitation set by skyrmion distortion will still remain in place and prevent skyrmions from moving faster than domain walls.}\\

Even nanoscale skyrmions like those reported by Romming {\it et al.} \cite{romming2013writing}, with arguably the highest rigidity of all skyrmions (due to small size) are not impervious to the effect, as Fig. \ref{fig:nonlinear}(d) shows: The mobility of the skyrmions (blue, orange and green data) initially follows the linear mobility curve for undistorted skyrmions that can be derived from the Thiele equation \cite{Buttner2018}. However, after a critical current density is reached, the skyrmion cannot withstand the enormous torques any longer and collapses long before reaching the 1-D domain wall limit (purple, red) in the material. The gray shaded area gives a phenomenologically determined stability regime for skyrmions, outside of which the skyrmion starts to loose its rigidity. Note that the slope of the Thiele mobility increases with decreasing magnetic field up to about 1~T, where it becomes too low to stabilize skyrmions in the material.


\subsection{Skyrmion Hall Effect and Gateability}
\begin{figure}[h]
    \centering
    \includegraphics[width=\columnwidth]{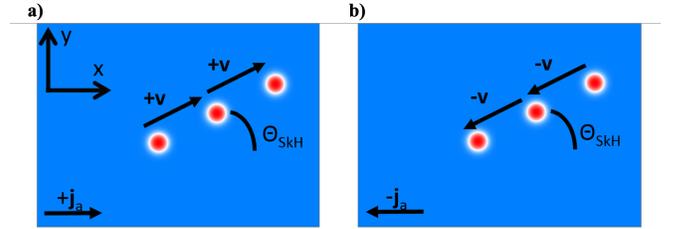}
    \caption{Sketch of the skyrmion Hall angle. Skyrmions, due to their topological nature, do not move parallel to a drive but at an angle, called the skyrmion Hall angle $\Theta_{SkH}$. This angle lets them drift towards the edge of a device, complicating their application. For compensated ferrimagnets and antiferromagnets, the two oppositely oriented sub-lattices can negate one another and reduce the skyrmion motion to one parallel to the driving current. Reproduced with permission from Phd thesis of K. Litzius, Johannes Gutenberg-Universität Mainz (2018). Open Access 2018. \cite{litziusPhD}.}
    \label{fig:skyrmionHall_sketch}
\end{figure}

\begin{figure*}[t]
    \centering
    \includegraphics[width=0.7\textwidth]{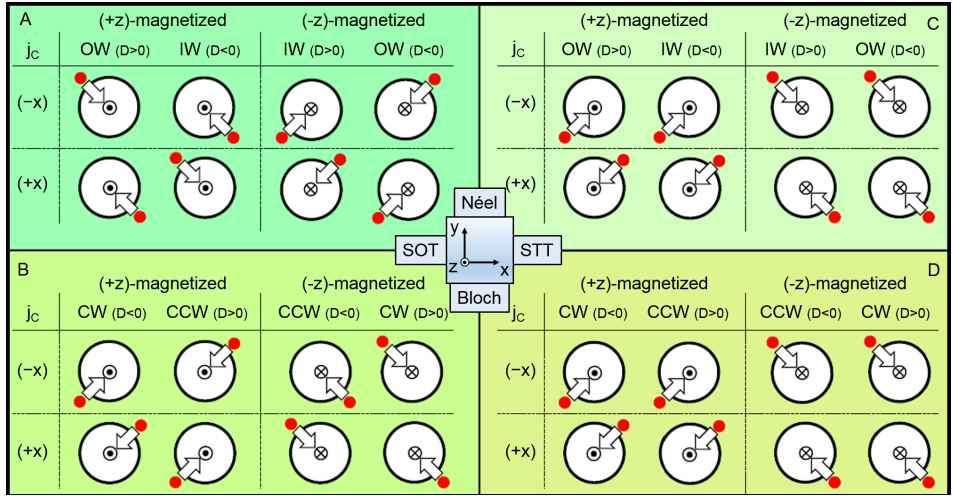}
    \caption{Characteristic motion directions for different skyrmion types and drives. The arrow originating at the initial skyrmion position indicated by the red dot gives the trajectory of the skyrmion. $j_C$ gives the charge current direction and the domain wall orientation of the skyrmions is indicated by OW (outward), IW (inward), CW (clockwise),and CCW (counter-clockwise). K. Litzius, Ph.D. thesis, Johannes Gutenberg-University Mainz, 2018. Open Access 2018. \cite{Honda2017, litziusPhD}}
    \label{fig:hall_motion}
\end{figure*}

One of the most defining characteristics of skyrmions is their topological property. As they are topologically non-trivial and can be mapped by a continuous transformation onto a sphere, these structures experience an uncommon response to applied currents. In contrast to topologically trivial domain walls, skyrmions exhibit the so-called skyrmion Hall effect; a force that drives them at an angle, the skyrmion Hall angle ($\Theta_{SkH}$) with respect to the applied driving current (Fig. \ref{fig:skyrmionHall_sketch}).
This curious property of skyrmion dynamics can complicate the applicability of skyrmions in data storage devices, especially in racetrack devices, as that relies on reproducibly moving skyrmions back and forth along the track. While this motion is technically possible and has in fact been reported before \cite{litzius2017skyrmion}, in the presence of strong current pulses along a narrow track, needed for fast storage and high bit density, skyrmions can crash into the track edges and get annihilated. 
We can predict the skyrmion Hall angle for any combination of skyrmion type, chirality, core polarization, and drive (Eq.~\ref{eq:skh}). Fig. \ref{fig:hall_motion} gives a summary of the most common combinations {shown as different quadrants for STT vs. SOT drives, and for N\'{e}el vs. Bloch skyrmions, for different magnetizations, current directions, for different signs of the DMI $D$.}

\begin{figure}[htp]
    \centering
    \includegraphics[width=\columnwidth]{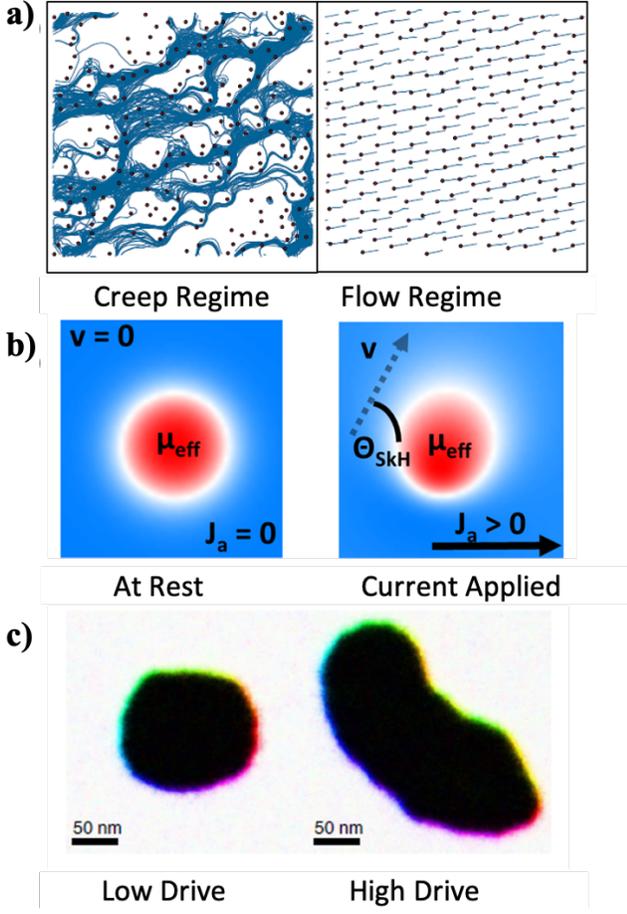}
    \caption{Three mechanisms that change the skyrmion Hall angle. (a) Pinning sites in the material randomize the skyrmion trajectories at low current densities and thus reduce the skyrmion Hall angle. This effect is mostly present in the creep, and depinning regime. C Reichhardt and C J Olson Reichhardt 2016 New J. Phys. 18 095005; licensed under a Creative Commons Attribution (CC BY) license. \cite{Reichhardt2016}. (b) At high drives, skyrmions can be deformed by the field-like component of the spin-orbit torque, thus changing their effective size and the skyrmion Hall effect. Reproduced with permission from K. Litzius \textit{et al.} Nature Phys 13, 170–175 (2017). Copyright 2017, Nature Publishing Group \cite{litzius2017skyrmion}. (c) Additionally, bubble skyrmions can experience DW oscillations and size deformations that can alter the skyrmion Hall angle. Reproduced with permission from K. Litzius \textit{et al.} Nat Electron 3, 30–36 (2020). Copyright 2020, Nature Publishing Group \cite{Litzius2020}.}
    \label{fig:skyrmionHall}
\end{figure}
\begin{figure}
    \centering
    \includegraphics[scale=0.5]{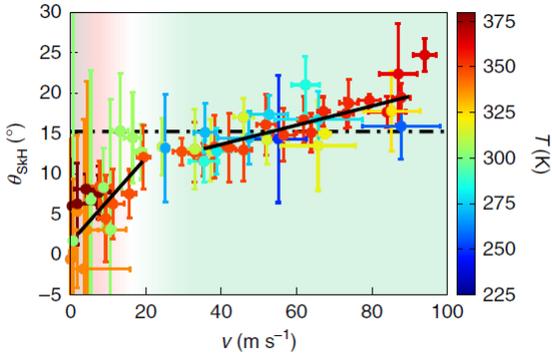}
    \caption{Temperature and drive dependence of the skyrmion Hall angle. No significant temperature dependence of the skyrmion Hall effect was found when plotted against the skyrmion velocity. Note that the latter can experience temperature dependencies due to {randomization of the} torque and conductivity. Two regimes, corresponding to creep and depinning/flow regime, respectively, are visible. Reproduced with permission from K. Litzius, \textit{et al.} Nat Electron 3, 30–36 (2020). Copyright 2020, Nature Publishing Group \cite{Litzius2020}.}
    \label{fig:skyHall_temperature}
\end{figure}
From the previous models, one might expect a constant skyrmion Hall angle for any drive since it appears not to depend on the applied current density. However, a series of experiments \cite{litzius2017skyrmion, jiang2017direct, juge_2019_current} have recently shown that this, in fact, is not the case and that the skyrmion Hall angle in ferromagnets depends strongly on the applied drive. Several mechanisms have been proposed to explain this deviation from the Thiele model. Most notably, the theory by Reichhardt \& Reichhardt \cite{Reichhardt2016} that proposes pinning effects to randomize the skyrmion trajectories in the creep and depinning regime (see Fig. \ref{fig:skyrmionHall}(a)). Other theories have proposed mechanisms that cause skyrmion deformations in the flow regime and could also contribute to a change in the skyrmion Hall angle \cite{litzius2017skyrmion, Litzius2020}. These deformations could either originate in a field-like spin-orbit torque contribution (Fig. \ref{fig:skyrmionHall}(b)) \cite{litzius2017skyrmion}, or in material inhomogeneities that induce random fluctuations in the skyrmion, which change the dynamics in a non-rigid way that is not included in the Thiele model (Fig. \ref{fig:skyrmionHall}(c)) \cite{Litzius2020}.

Aside from the drive dependence of the skyrmion dynamics, no direct influence of thermal effects on the skyrmion Hall angle has been found (Fig. \ref{fig:skyHall_temperature}) over a wide range of temperatures.

Since the skyrmion Hall angle is typically an undesirable property of skyrmionics, much work has been put into reducing its influence. The most promising approach is to employ compensated ferrimagnetic and anti-ferromagnetic materials, in which the skyrmion Hall angle cancels out due to the existence of sub-lattices of opposite polarization ($c_\mathrm{p}$). In the former, a reduction of the skyrmion Hall angle has already been observed experimentally \cite{Woo2018}, which makes these material classes currently the most promising candidates for skyrmion-based data storage devices. {There are other ways to create {\it{dynamic}} compensation points, typically by using graded material compositions, which can lead to automatic self-focusing of propagating skyrmions into quiescent lanes where the Magnus force cancels (Fig.~\ref{fig:hybrid}).}

\subsection{The role of Defects and thermal Diffusion}
Due to thermal effects, skyrmions in a race track can undergo diffusive displacement. If the racetrack is defect-free with little to no pinning sites, a skyrmion would be extremely susceptible to such thermal effects. This kind of random walk can be beneficial in stochastic applications of skyrmions, discussed later---but potentially destructive for some applications of skyrmions that encode analog information into their coordinates, as discussed in Sec.~IV. {Furthermore, skyrmions show inertia-driven drift shortly after a current pulse is removed, rather than stopping immediately (Fig.~\ref{fig:nonlinear}(b))}. 
One way to control such undesirable motion is by engineering confinement barriers such as point defects with a different anisotropy or notches etched into the racetrack (missing material). {This would increase the periodic pinning of skyrmions uniformly along the racetrack, instead of a random creep}. But it should be noted that the pinning must be small enough for the skyrmion to be movable by moderate magnitudes of current. 

The mean squared displacement (MSD) of skyrmion due to diffusion can be described as \cite{brwonianmassive,prlbrwonian} $\langle(r(t)-r(0))^2\rangle =4D_{diff}t$ where $r(t)$ is skyrmion position after time $t$. $D_\mathrm{diff}$ is the diffusion constant: 
\begin{equation}
D_\mathrm{diff} = k_BT\displaystyle\frac{\alpha \mathcal{D}_{xx}}{(4\pi S_N)^2+(\alpha \mathcal{D}_{xx})^2} 
\end{equation}
To calculate the localization time with a certain probability, we can use the Gaussian distribution:
\begin{equation}
    P_\mathrm{acc} = \int_{-L/2}^{L/2}dr\displaystyle\frac{1}{\sqrt{8\pi t D_\mathrm{diff}}}e^{\displaystyle -r^2/(8tD_\mathrm{diff})}.
    \label{gaussian}
\end{equation}
For a desired accuracy ($P_\mathrm{acc}$) and localization time (t) and length (L), we can get the required $D_\mathrm{diff}$.
To get a sense of magnitude, if we require that the center of a skyrmion remains localized to within $\pm$ 40~nm with probability $1-10^{-8} = 0.99999999$ for 1 hr, the diffusion constant must be less than $10^{-10}$~$\mu\text{m}^2/\text{s}$. The table below (Table~\ref{fig:diff}) shows the relation between diffusion constant and confinement lifetime.

\begin{table}[b]
\begin{center}
\caption{Table of diffusion constants and corresponding $\pm 40$ nm skyrmion localization times at $1e^{-8}$ inaccuracy.\\}
\begin{tabular}{l|l}
    \toprule
    $D_{diff}~(\mu m^2/s)$ & Localization time (t)\\[1mm]
    \hline\hline \rule{0mm}{3mm} 
    $1e^{-2}$\cite{prlbrwonian} & Volatile (ms) \\[1mm]
    \hline\rule{0mm}{3mm}
    $1e^{-8}$(CoGd)~\cite{BeachPrivate} & ~$10^3$ (s) (Cache) \\[1mm]
    \hline\rule{0mm}{3mm}
    $1e^{-10}$ & ~1 Day \\[1mm]
    \hline\rule{0mm}{3mm}
    $1e^{-12}$ & ~1 Year \\[1mm]
    \hline\rule{0mm}{3mm}
    42 k$_B$T Barrier & ~10 Years (Hard drive) \\
    \bottomrule

\end{tabular}
\end{center}
\label{fig:diff}
\end{table}

\begin{figure}
 \centering
    \includegraphics[width =\columnwidth]{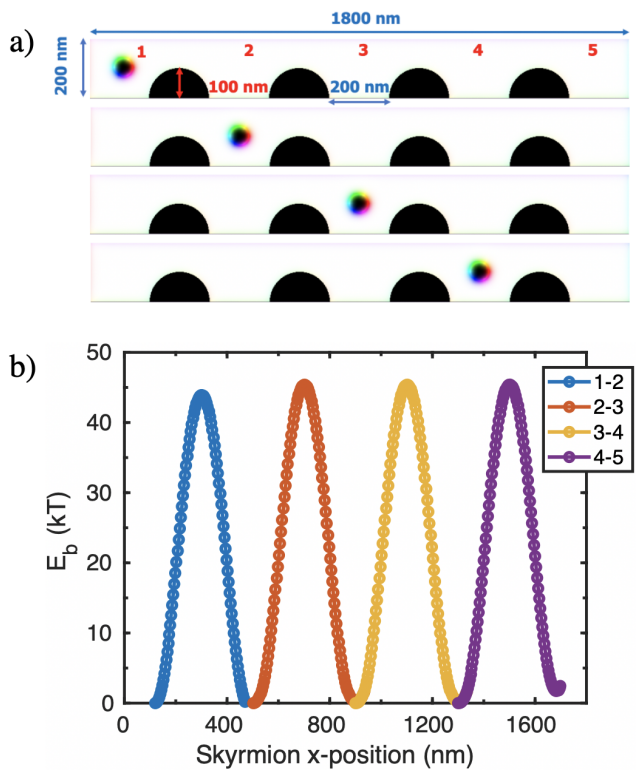}
    \caption{a) {The racetrack geometry with four identical semicircular notches (radius 100 nm) separated by 200 nm from each other.} The space between notches would be the places where the skyrmions would be positioned. b) Calculated minimum energy path for skyrmion passing a defect using {the parameters mentioned in the main text.} By using the string method \cite{defect}, the MEP from one side of the defect to the other side is calculated. The minimum unpinning current is calculated to be a 4 ns current pulse of $1.4\times10^{11}$ A/m$^2$}.
    \label{fig:notch}
\end{figure}

By placing a notch (Fig.~\ref{fig:notch}), it is possible to constrain a skyrmion's position. The skyrmion's positional lifetime (meaning the time it takes to go from one side to the other) can be calculated from transition state theory, $\tau_p = f_0^{-1} e^{E_b/k_BT}$, where $f_0$ is the attempt frequency. Approximately an $E_b$ of 30 $k_B$T will result in positional lifetime of seconds for an attempt frequency  $f_0$ of ~$10^{10}$ Hz \cite{bessarab_lifetime_2018,defect}. In the presence of a notch, the skyrmion needs to shrink in size in order to go past it, making it energetically unfavorable. The geometry and composition of the defect should be optimal so that the skyrmion can be localized in equilibrium and driven past it with a modest drive current without being annihilated at the notch or the racetrack edge. 

 As seen in Fig.~\ref{fig:notch}, the skyrmion positional lifetime can be long enough (years) for long-term memory applications and at the same have a moderate unpinning current. In Fig.~\ref{fig:notch}, the material parameters are {$M_s=100$ kA/m, $A_{ex}=12$ pJ/m, $K_u = 50$ kJ/m$^3$, 200 nm racetrack width, 5 nm thickness and 100 nm notch radius.} The notch is created by removing relevant parts of the racetrack in our model. By using a few of these notches consecutively, we find that it is possible to localize the skyrmion, in effect discretizing the skyrmion position. The skyrmion would still execute random rattles about its localized position between two neighboring notches, but it can not go over to the next block without a current pulse.
 
{As was mentioned in section III.F, to determine $E_b$, the minimum energy path (MEP) between two sides of the notch needs to be calculated. To find the MEP here, we used the string method \cite{String}. Similar to the MEP calculations for the skyrmion lifetime, an initial path needs to be determined by hand.} {As discussed earlier, in the skyrmion lifetime calculations} the initial path is between the skyrmion and the ferromagnetic background ground state. {For the} positional lifetime, however, the skyrmion path is between two sides of the notches. {The path is divided into multiple images of the system to define the initial path}. Then at each iteration, the path is moved toward the minimum energy and reparametrized. This is repeated until the path converges close enough to the MEP. 

\section{Skyrmionic Device Applications}\label{devices}
The original racetrack idea \cite{Parkin190} relied on a sequence of domain walls separating regions of opposite polarization, encoding the 0 and 1 bits. These domain walls on racetracks can have very high density, and the racetracks, in principle, can be stacked in 3-D. However, domain wall motion is restricted by pinning and creep. For skyrmions, information is contained in their absence or presence, but it relies once again on the ability of individual skyrmions to execute the particulate, quasi-ballistic, deterministic driven motion. While conventional densely sequenced DW/skyrmions-based memory is a possible application, there are other novel applications that can leverage the size and mobility advantages of such topological magnetic textures. We now discuss some of the device concepts involving skyrmions, including key operations needed such as read, write, erase and move. These applications could be key to building native memories and accelerators for applications such as temporal logic that leverages temporally encoded data for processing, stream-based computing that uses continuously streaming bits and is particularly oriented towards signal processing, for building spiking neural networks, ultra-compact reservoir computing, and nano-oscillators. While all of these can be built using conventional CMOS technology, skyrmions may provide scalability due to the compactness of the circuits and low power consumption due to the physics of the skyrmion motion mapping onto the application model.


\subsection{Magnetic Memory for Temporal Computing}

\begin{figure}
    \centering
\includegraphics[width=\columnwidth]{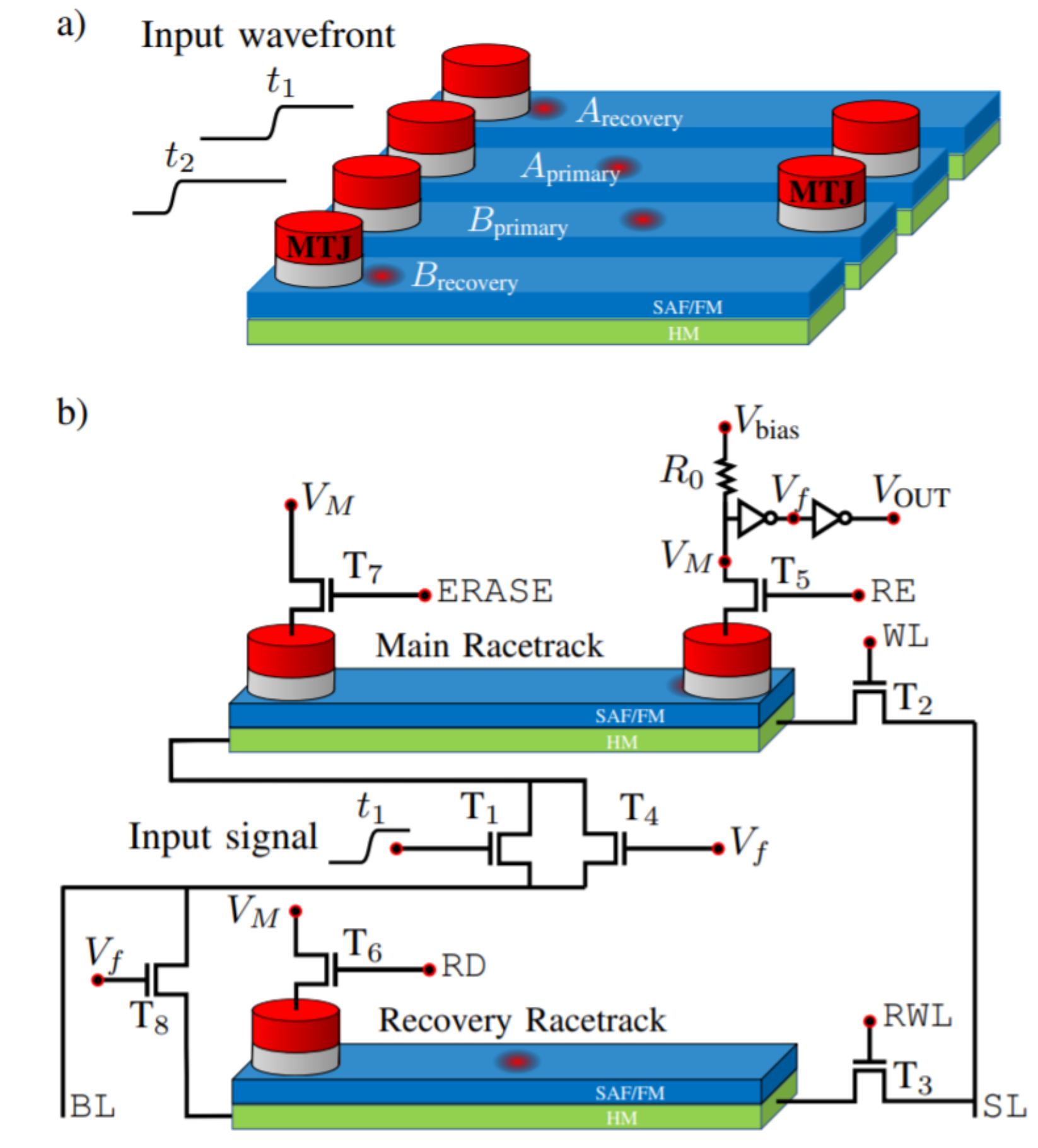}
   \caption{Temporal memory circuit illustration and concept: panel (a) shows the concept of N=2 cells single column temporal memory capable of storing and playing temporal wavefronts. A magnetic tunneling junction is placed for skyrmion detection. Panel (b) shows the
detailed description of the temporal memory circuits in each cell.   The Control lines consist of source line  (SL), the bit line  (BL),   write line  (WL),  recovery write line (RWL), read enable(RE), recovery line (RD), and erase line (ERASE). H. Vakili \textit{et al.}, IEEE Journal on Exploratory Solid-State Computational Devices and Circuits, vol. 6, no. 2, pp. 107-115, 2020; licensed under a Creative Commons Attribution (CC BY) license. \cite{wavefront}. \textcolor{black}{Note that only a single skyrmion per racetrack is used, with multi-bit information stored in the position of the skyrmion.
Having a single skyrmion per racetrack increases its reliability and control over a packed, interacting array of skyrmions.}
}
      \label{fig:CKT}
\end{figure}

\begin{figure*}
    \centering
    \includegraphics[width=\textwidth]{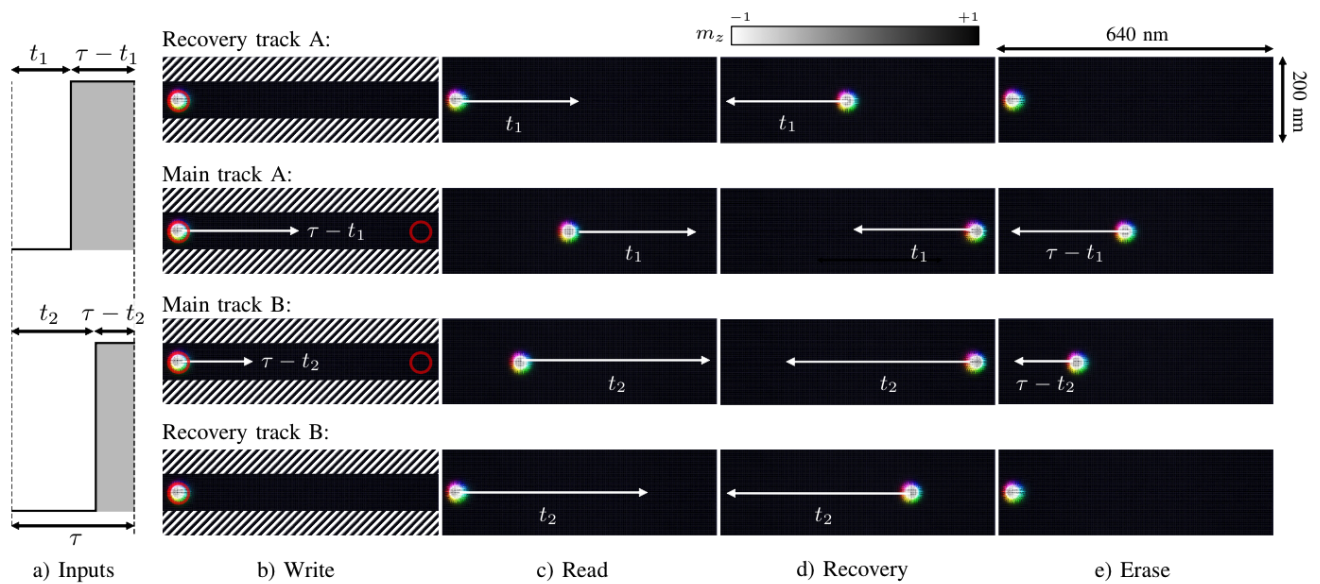}
   \def\svgwidth{\linewidth}
    \caption{(a) An example of two wavefronts with disparate duty cycle fed to the temporal memory cell. (b-e) Micromagnetic simulation of four operational phases of the temporal memory in magnetic racetracks. The figures show a background of up-spin magnetic material (black) with a down-spin skyrmion (white); color indicates the in-plane spin orientation. The location of the skyrmion at the beginning of each phase is shown, while the arrows denote the position of the skyrmion after finishing that phase. The magnetic tunnel junction (MTJ) placed for skyrmion detection is denoted by the red circle in the first column. The patterned area in the racetracks represents an energy barrier. Amplified repulsion from the edges prevents the pinning of skyrmions at the edges and skyrmion annihilation and helps in maintaining their expected trajectory towards the read MTJ. H. Vakili \textit{et al.}, IEEE Journal on Exploratory Solid-State Computational Devices and Circuits, vol. 6, no. 2, pp. 107-115, 2020; licensed under a Creative Commons Attribution (CC BY) license. \cite{wavefront}.}
\label{fig:wfenergy}
\end{figure*}

A recently proposed temporal computing scheme called race logic \cite{madhavan_race_2014,madhavan20174,RaceTrees} has shown acceleration of dynamic programming and machine learning algorithms. Information representation in race logic is different than the conventional Boolean approach as it represents \textcolor{black}{multi-bit} information \textcolor{black}{at the arrival times of} the digital rising edge of a wavefront in a wire \cite{madhavan_race_2014}. Such a temporal coding scheme potentially achieves orders of magnitude energy efficiency \cite{madhavan_race_2014} compared to traditional data encoding while accelerating the dynamic programming algorithm.  However, in Boolean computing, the analog temporal data needs to be converted to the digital domain. This domain translation imposes a significant area and energy penalty from analog-to-digital converter circuits, limiting the computational capability. Thus, there is a need for a native analog temporal memory to store and retrieve the analog temporal data to enable complicated processing with the temporal coding scheme. 

Assuming a quasi-ballistic motion of skyrmion with an applied drive current, the time duration of a drive current determines the distance traveled by the skyrmion on the racetrack. Therefore, we can design a scheme of storing the temporal data in the position of a single skyrmion along the racetrack. This temporal data can represent timing delays in race logic for pattern matching applications, as well as thresholds for regular or automata decision trees for parallel processing of near-sensor serial input data \cite{RaceTrees,Madhavan2020}. However, an adequate lifetime of skyrmions (Table V) will be needed to save on nucleation energy, as pre-written skyrmions can be reused rather than nucleating and destroying them repeatedly. \textcolor{black}{Such a scheme allows for occasional replenishment of skyrmions through added voltage-gated nucleation events, whose cost is readily amortized over several (ranging from hundreds to millions) operating cycles.} 

In skyrmion-based temporal memory \cite {wavefront} (Fig.~\ref{fig:CKT}), an arriving wavefront launches a pre-fabricated skyrmion positioned below the MTJ at the beginning of the racetrack. By turning on transistor $T_1$, a drive current is initiated in the heavy metal underlayer. This drive current is shut off when the wavefront completes its duty cycle. Thus, the analog arrival time {stamp} of the {rising edge of the} input signal gets encoded onto the {eventual} position {where the skyrmion comes to rest and is set by} the duty cycle (fraction of the total time when the signal is active) of the input signal. {The write operation is then the launch of the skyrmion and its subsequent placement by a current whose duration is set by the wavefront arrival time}. Due to the non-volatile nature of the magnetic skyrmion, there is no need to store the arrival time repeatedly, as it can be replayed back as needed. During this read operation, the skyrmion is driven to the read MTJ, consuming time set by the duration of the rising edge. This information, however, is lost once the skyrmion reaches the read MTJ at the end of the main racetrack. Thus, to replay the temporal data repeatedly, the proposed design \cite{wavefront} needs a recovery operation to avoid destructive read operation. The recovery operation needs an extra racetrack called the recovery racetrack (see Fig.~\ref{fig:wfenergy}) whose job is to store the positional value of the temporal data lost during the read operation {by launching a recovery skyrmion as soon as the read is executed, much like passing the baton during a relay race}. The proposed circuit {wavefront} is designed in such a way that the skyrmions in both the main and recovery racetrack travel the same distance during the read and recovery operations through synchronization of the drive currents. Due to the matched speed of the skyrmions in both racetracks, the position of the skyrmion representing the temporal data is restored and is ready to be replayed back repeatedly {(if the speeds do not match perfectly, this too can be accounted for in a calibration phase)}. The recovery racetracks only participate during the read and recovery operations. A reset operation is needed to store a new wavefront with a different duty cycle. The read MTJ stacks placed at the beginning and end of the racetracks provide a feedback signal to stop skyrmion movement after successful detection.

\begin{figure}
\centering
    \includegraphics[width=\columnwidth]{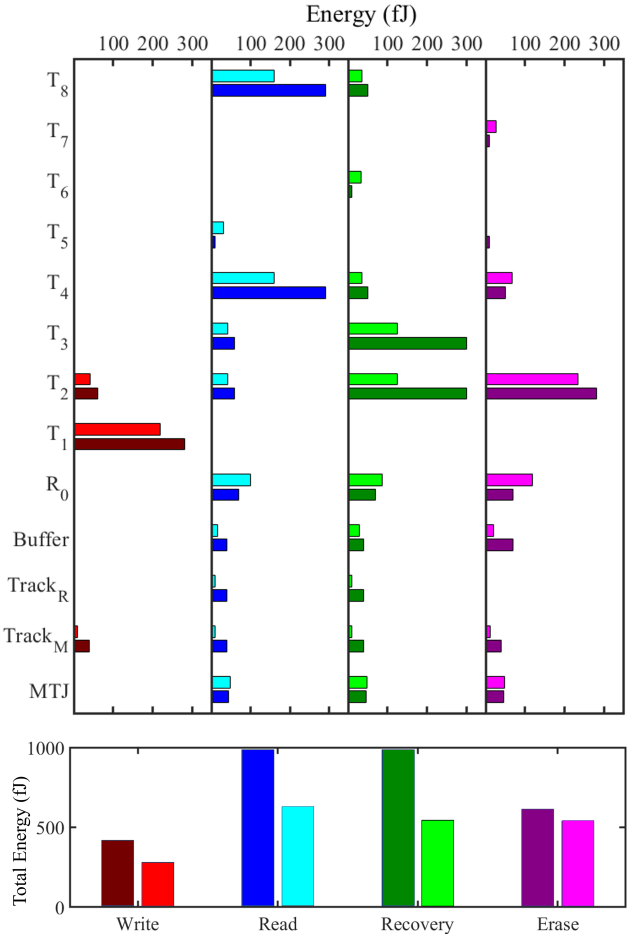}
    \caption{Energy breakdown of each circuit elements and total energy for 50 \% duty cycle. A set of 640~nm racetracks used to capture wavefront and provide the energy breakdown. The high overdrive voltage of transistors consume the largest energy in all operations. Both the racetracks and the associated transistors are active during the read and recovery operation, resulting in  consuming most energy. However, joule heating in the main (Track$_M$) and recovery (Track$_R$) racetracks contributes only a small fraction of the overall energy.{ Simulation results are shown in darker and brighter color bars  for 45~nm and 16~nm CMOS transistors, respectively. H. Vakili \textit{et al.}, IEEE Journal on Exploratory Solid-State Computational Devices and Circuits, vol. 6, no. 2, pp. 107-115, 2020; licensed under a Creative Commons Attribution (CC BY) license. \cite{wavefront}.}}
    \label{fig:full_energy}
\end{figure}

{Figure~\ref{fig:full_energy} shows the breakdown of energy dissipated in the circuit, resolved into individual components such as the 8 transistors $T_{1}\ldots T_8$,} {reference resistance $R_0$ (needed for voltage swings for reading Fig.~\ref{fig:CKT}), buffer used in the read circuitry (the inverter resistance)}, {the main and recovery track metal lines, and the read MTJ for a 640 nm racetrack at 50$\%$ duty cycle, assuming 45nm (dark bars) and 16nm (light bars) CMOS transistors. We can see that the read and recovery operations cost the most energy, and the bulk of this energy comes from the overhead transistor switching costs ($\sim$ 1 pJ overall) compared to the loss in the metal lines ($\ll 10$ fJ). Notably, doing the full operation with CMOS memory instead of its skyrmionic counterpart would consume well over a hundred transistors \textcolor{black}{to account for counters, storage latches, and decoders,} at considerably larger energy cost and overall footprint.} \textcolor{black}{Furthermore, there is increased efficiency due to our ability to playback the wavefront duty cycle at low damping (Energy-delay product another 32X lower than 1 GHz CMOS) for one cycle. If we do 10-100 reads per cycle, the product is almost 2500X lower. Finally, there is energy saving from repeated read operations due to the non-volatility of the skyrmions.}

\subsection{Reconfigurable Skyrmionic Logic}

A reconfigurable hardware fabric offers increased performance per Watt at a lower non-recurring design and fabrication cost than the Application Specific Integrated Circuits (ASICs). Field-Programmable Gate Arrays (FPGAs) are medium to large integrated circuits that can be programmed to repeatedly perform any desired digital logic without any extra engineering cost. A matrix of programmable logic blocks is interconnected by a routing fabric which is also programmable \cite{kuon2008fpga}. Static Random Access Memory (SRAM) Cells store the logic function data configurations in modern FPGAs. Moreover, SRAM cells also store the select bits from the multiplexers (MUXs). Though FPGAs' re-programmability and manufacturing simplicity are advantageous, the use of high cost per bit and volatile SRAM cells leads to large area overhead.\\
\begin{figure*}[htp]
    \centering
    \includegraphics[width=\textwidth]{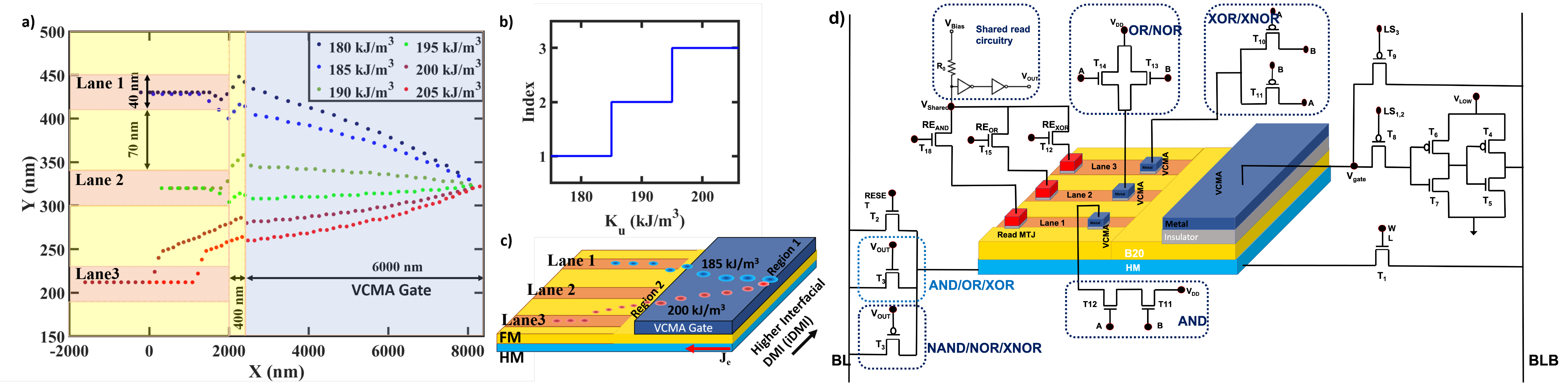}
    \caption{(a-c) Self-converging skyrmions can be used to design reconfigurable logic design. a) By varying the anisotropy, the skyrmion will end up in different lanes (b) Changing the anisotropy can be achieved by using VCMA gates, Which can be used to get a step-like function vs. $K_u$. (d) Skyrmionic programmable logic device using controllable convergence lane.  The programmable feature of the design allows it to be programmed for computing different logic functions such as AND, NAND, OR, NOR, XOR and XNOR operations. Level shifter (high to low) converts the applied voltage at the BLB to necessary step values of voltage needed at the large VCMA gate for lane selection purposes. The small VCMA gates placed in each lane control the motion of skyrmion, and the magnetic tunneling junction placed at the end of each lane detect the presence/absence of skyrmion. Reprinted with permission from H. Vakili \textit{et al.}, Phys. Rev. B 102, 17442, 2020. Copyright 2020 by the American Physical Society.\cite{hybridfocus, Programmable_hardware_spie}.  }
    \label{fig:hybrid}
\end{figure*}


Nowadays, FPGAs are used in a wide variety of applications such as compute accelerators, automotive, consumer electronics, etc., as the feature size reduction of CMOS technology is paving the way towards an increased number of logic blocks in FPGAs. Moreover, FPGAs are also used/offered by several cloud providers to accelerate bioinformatics, web search, and machine learning workloads \cite{caulfield2016cloud}. However, the operating of these devices is generally carried out under strict performance, power, and reliability constraints.
In this context, reconfigurable logic using skyrmionic racetracks carries the potential to bring non-volatility to FPGAs, thus obliterating the security concerns of programming the FPGA only at power ON. Moreover, reconfigurable logic using skyrmionic racetracks can be a promising alternative to the high-cost SRAM cells in FPGAs, as they can be much more densely packed than the SRAM-based designs exhibit reasonable propagation delay. 

Skyrmions and domain walls need spin current for their operations and can be used for ultra-small, low power, and nonvolatile memory or logic operations. Both skyrmions and DWs can have reasonably long lifetimes and high speeds. Skyrmions tend to have lower pinning than DWs, so the energy requirements can be lower. Nevertheless, one major difference between DW and skyrmion is that the DW can only move in the direction of the racetrack (1-D movement). In contrast, the skyrmion can move along both axes of the racetrack (2-D movement). By exploiting this extra degree of freedom and self converging nature of skyrmion {discussed shortly}, a reconfigurable logic device can be designed with only one nucleation site. \textcolor{black}{In \cite{zhang_magnetic_2015} by using the merging, duplication, and skyrmion's ability to deform to a double domain wall and back, a logic device has been proposed, which demonstrates some of the interesting dynamical behavior of skyrmions.}

Skyrmion-based reconfigurable logic has been proposed in \cite{luo_reconfigurable_2018} and in \cite {eds}. However, the device \cite{luo_reconfigurable_2018,eds} needs extra fine-tuning of the parameters to get the reconfigurability which is not the case for the device proposed in \cite{Programmable_hardware_spie}. By leveraging the interaction between the skyrmions combined with the electric-field-controlled magnetic
anisotropy (VCMA) effect, a reconfigurable logic gate is proposed in \cite {eds}. However, both the reconfigurable logic gates \cite{luo_reconfigurable_2018, eds} need at least two nucleation sites and more than one energy-hungry skyrmion nucleation to do logic computation.

Figure~\ref{fig:hybrid} shows an example of self-converging skyrmions, created using a compositionally graded structure that generates a compensation point where the transverse Magnus force cancels, and the skyrmion moves linearly along a 1-D racetrack. In the example shown, such compensation is achieved by using a varying DMI (e.g., Pt$_x$W$_{1-x}$ heavy metal underlayer) with a B20 magnet so that an injected skyrmion picks up growing interfacial DMI along with its current-driven transport and evolves into a Bloch-Neel hybrid that naturally reaches such a compensation. To make the converged lanes align with a pre-fabricated racetrack, we can place intermediate repulsive lanes with high anisotropy and further control electrostatically using a gate that tunes the voltage-controlled magnetic anisotropy (VCMA) across the interface between the magnet and a MgO overlayer. The lane selectivity with tuned anisotropy is shown in the inset at zero temperature (finite temperature gives stochastic wiggles around the transition points).

The reconfigurable logic circuit works as follows.
Based on the desired operation (AND/NAND, OR/NOR, XOR/XNOR), the corresponding voltage is applied to the large global VCMA gate. This would put the skyrmion into the appropriate lane. The small VCMA gates in each lane control the subsequent movement of a skyrmion. For the case of AND/NAND operation, the lane 1 VCMA gate is ON only when both inputs are 1, while for OR/NOR, the lane 2 gate is OFF only when both inputs are 0. For XOR/XNOR, the lane 3 VCMA gate is on only when exactly one of the inputs is 0, and the other is 1 (i.e., the two inputs are different). For the logic operation, the skyrmion is moved along the appropriate lane to the read MTJ at the end. Then for recovery, a current in the opposite detection is applied for a certain amount of time (depending on the lane length) to put the skyrmion back to its starting piston in the lane. The corresponding circuit needed for the operations is shown in Fig~\ref{fig:hybrid}(d)
\begin{figure}
   \begin{center}
    \includegraphics[width=0.5\textwidth]{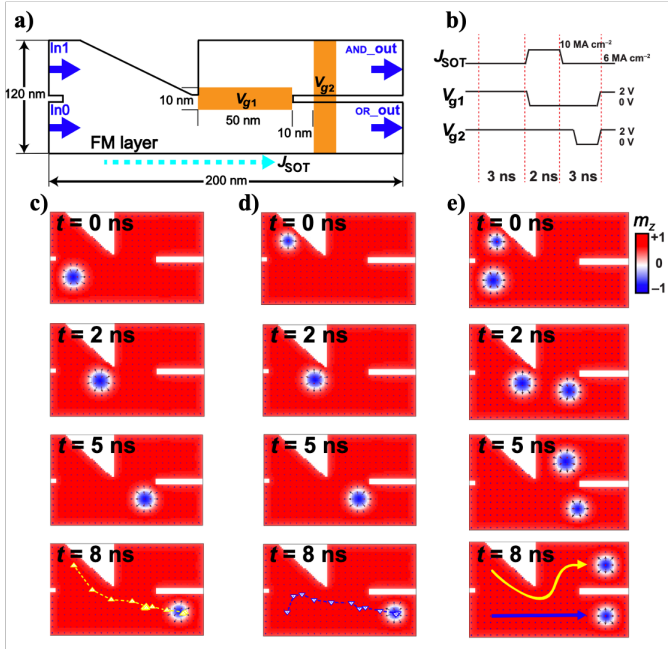}
	\end{center}
    \caption{Skyrmionic AND-OR logic device (a) Device structure with two input lanes and two output lanes. The two electrode gates (with voltages $V_\mathrm{g1}$ and $V_\mathrm{g2}$) are designed to modulate skyrmion motion through the VCMA effect. (b) Proﬁle of driving current density ($j_\mathrm{SOT}$) and positive voltages applied to achieve device function. (c-e) Three cases of inputs and their corresponding simulated processes and outputs. In the snapshots of the micromagnetic simulations, the color red represents spin up ($m_z = +1$), blue represents spin down ($m_z = −1$), and white represents spins that are horizontal in the plane ($m_z = 0$). Skyrmion trajectories are indicated in the magnetization snapshots at $t = \SI{8}{ns}$. Reprinted with permission from H. Zhang \textit{et al.}, Phys. Rev. Applied 13, 054049, 2020. Copyright 2020 by the American Physical Society.~\cite{Zhang2020Stochastic}.}
    \label{SC-AND-OR}
\end{figure}
A skyrmionic AND-OR gate is proposed by using two joint racetracks with a notch (Fig.~\ref{SC-AND-OR}(a)) under the assistance of voltage control magnetic anisotropy (VCMA)~\cite{Zhang2020Stochastic}. By accurately controlling the gate voltage $V_\mathrm{g1}$, $V_\mathrm{g2}$ and the current density $j_\mathrm{SOT}$ with time (Fig.~\ref{SC-AND-OR} (b)), the motion of the skyrmions can be guided along the racetracks. In Fig.~\ref{SC-AND-OR}(c)-(d), when only a single skyrmion arrives at the notch, it will go to the bottom track no matter where it originated. In Fig.~\ref{SC-AND-OR}(e), when both tracks have a skyrmion, they will both pass through the notch. The bottom skyrmion first enters the bottom track, and the upper skyrmion walks around the notch and enters back to the upper track. Therefore, this device achieves the functionalities of AND (upper track, which registers a signal only for $(1,1)$, i.e., skyrmion starting in both tracks) and OR (bottom track, registering a null only for $(0,0)$, no skyrmion in either track). By integrating the skyrmionic reshuffler with this AND-OR gate, a skyrmion-based stochastic multiplier can thus be realized. 

\begin{figure}
   \begin{center}
    \includegraphics[width=0.5\textwidth]{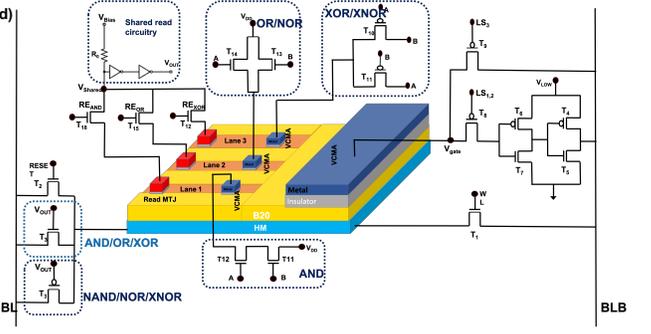}
	\end{center}
    \caption{(Top left) Stochastic bitstream generated with probability $p =0.325$. (Top right) Stochastic bitstream generated with probability $q =0.55$. (Bottom left) Signal obtained by using an AND operation on the two bitstreams. (Bottom right) Estimated probabilities $p_1$, $q_1$ and $p_1\wedge q_1$ from the bitstreams run through binary counters, compared against the expected values of $p$, $q$ and $p.q = 0.1788$, showing that AND acts as a multiplier.
    }
    \label{stoccom}
\end{figure}
\subsection{Skyrmionic Decorrelators for Stochastic Computing }
Stochastic computing utilizes the statistical nature of noisy signals to achieve a collective result from an ensemble of readings. Unlike a Turing machine which uses deterministic devices to generate a deterministic output, stochastic computing uses probabilistic devices but generates a reliable output using systems-level techniques~\cite{Shanbhag2010Stochastic}. Compared with deterministic devices, stochastic devices consume much less energy while maintaining acceptable reliability of computation. Several complex circuit operations  have been proposed, such as square rooting~\cite{toral2000stochastic}, polynomial arithmetic~\cite{qian2008synthesis, li2009reconfigurable, qian2010architecture}, and matrix operations~\cite{mars1976high}, as well as the “tan-sig” transform function employed in neural networks~\cite{zhang2008stochastic}. {As Fig.~\ref{stoccom} shows, the AND operation on two time-sampled uncorrelated probabilistic bitstreams generates a multiplication of probabilities ~\cite{Pinna2018Skyrmion}. Similar simplified operations (sum + thresholding) arises from an OR gate} \cite{stochastic}.

\begin{figure}
   \begin{center}
    \includegraphics[width=0.5\textwidth]{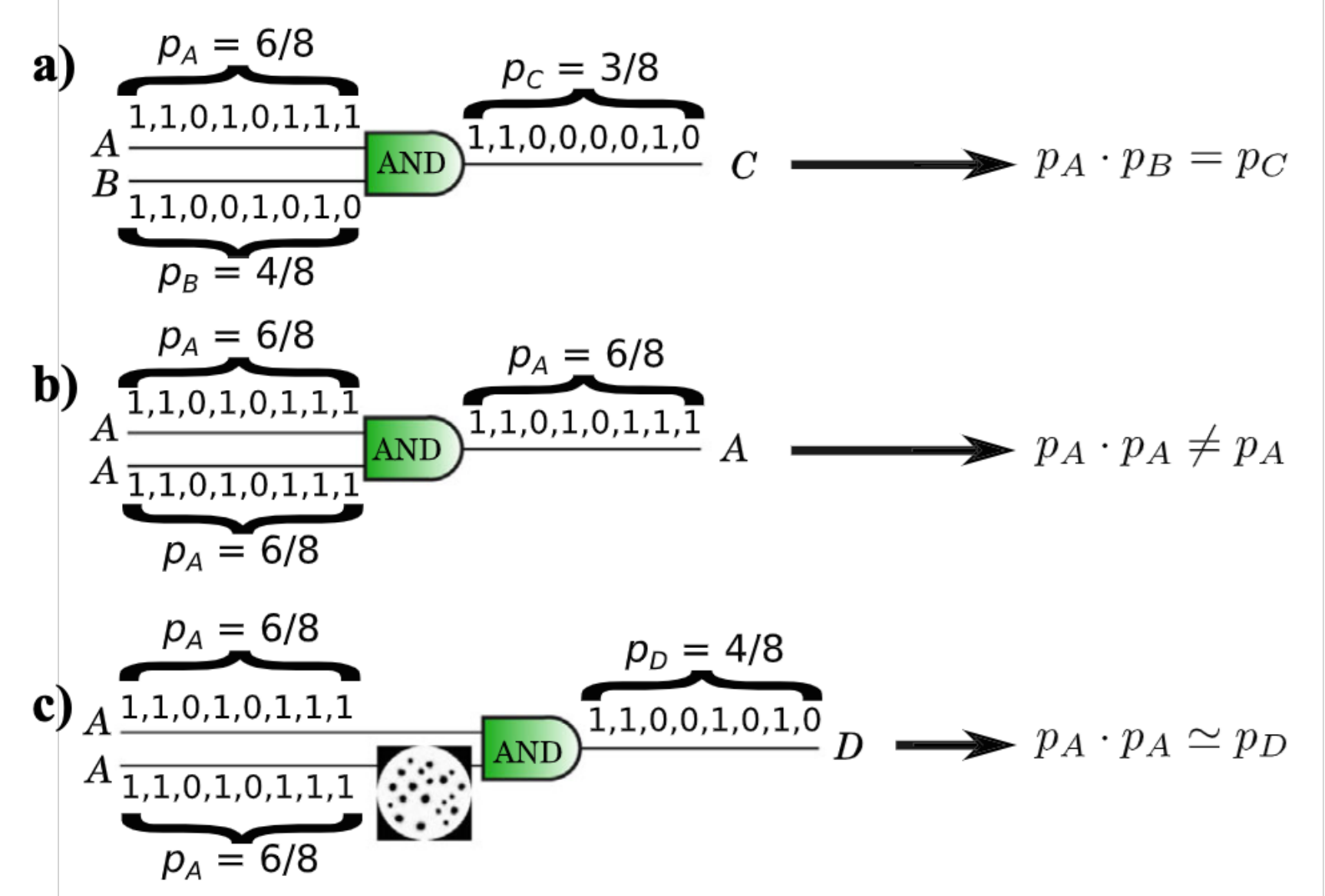}
	\end{center}
    \caption{(a) AND-gate implementation of the multiplication operation under the stochastic computing paradigm. The p-value of the output signal is equal to the product of the input signals’ p values. (b) Whenever correlations exist between the input signals, the AND gate does not perform the expected multiplication. (c) Forcing the input signals through a skyrmionic reshuffler allows for a correct multiplication operation even in the presence of strong correlations between the input signals. From Ref. Reprinted with permission from D. Pinna \textit{et al.}, Phys. Rev. Applied 9, 064018, 2018. Copyright 2018 by the American Physical Society.~\cite{Pinna2018Skyrmion}.}
    \label{SC-Rand}
\end{figure}

Essential to stochastic computing is the requirement that its input bitstreams are decorrelated.  For example in 
Fig.~\ref{SC-Rand}(a), we can see how
the two binary strings fed into a summator generate a multiplication. String $A$ encodes the number  $P_A = \frac{6}{8}$ as the probability of getting a 1 (recorded by its time average), while string $B$ encodes probability $P_B = \frac{4}{8}$. When passing through the summator (AND gate), the outcome probability of ``1'' is $\frac{3}{8}$, which is also the result of multiplication $P_A \cdot P_B$. However, this requires the two input strings to be decorrelated. We see that if the two input strings are perfectly correlated as in Fig.~\ref{SC-Rand}(b), the final output encodes $P_A$ rather than $P_A^2$. Specialized decorrelator circuits can be constructed that sample a bitstream and generate an uncorrelated bitstream with the same mean value. 

Broadly speaking, decorrelators can be categorized into three classes depending on their operating principle: (a) Regenerative decorrelator, which samples the input bitstream to first obtain a mean value and then generates a new output stochastic bitstream. This kind of decorrelator generates the best quality decorrelation but can stall computation due to the delay introduced in sampling. (b) Delay decorrelator that assumes a certain autocorrelation time of the bitstream source. The input bitstream is copied and ``delayed'' by a duration more than the autocorrelation time as the output bitstream. This ensures that the input and output bitstreams are uncorrelated due to the nature of the source signal. This is the preferred approach in pure CMOS-based decorrelators since we merely need to use a delay path (say a shift register); however, a good quality decorrelator requires a priori knowledge of the bitstream source's autocorrelation time. (c) Reshuffling decorrelator (Fig.~\ref{SC-Rand}(c)) that first captures the input bitstream and shuffles their order to generate the output stream, much like a deck of cards being shuffled after each game to ensure subsequent fair games. The quality of such a decorrelator depends on the length of the bitstream that can be recorded.

\begin{figure}
   \begin{center}
    \includegraphics[width=0.5\textwidth]{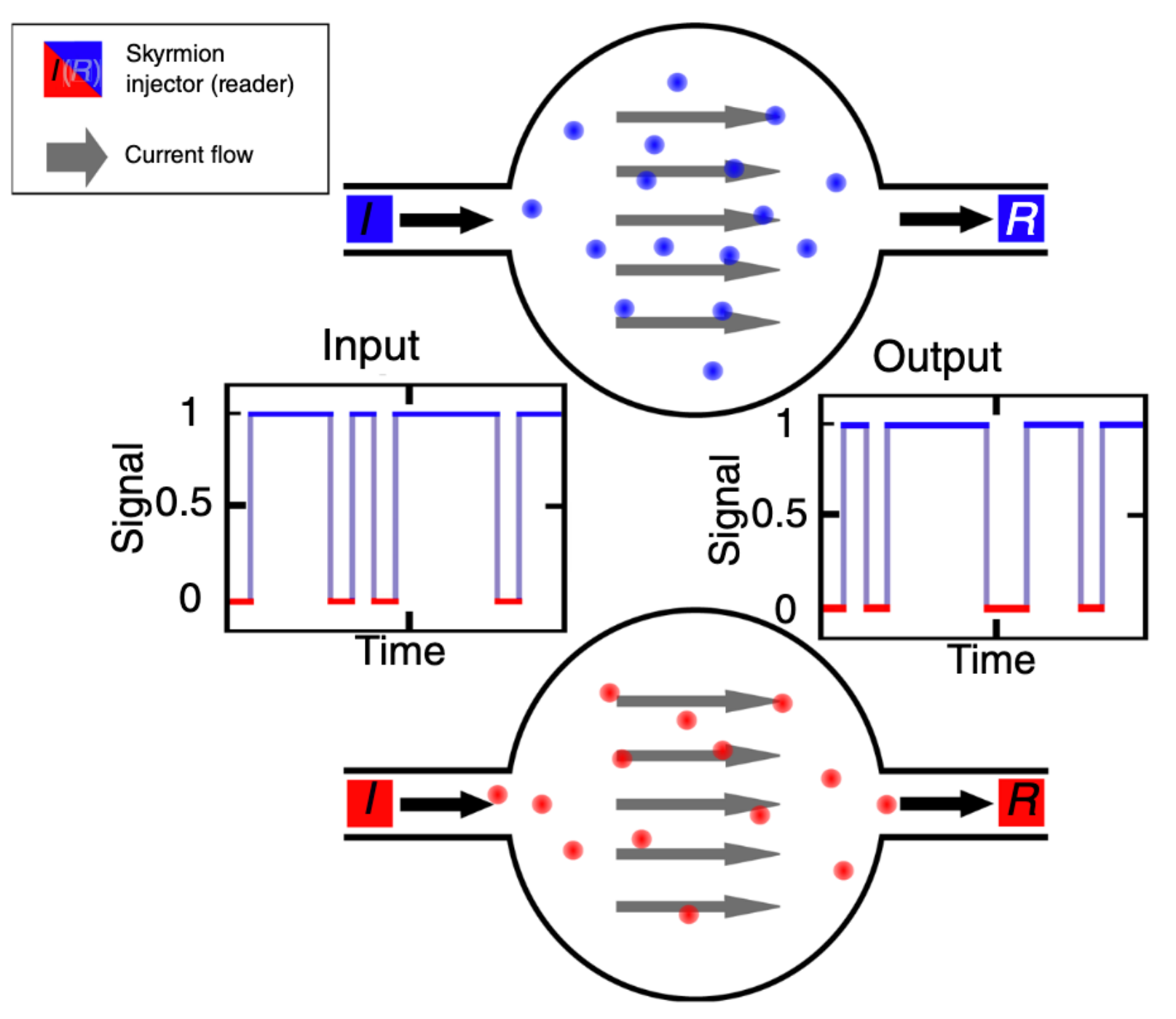}
	\end{center}
    \caption{The proposed device consists of two magnetic chambers into which skyrmions are injected depending on the state of an input telegraph noise signal. The net drift of the skyrmion particles due to a constant current flow along with the thermal diffusion in the chambers leads to an exit order that can be significantly different from that of entry. This behavior is employed to reconstruct a new outgoing signal with the same statistical properties as the first, as well as being uncorrelated from it. Reproduced with permission from D. Pinna \textit{et al.}, Phys. Rev. Applied 9, 064018, 2018. Copyright 2018 by the American Physical Society.~\cite{Pinna2018Skyrmion}.}
    \label{SC-shuffle}
\end{figure}

\begin{figure*}[htp]
    \centering
    \includegraphics[width=\textwidth]{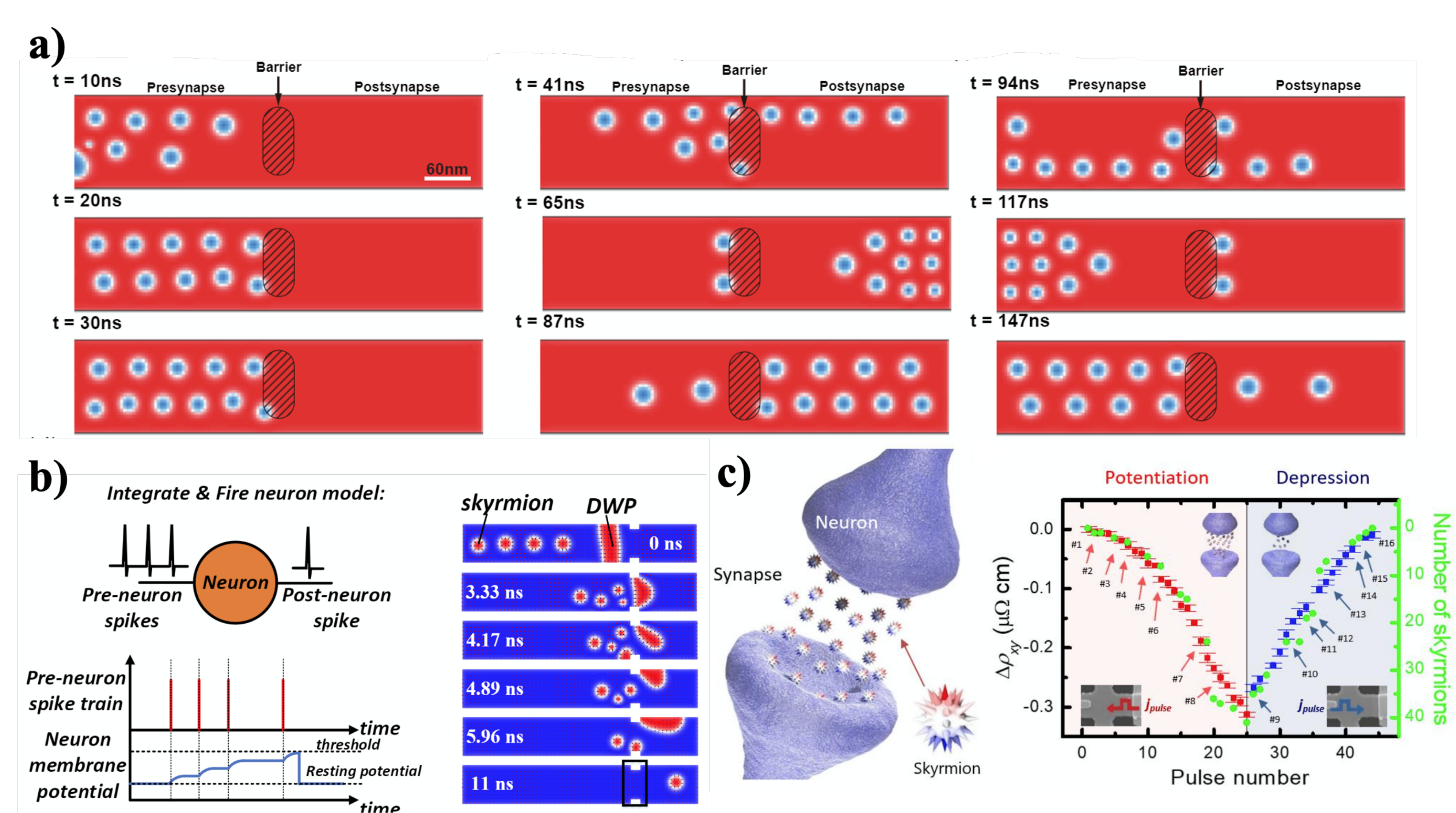}
    \caption{(a) The synapse proposed using a barrier in the middle of the racetrack. In the initialization mode, skyrmions are nucleated to populate the presynapse section. In the potentiation mode, skyrmions are pushed past the barrier, which results in a sudden change in the skyrmion population in the postsynapse area. Reproduced with permission from  Nanotechnology 28 08LT02 (2017). Copyright 2017, IOP Publishing. \cite{Huang_2017}(b) A proposed skyrmion-based neuron. In this case, notches provide the barrier for the domain wall. As the number of skyrmion behind the domain wall increases (presynapse area), their dipole-dipole interaction with the domain wall increases. When the number of skyrmions which is based on the number of spikes, reaches a certain value, the skyrmion and domain wall can pass the notch. \cite{He2017DevelopingAS}(c) The experimental realization of a skyrmion-based synapse, in this model, in the potentiation mode, for each spike, a nucleation current pulse is applied, which nucleates skyrmions and populates the racetrack. For the depression mode, for each applied current, pulse skyrmions get annihilated, which depopulates the racetrack. Reproduced with permission from  K.M. Song \textit{et al.} Nat Electron 3.  148–155 (2020). Copyright 2020, Nature Publishing Group. \cite{song_skyrmion-based_2020}}
    \label{fig:synapse}
\end{figure*}
It has been proposed that a magnetic chamber, as shown in Fig.~\ref{SC-shuffle} can play an efficient role as a skyrmion-based reshuffling decorrelator. This device is demonstrated by modeling the ensemble dynamics in a collective coordinate approach where skyrmion-skyrmion and skyrmion-boundary interactions are accounted for phenomenologically. The skyrmions' order gets thermally reshuffled due to the diffusive 2-D dynamical nature of interacting skyrmions at finite temperature. {The output stream decorrelates from the input while maintaining the same probability.} {The strength of stochastic computing is its independence of the precise placement of 1s and 0s, relying just on the relative frequency of their occurrence. There are, however, other considerations pertaining to our ability to encode an analog number accurately within a given finite-sized bitstream.} Nonetheless, by providing a compact natural decorrelator driven by true random thermal noise, skyrmions can enable large correlation-free computational graphs for stochastic computing.

\subsection{Artificial Skyrmionic Synapses and Neurons }
An artificial neuron is one of the fundamental computing units in brain-inspired artificial neural networks. 
To implement the non-linear neuron activation function, various designs using standard CMOS have been proposed. These CMOS-based artificial neurons typically consist of a large number of transistors, which require a large area and power consumption. Skyrmions can provide an alternative to efficiently implement neuronal behavior.
In a nervous system, synapses carry out the role of passing electrical or chemical signal signals between neurons. One of the main characteristics of biological synapses is the synaptic weights which are no longer binary. \textcolor{black}{Such initiation, potentiation, and depression behavior are} easy to realize in software but harder to implement in digital CMOS-based circuits.

Various methods have been proposed to implement synapses and neurons using skyrmions (Fig~\ref{fig:synapse}(a)). One approach is using a defect/notch in the racetrack. The notch acts as an energy barrier, which a skyrmion or domain wall needs to climb over to pass through. For the initiation mode, skyrmions are nucleated and trapped behind the energy barrier of the notch. For the potentiation mode, a large enough current propels the skyrmions past the defect. This behavior mimics the integrate-and-fire behavior of neurons. In the depression mode, the skyrmions are in the postsynapse area, and a current is applied that is opposite to the current direction used for potentiation mode. The skyrmions then get moved back from the postsynapse area to the presyanpse area. \cite{Huang_2017}. The synaptic weight for each mode (potentiation or depression) is determined by the populations of skyrmions in presyanpse or postsynapse areas. Other variations of this approach have been used to mimic integrate-and-fire neurons  \cite{He2017DevelopingAS,snc_he}. One such proposal combines the movement of a skyrmion along with a domain wall. In this case, each spike corresponds to an added skyrmion behind the domain wall. When a sufficient number of skyrmions build up behind the domain wall, a skyrmion will pass through to the other side of the wall. This effect also mimics the integrate-and-fire neuron activation since the device `fires' only after a sufficient number of skyrmions have been accumulated or `integrated'. The role of notches in this proposed device is similar to the notches described earlier in Section \ref{sectionSkyrmionDynamics}C. Results from micromagnetic simulations show up to $98\%$ accuracy for the proposed device.

Recently an experimental realization of a skyrmionic synapse has been reported \cite{song_skyrmion-based_2020}. In this work, during the potentiation mode, skyrmions are nucleated for each spike pulse, which then changes the resistance through the anomalous Hall effect and thereby the output voltage, based on the number of skyrmions in the racetrack. Similarly, for the depression mode, skyrmions are annihilated at a rate related to the depression potential. Based on the number of nucleation/annihilation pulses, the number of skyrmions corresponding to the synaptic weight can be controlled. The accuracy of this skyrmion-based synapse was reported to be $89\%$ in experiments, as compared to $94\%$ from software simulation of the synapse. {Clearly, more work needs to be done to make such neuronal devices competitive.}


\subsection{Reservoir Computing using Skyrmions}
\begin{figure}
    \centering
    \includegraphics[width=\columnwidth]{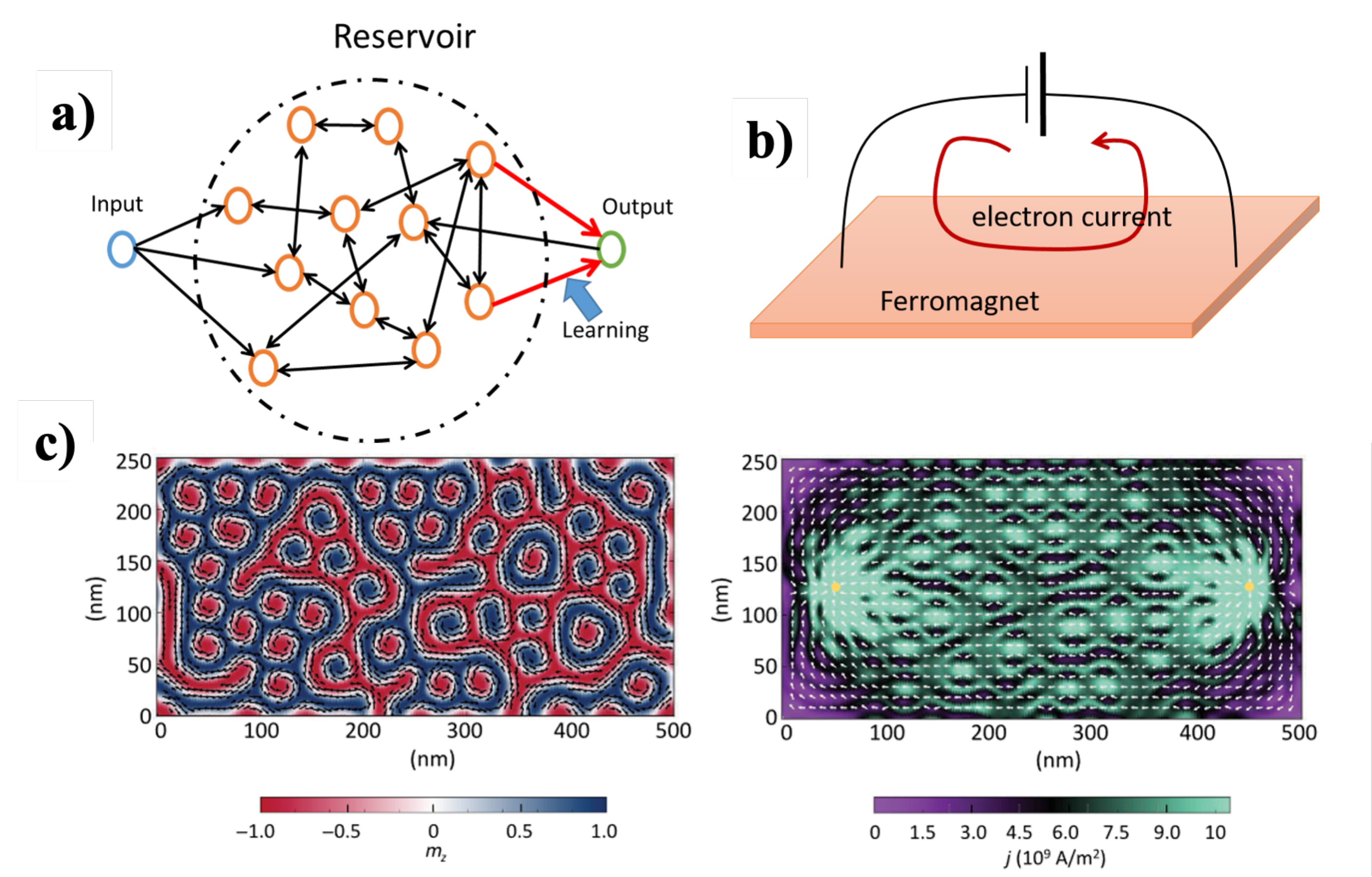}
    \caption{(a) General schematic of a Reservoir Computer, which consists of randomly, sparsely, and recurrently connected non-linear activation units. The input signal is imposed on the reservoir, while the output is developed by sampling the reservoir states. Only the output is adjusted during the learning process. (b) Schematic of skyrmion fabric-based Reservoir Computer. The ferromagnet hosts multiple skyrmions and stripe phases, and the output of the reservoir is obtained by measuring the resistivity of the film, which changes depending on the skyrmion texture due to the anisotropic magnetoresistance (AMR) effect. (c) Simulation of a skyrmion texture (left panel) and current flow (right panel) through it. Reproduced with permission from G. Bourianoff \textit{et al.}, AIP Advances 8, 055602 (2018), with the permission of AIP Publishing. \cite{ReservoirSkyrm2}).}
    \label{fig:Skyrm_res}
\end{figure}

Reservoir computing is a paradigm of computing that intersects neural networks, dynamical systems, state-space filters, and random networks. The central entity in this mode of computing is the notion of a ``reservoir'' which is a collection of units with non-linear activation functions (or neurons from a neural network perspective). This collection is sparsely and randomly connected with recurrence, i.e., potentially many feedback paths (Fig. \ref{fig:Skyrm_res}a). Reservoir computing has been developed in two flavors independently, viz., the echo-state network (ESN) and liquid state machines (LSM). The essential difference between the two modes is that ESN works with continuous signals, akin to conventional artificial neural networks, while LSM works with spiking signals similar to biological neural networks and is proposed as a biological model of autonomous multi-scale learning.

Reservoir computing provides spatio-temporal inferencing capabilities, i.e., it deals with data that spans both space and time domains (say, a video or multi-dimensional biological signal like EEG). Its recurrent nature allows for an input signal onto the reservoir to propagate through the network at various speeds and activate the neurons at differing times. This enables developing spatio-temporal correlations, as signals from the ``past'' may be correlated with the ``present'', because the ``echo'' or the signature of the input signal persists within the reservoir and is evident in the state vector of the reservoir. By sampling these states, it is possible to make inferences about the input signal both spatially as well as temporally. The learning aspect in reservoir computing arises in the readout operation of the reservoir, which is generally a linear weighted sampler over the reservoir state vector (i.e., a list of activation or non-activation of the reservoir units at any given time). We can train the sampler weights over the reservoir states to make a particular inference. We can also attach multiple different samplers to obtain multiple inferences since the reservoir remains unperturbed by a readout process. It should be noted that in reservoir computing, the interconnections within the reservoir themselves are not tuned or modified, unlike other models of recurrent neural networks such as GRU or LSTMs, which allow for easy one-shot learning approaches similar to extreme learning machines.

In \cite{ReservoirSkyrm1,ReservoirSkyrm2} authors propose to exploit extended skyrmion textures or fabrics such as skyrmion lattices or stripe states on a magnetic film (Fig. \ref{fig:Skyrm_res} (b,c)) as the reservoir while the readout or sampling is performed by measuring non-linear interaction between the randomized skyrmionic texture and the current through the film via the anisotropic magnetoresistive (AMR) effect. The current flows through the path of least resistance along the various domain walls in the disordered self-organized domain structure (Fig. \ref{fig:Skyrm_res}(d)). This plays the same role as an input signal propagating through a random reservoir via multiple paths. This skyrmionic texture can also be modified by the applied current, which plays the role of an input signal changing the state of the reservoir.

Such a skyrmion fabric can potentially host a very large density of closely packed interacting reservoir nodes whose dynamics and evolution are determined by the interplay between exchange, DMI, dipolar, thermal interactions, and an applied input current, thus eliminating the area expense of laying down a complex pattern of interconnections between different nodes of the reservoir, if they are built from discrete elements. As initial simulation studies show promising results, however, practical engineering challenges of readability and distinguishability of reservoir states through the AMR effect is the major hurdle towards a feasible design and implementation.

\subsection{Skyrmion based Spintronic Oscillators}

Spintronic oscillators use the built-in precession or gyration of a magnetic state to build an oscillator. This motion can then be detected through the magneto-resistive effect, typically using MTJs. Most commonly used spintronic oscillators employ either spin-transfer or spin-orbit torques along with a magnetic field to generate two compensating torques that do not allow the free layer magnet to relax into one of its stable states. The frequency of oscillation of such devices is highly tunable, which makes them a great candidate for ultra-small and versatile frequency sources. However, these spin-torque nano oscillators (STNO) suffer from high phase noise and large linewidths primarily due to their small volumes and non-deterministic multi-domain behavior.

\begin{figure}
    \centering
    \includegraphics[width=\columnwidth]{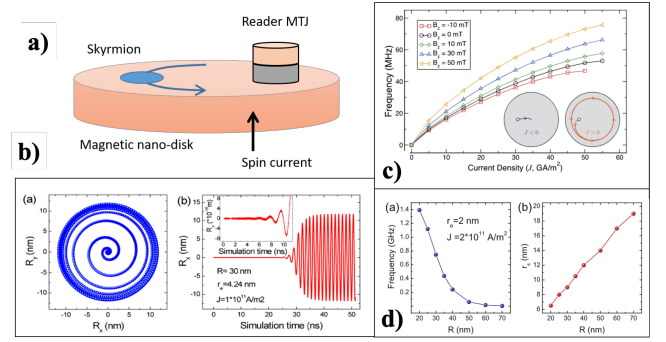}
    \caption{(a) Schematic of a skyrmionic nano-oscillator (SkNO). Skyrmions are placed in a wide nano-disk, whereupon applying a spin torque through the injection of a spin current drives the skyrmion in a circular orbit through the disk. The readout is done through a small MTJ structure placed on the nano-disk, which samples the skyrmion every period of the gyration orbit. (b) Simulation showing the trajectory of the skyrmion on the nano-disk and the motion through the x-axis (c) Dependence on the central oscillation frequency with current density for various applied magnetic field strength (F. Garcia-Sanchez \textit{et al.}, New J. Phys. 18 075011, 2016; licensed under a Creative Commons Attribution (CC BY) license.\cite{Garcia_Sanchez_2016}) (d) Dependence of the central oscillation frequency with the nano-disk radius, given a fixed current density and (left panel) injection point contact radius and (right panel) skyrmion radius (b,d  from S. Zhang \textit{et al.} , New J. Phys. 17 023061, 2015; licensed under a Creative Commons Attribution (CC BY) license. \cite{Zhang_2015}).} 
    \label{fig:skno}
\end{figure}

One potential solution to such problems is the use of a controlled multi-domain magnetic texture such as a skyrmion and its gyration around an axis to build an oscillator \cite{Zhang_2015,Garcia_Sanchez_2016}. In these structures, the skyrmion exists on an extended nano-disk where it can gyrate around the central geometric axis (Fig. \ref{fig:skno}(a)). The spin-torque pushes the skyrmion towards the boundary in this geometry while the boundaries repel them. As a result, the skyrmion moves along stable circular boundaries of the nano-disk under the spin-torque drive. By placing a small MTJ based reader that covers only the part of the nano-disk, it is possible to read out an oscillatory electrical response (Fig. \ref{fig:skno}(b)). \textcolor{black}{It is important to note that the combination of Magnus force and edge repulsion makes the device work. As the Magnus force pushes the skyrmion to the edge of the circle, the repulsive force from the edge makes it move parallel to the edge, which results in a circular motion.}

The frequency of oscillation of the skyrmionic nano-oscillator (SkNO) depends on not only the magnitude of the driving spin-torque (Fig. \ref{fig:skno}(c)) but also the skyrmion size, nano-disk radius (Fig. \ref{fig:skno}(d)), and the drive current, providing a rich phase space for oscillator design. It was also reported in \cite{Zhang_2015} that the linewidth was less than a $1$ MHz with a central frequency of nearly $800$ MHz, whereas \cite{Garcia_Sanchez_2016} reported a somewhat lower center frequency of a few 100s of MHz. Analysis from both papers indicated that skyrmion motion could be fairly robust against moderate disorders such as grain structure of the magnetic nano-disk and other imperfections such as bubbles, dots, and notches. Using additional skyrmions on the nano-disk can increase the frequency to more than 1 GHz by reading these multiple skyrmions in a single gyration \cite{Zhang_2015}. 

\section{Technology Challenges}\label{challenge}
Along the path towards a reliable low cost and energy-efficient skyrmionic device technology, several challenges need to be overcome. Depending on the target application, these can be of varying importance. For conservative applications based on shuttling skyrmions such as our aforementioned temporal memory, any nucleation of skyrmions needs to be deterministic both in position and in number. On the other hand, stochastic applications such as skyrmion reshufflers are not sensitive to the position and quantity of nucleated skyrmions, although they could be sensitive to other aspects such as the rate of nucleation.

{To start with, the energy advantages of a skyrmionic circuit need to be evaluated in totality, including the control and read circuitry. As pointed out near the introduction, a humongous chunk of the energy bottleneck with magnetic devices lies in the overhead. We saw that with our temporal memory analysis, where the energy consumed in the racetrack was $\ll 10 $ fJ, while that in the CMOS transistors used to turn the currents on and off consumed $\sim 1$ pJ. This also means we need to evaluate the compatibility of the various components --- for instance, impedance matching between the transistors and the heavy metal underlayers or the ability for these transistors to provide adequate drive current or voltage (e.g., some of the logic applications need $\sim 2$ V VCMA gating, which is more than the nominal voltage of 180 nm transistor).}

{Adequate drivability is achievable by using an electronic read unit to detect skyrmions for each logic operation. This in itself is challenging given the small footprint of a skyrmion relative to an MTJ, exacerbated in ferrimagnets. Moreover, after reading the data, an amplifier will be needed to have a reliable output resulting in extra energy and area overhead.}

{As with many emerging technologies, appropriate gate design for suitable electrostatic or anisotropy control, as well as material integration (channel, oxide, heavy metal), will need to be perfected. Drivability will require an optimized material thickness that allows long lifetimes and strong interfacial DMI and SOT coefficients. {In Fig.~\ref{SC-AND-OR}} we have assumed two-input gates, so the geometry needs to be altered for a larger number of inputs to avoid cascading becoming prohibitively costly. However, altering the device geometry may result in reliability issues. It is recommended to pay the cost penalty rather than playing with the device geometry to perform reliable logic operations. Some of the logic setups rely on precise synchronization of skyrmion arrival times, which is, of course, challenging. Finally, reliable nucleation of skyrmions will be an issue for all devices, especially ones that rely on re-nucleating skyrmions rather than reshuffling them around.
A dis-uniform defect density can annihilate one of the skyrmions in a two-lane AND-OR device (Fig \ref{SC-AND-OR}) or alter the arrival sequence in a temporal memory device (Fig.~\ref{fig:wfenergy}). This is maybe less relevant for stochastic computing applications depending on their error thresholds. For the latter, reshuffler-based applications (Fig.~\ref{SC-shuffle}), the synchronization between input stream, nucleation, and expel speeds needs to be ensured to avoid bottlenecks that can delay the overall performance by a handsome margin and hamper its reliability. There are other considerations---for instance, higher current densities might enforce higher degrees of entanglement among bitstreams.}

{We briefly summarize some of these challenges below. 
}

\begin{figure}
 \centering
    \includegraphics[width=\columnwidth]{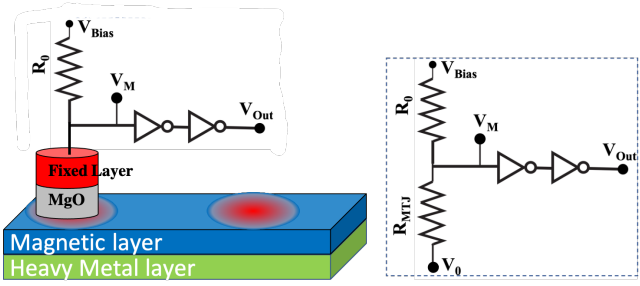}
    \caption{MTJ electrical detection unit with the equivalent circuit on the right. {The $R_\mathrm{MTJ}$ changes due to TMR, causing a swing in the $V_M$ value which can  be used for reading the data.} $V_\mathrm{Bias}$ is the applied voltage to the MTJ and $V_0$ is the racetrack voltage. {The maximum voltage swing would be for $R_0 \approx \sqrt{R_{Skm}R_P}$. Where $R_\mathrm{MTJ}\;(R_P)$ is the MTJ resistance in skyrmion (parallel) state. Typical value for $R_\mathrm{MTJ}$ would be few $k\Omega$ depending on the thickness of the insulator layer (e.g. MgO thickness of the MTJ)}. \textcolor{black}{The two inverters between $V_M$ and $V_{Out}$ act as an amplifier.}}
    \label{fig:MTJ}
\end{figure}

\textbf{Readability:} To use skyrmions or DWs in a digital circuit, we need to be able to electrically detect these magnetic excitations. The most practical method for detecting isolated skyrmions is using MTJs. When a skyrmion or DW reaches an MTJ, the corresponding tunnel magnetoresistance (TMR) generates an output voltage swing which needs to be amplified to get a 0/1 in the output node of the circuit. In order to have a large enough voltage swing without any erroneous electrical detection, the TMR needs to be quite large, $50~\%$ in our proposed race logic and reconfigurable hardware applications. Owing to the small size and chiral spin texture of skyrmions, the effective TMR is set by an overall filling factor and can be significantly smaller than the TMR for a magnetic pillar. For a skyrmion of the same size as the MTJ, {this fill factor can be as small as $30\%$ (TMR$_\mathrm{effective}=(\mathrm{fill}\;\mathrm{factor})\times \mathrm{TMR}$). The TMR can be smaller for ferrimagnets or antiferromagnets as the two sub-lattice spins can oppose each other. \textcolor{black}{TMR ratios of above 100 have been theoretically predicted for antiferromagnets \cite{ATMR_theory,AFM_spintronics}, but so far no experiment has demonstrated large TMR ratios for AFMs. The anisotropic magnetoresistance (AMR), which varies as $\sim (\bold{m \cdot j_c})^2$, $j_c$ being the applied current density, has also been investigated for canted magnetic structures such as skyrmions \cite{AFM_spintronics}. It is predicted to achieve above 100 percent magnetoresistance (MR) ratio \cite{AFM_AMR}, though the experimental demonstration of it at room temperature is yet to be achieved. Multilayer spin valve structures such as synthetic antiferromagnet (SAF) or an intermediate ferromagnetic layer \cite{multi_spinvalve, SAF_exp_cofeb} are other options. }
By tuning the bias voltage and reference resistance of the read unit (Fig. \ref{fig:MTJ}), erroneous data detection seems preventable at a somewhat low effective TMR ($\sim 50~\%$), but the circuit might not be robust enough against variations. In a synchronously clocked system, perhaps where notches quantize the racetrack in the magnetic substrate, a more sensitive detection mechanism like a Pre Charged Sense Amplified (PCSA) \cite{PCSA2009} could be adopted if indeed very low effective TMR becomes a persistent problem.}

{An alternate way to detect a skyrmion is to look for chiral signatures in a Hall measurement.}
In magnetic materials there are three components to the transverse resistivity, $\rho_{xy} = \rho_{xy}^\mathrm{OHE}+\rho_{xy}^\mathrm{AHE}+\rho_{xy}^\mathrm{THE}$. The first one, $\rho_{xy}^\mathrm{OHE}$, is the ordinary Hall effect (OHE) which is proportional to the out-of-plane magnetic field. The second one, $\rho_{xy}^\mathrm{AHE}$, is the anomalous Hall effect (AHE) which arises from spin-orbit torque interactions and occurs in systems with broken time-reversal symmetry such as ferromagnets. AHE would be proportional to the out-of-plane magnetization $m_z$. The last one, the topological Hall effect (THE), is induced by the skyrmion's chiral spin texture. In a magnetic layer with existing skyrmions, an electron traveling adiabatically will feel an emergent magnetic field arising from the spin texture of the magnetic layer, leading to a corresponding change in the phase of the electron wavefunction. This phase causes a voltage difference across the width of the racetrack. This effect, THE, is reported to be dominant in B20 skyrmion lattices, but for isolated skyrmions, the AHE becomes the dominant contribution to the transverse Hall resistivity.  
The magnitude of $\rho_{xy}$ is small for a single skyrmion, but for a larger number of skyrmions, it can have the necessary voltage swing for it to be useful in an electrical circuit. The $\Delta \rho_{xy}$ reported for a single skyrmion can be as small as $3.5\pm 0.5~n\Omega$-cm for sub 100 nm skyrmions in racetrack width of $1\;\mu m$ \cite{maccariello_electrical_2018}. For a current density of 1 TA~m$^{-2}$ and width of $1~\mu$m, the voltage swing would be 3.5 $mV$.\\

\textbf{Deterministic Nucleation/Annihilation:} {As we discussed in Sec.~II.E, there are multiple mechanisms available to nucleate and annihilate skyrmions from external magnetic fields to current pulses to defect sites with locally reduced magnetic anisotropy \cite{romming2013writing}. The challenge, however, is to have deterministic nucleation of skyrmions since the methods discussed can lead to inadvertent nucleation of multiple skyrmions~\cite{maccariello_electrical_2018,Caretta2017}. Moreover, the position of the nucleated skyrmions needs to be deterministic \cite{woo_deterministic_2018}, meaning that the pinning sites for skyrmion nucleation need to be controlled. For applications where single skyrmions are used to encode data \cite{wavefront,luo_reconfigurable_2018,Programmable_hardware_spie}, the unpredictability of location or number of skyrmions would make the device unworkable. One way around is to design the device so that it can be conservative in the number of skyrmions \cite{wavefront,joe_Friedman}, meaning no need for annihilation in the first place. The benefit of conservative devices is that the skyrmions can be pre-fabricated in a controlled environment.
It is also important to avoid added nucleation. Current pulses of magnitude above $j_c \approx10^{12}$A/m$^2$ are typically used to nucleate skyrmions, in fact, even a bit lower sometimes due to thermal effects \cite{buttner_field-free_2017, woo_deterministic_2018}. This means the operational current, $j$ of the device, needs to be well below the nucleation/annihilation limit. On the other hand, the current has to be large enough to unpin the skyrmion. For the applications where the density of skyrmions is important, for example, neuromorphic applications \cite{stochastic}, the deterministic nucleation/annihilation of individual skyrmions would influence the ratio at which they are created or deleted, and thus the overall probability of ones and zeros. This would also be sensitive to the magnitude of current density and defects.}
\\
\indent\textcolor{black}{In addition to deterministic challenges, skyrmion nucleation can be energetically costly \cite{Zhang_2020}, as the current pulse $I_n$ for nucleation needs to be larger than the driving current $I_d$. The power consumption would be $(I_n/I_d)^2$ times higher, which combined with nucleation time would make each creation/annihilation process quite energy-hungry, arguing for a conservative design that shuttles rather than creates/destroys skyrmions. Our pulse duration studies discussed in section II.E place the nucleation time at $\sim 1-10$ ns, comparable to drive time. Reference \cite{buttner_field-free_2017} uses a current pulse of $~10^{12}$A m$^{-2}$ over 6 ns for nucleation, while the drive time we estimated for race logic was under 10 ns.  Reusing pre-fabricated skyrmions would 
make  the recovery operation in race logic 1-2 orders of magnitude more energy-efficient than nucleating skyrmions each cycle.}

\textbf{Positional Stability:} {Large skyrmions generally have larger lifetimes and more diffusion, while small ones get easily pinned by defects and go through a thermally activated creep regime in transport. The amount of diffusion needs to be tightly restricted to service the needs of a given technology (Table V).}

 In order to ensure the reliability of the magnetic memory, {pinning sites need to be engineered lithographically}. The pinning energy barrier needs to be large enough to stabilize skyrmion location and, depending on the application, similar to the lifetime energy barrier, around 30 to 50 $k_BT$. Much like the lifetime, there is a trade-off between having adequate skyrmion stability (higher barriers) vs. modest drive currents that can unpin and move them efficiently without blowing up the energy budget, crashing them into the sidewalls, or otherwise annihilating them in the process. {In addition, if the driving current is too large, it can nucleate unwanted skyrmions.} In the case of domain walls, similar positional stability is needed, but unavoidable lithographic edge roughness can cause significant DW pinning in the racetrack. Controlling the amount of pinning for skyrmions might be easier to achieve as they can avoid edges, implying a distinct advantage in terms of material engineering. {Figure~\ref{fig:notch} shows simulated notch geometries that create a $\sim 40$ k$_B$T energy barrier yet allow a $\sim 50$ nm skyrmions to unpin at a modest 4~ns current pulse of $j_p = 1.4 \times 10^{11}$ A/m$^2$. This would mean the operational range of driving current has to be between $j_c$ and $j_p$. With the parameters used in Sec.~III.C, that would be $1.4 \times 10^{11}$<j<10$^{12}$A/m$^2$.} 

\textbf{Integration Challenge:} {The magnetic racetracks need to be integrated with silicon CMOS transistors that constitute the control circuitry for memory/logic applications. The magnetic racetracks we have been studying have heavy metal underlayers, whose considerably smaller resistance compared to the CMOS control circuitry limits the skyrmion drive current. Using larger transistors will reduce their resistance and allow a high drive current. Since the control circuitry consumes the bulk portion of the overall energy (Fig.~\ref{fig:full_energy}) \cite{wavefront}, the reduction of transistor resistance will reduce the overall energy consumption by balancing the impedance between the racetrack and the control circuitry and allow higher frequency operation enabled by higher skyrmion speeds, albeit with an associated area overhead for the larger transistors. }The difficulty in designing a high-speed skyrmionic device is {the large current needed to move the skyrmion at high speed}. Due to the impedance mismatching between the racetrack and CMOS control circuitry, only a tiny fraction of current is fed to the racetrack limiting the scope of supply voltage reduction.

{For conventional racetrack memories, the energy consumption ratio between the CMOS and the racetrack underlayer will depend on the number of skyrmions in the racetrack and the number of read stacks placed on the racetrack. As the applied current to the racetrack underlayer will move all the skyrmions present in the racetrack, the driving energy consumption ratio of CMOS/racetrack will be lower for a higher number of skyrmions in a racetrack. Thus, longer racetracks and a higher number of small-sized skyrmions will lower energy consumption. However, smaller skyrmions are harder to detect, and longer wires are more susceptible to noise, creating erroneous output. Thus, to achieve energy efficiency without affecting reliability (a trade-off we emphasized near the start of this article), choosing the optimal racetrack length and skyrmion size will be vital. However, it is worth mentioning that the skyrmion nucleation energy will be a deciding factor of the overall energy consumption for conventional multi-skyrmion racetrack memory.
}

{There is also an issue of current sharing between metallic magnets with the heavy metal underlayer, which can be mitigated by using insulating magnetic oxides such as spinels and ferrites. The physics gets more involved in antiferromagnetic spin dynamics and will need to be explicitly re-evaluated for skyrmion transport.}\\\\

\textbf{Speed Stability:} \textcolor{black}{As was mentioned in Section I.C skyrmions need to be fast while requiring a small drive current to guarantee energy efficiency for skyrmionic devices. However, there are two additional constraints on the range of drive current besides the energy efficiency requirements. As seen in Fig.~\ref{fig:nonlinear}, if the drive current is too large, it can distort and dissolve the skyrmions. Additionally, as discussed, to have positionally stable skyrmions, engineered pinning sites are required. The minimum unpinning current of the skyrmions would be the lower limit of the drive current flowing in an operable skyrmionic device. This, in turn, is determined by the required positional lifetime of the skyrmions. The maximum current before skyrmion is dissolved would be the higher limit. The region between these two limits is the "Goldilocks" regime for the drive current and pinning strength of defects to have a reliable skyrmionic device. Moreover, due to the distortion of the skyrmions at large current, the skyrmion speed vs. current will be nonlinear, which has to be taken into account in designing a skyrmion-based device.}

\section{Summary/Outlook}
Recent experiments have unearthed a wealth of solitonic excitations in thin magnets, from skyrmions to skyrmioniums and chiral bobbers \cite{GOBEL2020}. What makes these entities interesting is that they can be scaled down to fairly small sizes through material engineering, utilizing their topological barriers that work effectively against thermal fluctuations, especially if the sizes are not too small. Topology also covers their dynamical properties, as these excitations can work around defects by distorting their shapes at low current densities. The size scaling offers distinct advantages relating to high-density information encoding that can be challenging to accomplish with conventional magnets. Their quasi-linear current-driven motion and tunable properties would allow certain applications such as temporal racetrack memory for low-density skyrmions, while in low current thermally diffusive setups, they could act as decorrelators in stochastic computing.

{There are, however, cautionary lessons, as the topological protection is not absolute. The energy barrier for small isolated metastable skyrmions in thin films can be small, while thicker films tend to make drivability challenging. While topology also offers some dynamical resilience, it comes at the cost of added topological damping, Magnus force, and overall slowdown compared to domain walls at large current densities. Finally, device applications need to be carefully evaluated for the entire circuit, including individual overhead costs for acceptable electronic read-write-erase-recovery operations.}

{Key milestones along this path will include managing the energy cost and predictability of skyrmion nucleation, adequate TMR across MTJs, and both static (lifetime) and dynamic (positional) stability at room temperature. Key experiments along these lines will need to be executed, demonstrating projected lifetimes $\sim$ days or years for $\sim 10-20$ nm skyrmions (Fig.~\ref{fig:skm-life}), adequate positional localization (Table V) with controlled diffusion or pinning along a racetrack (Fig.~\ref{fig:notch}),  subsequent release with temporal wavefront arrival under modest current densities $\sim 10^{11}$ A/m$^2$ over $< 10$ ns, as well as successful read (effective TMR including skyrmion shape function $> 50\%$), recovery and reset (erase), in order to capitalize on the intrinsic strengths of these topologically protected computing bits.}
\section{Acknowledgments}
This work is funded by the DARPA Topological Excitations in Electronics (TEE) program (grant D18AP00009). We acknowledge useful discussions with Mark Stiles, Advait Madhavan, Matthew Daniels, Brian Hoskins, Joe Friedman, Jayasimha Atulasimha, Hans Nembach, Kyung-jin Lee, Vivek Amin, and Mohit Randeria.
\section{Data Availability}
The data that support the plots within this paper and other findings of this study are available from the corresponding author on reasonable request.
\section{Appendix}
\appendix
\section{Skyrmion topology and Poincar\'{e} map classification}
{
\textcolor{black}{Alternative to the energy arguments in the text the skyrmion winding number $N_{sk}$ and domain angle $\psi$, can be derived from analysis of the Poincar\'{e} maps for DMI vector field.} $N_{sk}$ sets both the exchange $E_{ex}$ and the symmetry breaking DMI $E_\mathrm{DMI}$ energy barriers (Eq.~\ref{eneqns}) that generate the requisite spin winding against the background spin distribution.
A convenient way to see the topological classifications of skyrmions arising from different forms of the DMI is by plotting a 2-D cross sectional cut in the x-y plane. The different skyrmionic forms map onto the well known classifications of linear systems (Fig.~\ref{fig:top}),  summarized by stable solutions to the 2-D differential equation set
$\displaystyle\frac{d}{dt}\left(\begin{matrix}x \\ y\end{matrix}\right) = A\left(\begin{matrix}x \\ y\end{matrix}\right)$
where $t$ is a parameter that takes us along a solution curve. The shapes of the curves are
set by the Trace (tr) and Determinant (det) of the matrix A and can be plotted as so-called Poincar\'{e} maps \cite{goldstein} --- for instance, stars (tr$^2 = 4$ det $> 0$) that map onto N\'{e}el skyrmions, limit cycles (det $> 0$, tr $=0$) that map onto Bloch skyrmions, spirals ($0 < $ tr$^2 < 4$ det) that map onto a combination of N\'{e}el and Bloch, and saddle points (det $< 0$) that map onto antiskyrmions.  

For skyrmions stabilized by DMI set by crystal structure (Fig.~\ref{fig:xals}), the symmetry breaking energy connecting two spins $\boldsymbol{S}_{i,j}$ takes the form
\begin{eqnarray}
    E_\mathrm{DMI} &=& \boldsymbol{D}.(\boldsymbol{S}_i\times\boldsymbol{S}_j), \text{with} \\
    \label{EDMI}
    \boldsymbol{D}&=&\begin{cases}
    \hat{z}\times \boldsymbol{r}_{ij}~~ \text{Interfacial}\\
    \boldsymbol{r}_{ij}~~ \text{B20}\\
    \sigma_z.\boldsymbol{r}_{ij}~~\text{$D_{2d}$}
    \end{cases}
\end{eqnarray}
The first is set by a symmetry breaking field perpendicular to the interface between a heavy nonmagnetic metal and a magnetic film, the second is aligned within a non-centrosymmetric magnet between the spin carrying atoms, and the third by the elongation of the magnetic unit cell along with the $\hat{y}$-direction. 
The vector field for the $\boldsymbol{D}$ can then be rewritten in 2-D as:
\begin{eqnarray}
\boldsymbol{D}=    
    \begin{cases}
     \pm \begin{pmatrix}
    -y\\
    x
    \end{pmatrix} & \text{Interfacial, ~N\'{e}el~skyrmions}\\
     \pm \begin{pmatrix}
     x\\
     y
    \end{pmatrix} & \text{B20, ~Bloch~skyrmions}\\
     \pm \begin{pmatrix}
     x\\
     -y
    \end{pmatrix} & \text{$D_{2d}$, ~antiskyrmion}
    \end{cases} 
\end{eqnarray}
From the $E_\mathrm{DMI}$ in order to get lowest DMI energy, the vector field of each $\boldsymbol{S}$ must be in a plane perpendicular to $\boldsymbol{D}$ so that $\boldsymbol{S}_i\times\boldsymbol{S}_j$ is parallel/antiparallel to 
$\boldsymbol{D}_{\bot }$, the vector perpendicular to the DMI vector $\boldsymbol{D}$:
\begin{equation}
\frac{d}{dt}\left(\begin{matrix}x \\ y\end{matrix}\right) = \boldsymbol{D}_{\bot }=    
    \begin{cases}
     \pm \begin{pmatrix}
    x\\
    y
    \end{pmatrix} & \text{N\'{e}el}\\
     \pm \begin{pmatrix}
    -y\\
     x
    \end{pmatrix} & \text{Bloch}\\
     \pm \begin{pmatrix}
    y\\
     x
    \end{pmatrix} & \text{antiskyrmion}
    \end{cases} 
\end{equation}
This would give A in our linear systems classification as:
\begin{equation}
A =    
    \begin{cases}
     \pm \begin{pmatrix}
    1 & 0\\
    0 & 1
    \end{pmatrix} & \text{N\'{e}el}\\
     \pm \begin{pmatrix}
    0 & -1\\
    1 & 0
    \end{pmatrix} & \text{Bloch}\\
     \pm \begin{pmatrix}
    0 & 1\\
    1 & 0
    \end{pmatrix} & \text{antiskyrmion}
    \end{cases} 
\end{equation}
{whose traces and determinants satisfy the 2-D Poincar\'{e} map classification rules above.}

We can quantify the winding of spins around the skyrmion with the winding number $N_{sk}$ introduced earlier (Eq.~\ref{eqnsk}). An electron moving around a closed loop on the Fermi surface picks up a geometrical Berry phase
related to the solid angle subtended at the center of the Fermi sphere, similar to the precession of a Foucault pendulum. This phase
is given by the flux of the magnetic field in k-space, $\phi_n = \oint \vec{B}_k\cdot d\vec{S}_k, ~~\vec{B}_k =
\vec{\nabla}_k \times \vec{A}_k, ~~\vec{A}_k = iu^*_k\vec{\nabla}_ku_k$, where $u_k$ is the Bloch part of the electron wavefunction. For magnets, {the corresponding phase space} $k \rightarrow \boldsymbol{m}$ where $\boldsymbol{m}$ is the normalized magnetic moment. We use a standard $2\times 1$ spinor representation $u_{\boldsymbol{m}} = \left(\begin{array}{c}\exp{[-i\Psi/2]}\cos{\theta/2} \\ \exp{[i\Psi/2]}\sin{\theta/2} \end{array}\right)$. The corresponding magnetic field then maps onto a magnetic monopole while the Chern number is set by the quantum of the Berry flux, $C_n = \phi_n/2\pi$. In 2-D $C_n$ simplifies to $N_{sk}$.

}
\section{Skyrmion Energy terms}
\begin{figure}
    \includegraphics[width=\columnwidth]{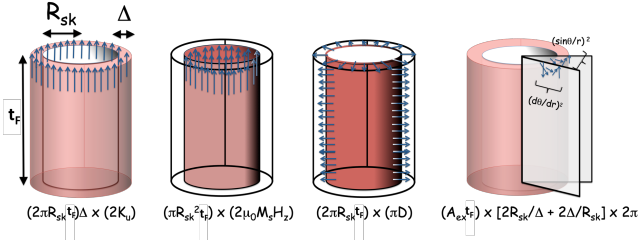}
    \caption{A schematic view of skyrmion as a cylinder.}
    \label{fig:enterms}
\end{figure}
In a 1-D continuum representation, the width of a domain wall arises from a competition between exchange stiffness $A_{ex}$ and uniaxial anisotropy $K_u$. Taking into account only the first two energy terms we get the energy per unit area:
\begin{equation}
E^\prime = t_F\int_{-\infty}^\infty dx[A_\mathrm{ex}(d\theta/dx)^2 - K_u\cos^{2}{\theta}].
\label{eq:energy}
\end{equation}
Minimizing this energy, i.e., invoking Euler-Lagrange's equations, leads to 
the differential equation
\begin{equation}
\frac{d^2\theta}{dx^2} = \frac{\sin{\theta}\cos{\theta}}{\Delta_0^2},
\end{equation}
where $\Delta_0 = \sqrt{A_\mathrm{ex}/K_u}$ is the domain wall width (exchange wants to spread the spins around, uniaxial anisotropy wants to align them). Solving this equation
with boundary conditions $d\theta/dx|_{x=\infty} = 0$ gives us the equation for a 1-D domain wall as $\theta(x) = 2\tan^{-1}\left[e^{\displaystyle (x-x_0)/\Delta_0}\right]$ spread over a thickness $\Delta_0$ around the position $x_0$. 
We can then approximate the various integrals per unit area as
\begin{eqnarray}
E^\prime_\mathrm{ex} &=& t_F\int_{-\infty}^\infty dxA_{ex}(d\theta/dx)^2 = 2A_{ex}/\Delta_0,\nonumber\\
E^\prime_\mathrm{ani} &=&  t_F K_u\int_{-\infty}^\infty dx\sin^2{\theta} = 2K_u\Delta_0,
\end{eqnarray}
where we  used the differential equation above to replace one $d\theta/dx = \sin{\theta}/\Delta_0$ term in each of the first two integrals. We can also calculate the 1-D equivalent of a chiral DMI energy as
\begin{equation}
    E^\prime_\mathrm{DMI} = -t_F \sin{\psi_0} D\int_{-\infty}^\infty dx d\theta/dx =-\pi t_F D \sin{\psi_0}
\end{equation}

Where $\psi_0$ is the domain wall angle, $\psi_0 = 0$ is a Bloch domain wall, and $\psi_0 = \pi/2$ is a N\'eel domain wall.
The full 3-D form of Eq.~\ref{eq:energy} with DMI corrections should be taken into account to describe 2-D skyrmions. Writing $\mathbf{m}$ in spherical cooridnates, $\mathbf{m} = \left(\sin\theta\cos\psi,\sin\theta\sin\psi, \cos\theta\right)$ and using  Euler-Lagrange's equation, the solutions for $\theta(r)$ can be {approximated as two opposing domain walls at antipodal locations}:
\begin{equation}
    \theta(r) = 2 \tan{}^{-1}\left[\displaystyle\frac{\sinh{(R_{sk}/\Delta)}}{\sinh{ (r/\Delta)}}\right],
\end{equation}
where $R_{sk}$ is the radius of the core. Substituting this form, we can evaluate all the energy integrals \cite{Thiaville,bogdanov2001chiral,Buttner2018}:
\begin{eqnarray}
E_\mathrm{ex} &=& (2\pi A_{ex}t_F)\left(\frac{2R_{sk}}{\Delta} + \frac{2\Delta}{R_{sk}}N_{sk}^2\right)f_\mathrm{ex}(\rho)\nonumber\\
E_\mathrm{DMI} &=& -\left(2\pi R_{sk}t_F\right)\pi Df_\mathrm{DMI}(\rho)\nonumber\\
E_\mathrm{ani} &=& \left(4\pi K_u t_F\right)R_{sk}\Delta f_\mathrm{ani}(\rho)\nonumber\\
E_\mathrm{Zeeman} &= & \left(\pi R_{sk}^2t_F\right)\times\left(2\mu_0M_SH_z\right)f_\mathrm{Zeeman}(\rho)\nonumber\\
E_\mathrm{demag} &=& -\left(2\pi \mu_0 M_s^2 t_F\right)R_{sk}\Delta\times\nonumber\\ &&(f_\mathrm{s}(\rho,t_{FM}/\Delta)-\cos{\psi}^2f_\mathrm{v}(\rho,t_{FM}/\Delta))\nonumber\\
{\rm{with}}~~~~D  &\equiv& \Bigl(D_\mathrm{int}\cos{\psi}-D_\mathrm{bulk}\sin{\psi}\Bigr),
\label{eneqns}
\end{eqnarray}
where $\psi$ is once again the domain tilt angle. $f_\mathrm{s}(\rho,t_{FM}/\Delta)$ and $f_\mathrm{s}(\rho,t_{FM}/\Delta)$ are surface and volume  demagnetization form factors respectively. For ultra thin film limit, $t_{FM}\ll \Delta,R_{skm}$, $f_\mathrm{s}\simeq f_\mathrm{ani}$ and $f_\mathrm{v}$ can be ignored\cite{wang_theory_2018}. The form factors for small size, obtained by fitting numerical simulations, are given by
\begin{eqnarray}
f_\mathrm{ex}(\rho) &\approx& \left[1+1.93\frac{\rho\left(\rho-0.65\right)}{\rho^2+1}e^{\displaystyle -1.48(\rho-0.65)
}\right]\nonumber\\
f_\mathrm{ani}(\rho) &\approx&  \left[1-\frac{1}{6\rho}e^{\displaystyle -\rho/\sqrt{2}
}  \right]
\nonumber\\
f_\mathrm{DMI}(\rho) &\approx& \left[N_{sk}+\frac{1}{2\pi \rho}e^{-\displaystyle\rho} \right]\nonumber\\
f_\mathrm{Zeeman}(\rho) &\approx& \left[1+\frac{\pi^2}{12\rho^2}-\frac{0.42}{\rho^2}e^{-\displaystyle \rho^2}\right]
\label{eneqnsfit}
\end{eqnarray}
\begin{table*}[t!]
\centering
        \caption{Heuslers with low damping and $M_s$ for skyrmions. Reproduced with permission from IEEE Transactions on Magnetics, vol. 56, no. 7, pp. 1-8, 2020 \cite{xie2020computational}. Copyright 2018 IEEE.}
        \input{Heusler_table2}
        \label{tab:Inverse_Heusler_HalfMetal}%
\end{table*}
We have deliberately separated out the large radius expressions, which are easy to understand intuitively (see Fig.~\ref{fig:enterms}), times dimensionless form factors that account for the small skyrmion adjustment. Note that even in the large skyrmion expression we get an extra exchange term $\propto N_{sk}{^2}\Delta/R_{sk}$ relative to a 1-D domain wall, that arises exclusively from the angular winding term $N_{sk}\langle \sin^2{\theta}\rangle/r^2$, i.e. the 2-D curvature, integrated between $R_{sk}\pm \Delta$, and replacing once again $\sin{\theta} \approx \Delta (d\theta/dr)$ when integrating along the radial coordinate. 
\\
Collecting terms, we see that for large skyrmions with unit vorticity number $N_{sk} = 1$, we 
get $E_{ex} \sim 4\pi  A_{ex}t_F(R_{sk}/\Delta + \Delta/R_{sk})$, $E_{ani} \sim t_F K_uR_{sk}$ and $E_{DMI} \sim - \pi t_FD(R_{sk}+\Delta)$. 

\section{Dynamics of Skyrmion and DWs}
{We now deal with the wall coordinate $X$ and the domain wall angle $\psi$, setting the skyrmion spin orientation within the rigid skyrmion (Thiele) approximation $\theta(t) = 2\arctan{\left[\exp{(x-X(t))/\Delta_0}\right]}$. Writing out the LLG equation, and integrating once again over the domain wall coordinate, we get the coupled equations 
\begin{align}
\frac{d\psi}{dt}+\frac{\alpha_\mathrm{eff}}{\Delta_0}\frac{dX}{dt} &= \gamma_\mathrm{eff}H_a, \\
\frac{dX}{dt}-\Delta_0\alpha_\mathrm{eff}\frac{d\psi}{dt} &= \Delta_0\gamma_\mathrm{eff}(H_D\sin{\psi}-H_k\sin{\psi}\cos{\psi}),
\end{align}
where for a two sublattice system with same thickness $t_F$, $H_a = {\pi\hbar\theta_{SH}j_\mathrm{HM}}\cos{\psi}/{4q\mu_0M_\mathrm{eff}t_F}$, $M_\mathrm{eff} = M_1(T)-M_2(T)$, $\gamma_\mathrm{eff} = M_\mathrm{eff}/(S_1-S_2)$, $\alpha_\mathrm{eff}=\alpha(S_1+S_2)/(S_1-S_2)$, with $S_i=M_i/\gamma_i$ for the two sub-lattice system. Solving for ${dX}/{dt} \equiv   \boldsymbol{v}_{DW}$ with ${d\psi}/{dt} = 0$ (no precessional motion or Walker breakdown) and $H_D \gg H_K$ we get the equation~\ref{eqdw} in main text:}
\begin{equation}
    v_\mathrm{DW} = \frac{\pi}{2}\frac{D_{int}j_\mathrm{HM}}{\sqrt{(S_0(T)j_{hm})^2+(\alpha S(T)j_0)^2}}.
    \label{eqdw2}
\end{equation}
Note that with AFMs where $S = 0$, it is theoretically possible for a DW to reach very high speeds. {In Eq.~\ref{eqdw} ${v}_\mathrm{DW} \propto j_\mathrm{HM}/\alpha$ at high currents and low damping so that in an antiferromagnet the speed would become very large.} At that point, the broader assumption of skyrmion or DW rigidity will no longer hold and the structures start to contract, assuming they are still stable energetically. For speeds $v'_\mathrm{DW}$ comparable to magnon speeds $v_m$, a Lorentz contraction of domain wall width happens, $\Delta'_0 = \Delta_0\sqrt{1-(v'_\mathrm{DW}/v_\mathrm{m})^2}$ \cite{Haldane}. Substituting in Eqs. 14 and 15, the speeds become \cite{domainwallrelativistic}:

\begin{align}
{v'}_\mathrm{DW} &= {v}_\mathrm{DW}\sqrt{1-(v'_\mathrm{DW}/v_\mathrm{m})^2},
\end{align}
where ${v}_\mathrm{DW}$ is the non relativistic DW speed. {Similarly for AFM skyrmions the speed is ${v}_{sk} \propto j_\mathrm{HM}/\alpha$ high current low damping would be very large. At speeds near magnon speeds \cite{skyrmionrelativistic} we have to revisit our assumption of circular skyrmions and assume an elliptical shape with the two diameters functionally dependent on the skyrmion speed.} We can derive the relativistic corrections using staggered coordinates, set by the N\'{e}el vector $\boldsymbol{n\equiv (m_1-m_2)/|m_1-m_2|}$:\begin{align}
n &= (\sin{\theta}\cos{\psi} \sin{\theta}\sin{\psi},\cos{\theta})\nonumber\\
\theta &= 4\tan^{-1}[exp(\sqrt{X^2+Y^2})]\nonumber\\
\psi &= \tan^{-1}(Y/X)\nonumber\\
Y &= y/D_y\nonumber\\
X &= (x-x(t))/D_x
\label{eqneel}\end{align}$D_x$ and $D_y$ are the majority and minority axial lengths for a skyrmion ellipse. {We can break the forces acting on a skyrmion into two categories. Conservative forces derive from a potential gradient and give zero net work when integrated around a closed path ($\oint \boldsymbol{\nabla}{U}\cdot d\boldsymbol{l} = 0$). In contrast, non-conservative forces such as spin torque and damping can siphon energy in or out of the system all along the closed path and do not sum out to zero. The former is described by a Lagrangian $L$, the difference between kinetic and potential energy, while the latter is describe by a Rayleigh dissipation density $R_0$. In staggered coordinates, we write them as:}
\begin{align}
   L &= \frac{a_{m}}{2}\dot{\boldsymbol{n}}^2-\frac{A_\mathrm{AFM}}{2}\Sigma\partial_i\boldsymbol{n}^2+K/2(n_z)^2-\nonumber\\
   &\frac{D_\mathrm{int}}{2}\boldsymbol{n.[(\hat{z}\times\nabla)\times n )]}\\
   R_0 &= \alpha\frac{M_s}{\gamma} \dot{\boldsymbol{n}}+\tau_{sh}(\hat{y}\times \boldsymbol{n})
\end{align}
where $a_{m} = (M_s/\gamma)^2 \chi$,  $\chi$ is the magnetic susceptibility, $A_\mathrm{AFM}$ is the exchange stiffness, $K$ the uniaxial anisotropy and $D_\mathrm{int}$ the interfacial DMI. {$\tau_{sh}$ is the coefficient of the spin orbit torque for current propagating in the $\hat{x}$ direction defined as: $\tau_{sh} = \gamma a/t_{F}=\displaystyle\frac{\gamma\hbar}{2q M_s t_F}\theta_{SH}j_\mathrm{HM}$}.
\\
Solving Euler-Lagrange equation including Rayleigh dissipation, $\displaystyle\frac{\delta L}{\delta q}-\frac{d}{dt}\frac{\delta L}{\delta \dot{q}}+\frac{\delta R_0}{\delta \dot{q}}=0$ where the generalized coordinate set $q\in \{D_x,D_y,x(t)\}$, and using Eqs.~ \ref{eqneel} we get:
\begin{align}
    D_x &= \frac{D_0}{2}\Biggl[1+\sqrt{1-(v_{sk}/v_m)^2}\Biggr]\\
    D_y &= \frac{D_0}{2}\Biggl[1+\frac{1}{\sqrt{1-(v_{sk}/v_m)^2}}\Biggr]
\end{align}
where $D_0 = 2R_{sk}$ is the diameter of a circular skyrmion. We can see that as $v_{sk}$ increases, $D_x$ reduces while $D_y$ increases, and the skyrmion gets more elliptical. For SOT driven skyrmion we get:
\begin{align}
    v'_{sk}=v_{sk}\Biggl[\frac{1+\sqrt{1-(v'_{sk}/v_m)^2}}{2}\Biggr]
\end{align}
{The end result is that distortion limits relativistic skyrmions in antiferromagnets to the same bounds that limit 1-D domain walls.}
\bibliography{bibfile} 
\end{document}

%% file: Heusler_table2.tex
 \begin{tabular}{c|c c c c c c c c|c c c c c c}
    \toprule
              & \multicolumn{8}{c|}{Cubic phase}  & \multicolumn{2}{c}{Tetragonal phase}   \\ 
      \cline{2-11}
      $\mathrm{XYZ}$   & $a_\mathrm{Cal}$ & $a_\mathrm{Exp}$  & $M_{S}$ & $\Delta E_{\rm HD}$ & $T_{N}(\mathrm{Exp})$ &$T_{N}(\mathrm{Cal})$ & $\alpha$   &Electronic & $a, c$ & $\Delta E$  \\
      & \multicolumn{2}{c}{$(\mathring{A})$}  & (emu/cc) & (eV/atom) &   \multicolumn{2}{c}{(K)} & ($10^{-3}$)   &  ground state  & $(\mathring{A})$ & (eV/atom)    \\
    \hline
    \midrule
    Mn$_{2}$CoAl    & 5.735  &5.798\cite{PhysRevLett.110.100401}   & 393.5  & 0.036 & 720\cite{PhysRevLett.110.100401} & 845 & 4.04 &   HM & 3.76, 6.68 & -0.05\\
     \hline
    Mn$_{2}$CoGa    & 5.76  &5.86\cite{PhysRevB.77.014424}   &  389.1  & 0 & 740\cite{UMETSU2015890} & 770 & 2.18 &   NHM & 3.71, 7.13 & -0.0103 \\
     \hline
    Mn$_{2}$CoSi    & 5.621  &    &  627.3  & 0.018 &  & 460 & 3.01 &   HM & & \\
     \hline
    Mn$_{2}$CoGe    & 5.75  & 5.80\cite{PhysRevB.77.014424}   & 590.6   & 0.03 &  & 471& 4.97 &   HM & 3.75, 6.84 &  0.0144\\
     \hline
   Mn$_{2}$FeAl    & 5.75  &    & 195.3   & 0.008 &  & 380 & 8.14&   NHM & 3.67, 7.28 & 0.0026\\
     \hline
    Mn$_{2}$FeGa    & 5.79  &  &  198.5  & 0.018 &  & 496 & 7.37 &   NHM & 3.68, 7.29 & 0.0331 \\
     \hline
    Mn$_{2}$FeSi    & 5.60  &    &  424 & 0    &  & 71 &3.98 &   NHM & 3.56, 7.26 & -0.071 \\
     \hline
    Mn$_{2}$FeGe    & 5.72  &    & 399.1   & 0    &  & 210 & 3.04& NHM & 3.62, 7.45 & -0.0164\\
     \hline
   Mn$_{2}$CuAl   & 5.89  &   & 33.4  & 0.042 &  & 1145 & 1.84 &      Metal & & \\
    \hline
    Mn$_{2}$CuGa    & 5.937  &   & 57.86  &  0.0208 &  & 1242 & 1.59 &      Metal & & \\
     \hline
    MnCrAs    & 5.51  &  &0     & 0.083    &  & 985 & 1.23 &  HM & & \\
    \bottomrule
    \end{tabular}%